\documentclass[12pt]{article}
\usepackage{amssymb}
\usepackage{color}
\usepackage{graphicx,psfrag}
\usepackage{amsmath}
\usepackage{hyperref}
\usepackage{mathrsfs}
\usepackage{booktabs}
\usepackage{colortbl}
\usepackage{amsfonts}

\newcommand{\uv}{_{\mathrm{UV}}}
 
 \newcommand{\oru}[2]{#1^{(#2)}}

\newcommand{\dd}{\mathrm{d}}
\newcommand{\hf}{{1\over 2}}

\newcommand{\s}{{\star}}
\newcommand{\enn}{\tilde{{\star}}_n}

\newcommand{\E}{\Delta_K}
\newcommand{\n}{\nabla}

\begin{document}

\numberwithin{equation}{section}

\begin{titlepage}

\setcounter{page}{1} \baselineskip=15.5pt \thispagestyle{empty}

\bigskip\
\begin{center}
{\Large\bf A Toolkit for Perturbing Flux Compactifications}
\vskip 15pt
\end{center}
\vspace{0.5cm}
\begin{center}
{\large Sohang Gandhi,${}^1$ Liam McAllister,${}^1$ and Stefan Sj\"{o}rs${}^2$}
\end{center}

\vspace{0.1cm}
\begin{center}

\vskip 4pt
\textsl{${}^1$ Department of Physics, Cornell University,
Ithaca, NY 14853 USA}\\
\vskip 4pt
\textsl{${}^2$Oskar Klein Center for Cosmoparticle Physics and
Department of Physics,\\ Stockholm University, Albanova SE-106 91 Stockholm, Sweden}
\end{center} %\vfil
%\vspace{0.8cm}
%\vspace{1.2cm}
{\small  \noindent  \\[0.2cm]
\noindent
We develop a perturbative expansion scheme for solving general boundary value problems in a broad class of type IIB flux compactifications.
The background solution is any conformally Calabi-Yau compactification with imaginary self-dual (ISD) fluxes.
Upon expanding in small deviations from the ISD solution, the equations of motion simplify dramatically: we find a simple basis in which the $n$-th order equations take a triangular form.
This structure implies that the system can be solved iteratively whenever the individual, uncoupled equations can be solved.
We go on to demonstrate the solution of the system for a general warped Calabi-Yau cone: % Liam
we present an algorithm that yields an explicit Green's function solution for all the supergravity fields, to any desired order,
in terms of the harmonic functions on the base of the cone.
Our results provide a systematic procedure for obtaining the corrections to a warped throat geometry induced by attachment to a compact bulk.
We also present a simple method for determining the sizes of physical effects mediated through warped geometries.
}

 \vspace{0.3cm}

\vspace{0.6cm}

\vfil
\begin{flushleft}
\small \today
\end{flushleft}
\end{titlepage}

\newpage
\tableofcontents
\newpage

\section{Introduction}
\label{sec:intro}

Flux compactifications of type IIB string theory provide a promising framework for phenomenological and cosmological models in string theory, but the study of general compact spaces remains difficult.  Warped throat regions, which arise naturally in this setting, are comparatively tractable: a throat region can be approximated by a
% May 31 2 PM
portion of a noncompact warped cone, and explicit computations performed in the local model then serve to characterize the corresponding sector of the four-dimensional effective theory.

A significant challenge in this context is that the best-understood warped throat solutions, such as the Klebanov-Strassler throat \cite{KS}, are noncompact and supersymmetric,  while realistic model-building with dynamical four-dimensional gravity requires a finite throat region subject to supersymmetry breaking.  It is therefore important to understand finite, non-supersymmetric warped throat regions of flux compactifications with stabilized moduli.

To first approximation, a finite warped throat can be replaced by a finite segment of a noncompact warped cone, terminating in the ultraviolet (UV) at some finite value of the radial coordinate, $r=r_{UV}$, where the throat is glued into a compact space.  We seek here to understand corrections to this approximation generated by  compactification.  From the viewpoint of the supergravity fields in the throat, the properties of the bulk space
% Including the configurations of D-branes, orientifold planes, fluxes, and quantum effects in the bulk
determine boundary conditions on the gluing surface, or UV brane.  For a given compact space, one could in principle pursue a solution for the throat fields with the corresponding boundary conditions, in a perturbation expansion around the solution obtained in the noncompact limit that decouples the bulk sources.  A significant simplification is that the solution in a region at radial location $r_{\star} \ll r_{UV}$ is accurately described by the finite set of modes that diminish least rapidly towards the infrared (IR).  In the dual field theory, this is just the statement that in the deep IR, a description in terms of the handful of most relevant operators is sufficient. However, even after making use of this radial expansion, the equations of motion are coupled in a complicated way,
% May 31
making an analytic solution impractical in general.

Our starting point is the observation that in an interesting class of compactifications, an additional expansion is available.  In the scenarios \cite{KKLT,BBCQ,Braun} for K\"ahler moduli stabilization, the solution is nearly conformally Calabi-Yau, with fluxes that are nearly imaginary self-dual (ISD).  We can therefore formulate a double perturbation expansion whose small parameters are
%the ratio
$r_{\star}/r_{UV}$,
%of the radial location $r_{\star}$ of the region of interest and the radial location $r_{UV}$ of the UV brane,
and the size of the deviations on the UV brane from the ISD, conformally Calabi-Yau solution.  For brevity we will refer to these as the radial expansion and the ISD expansion.

Upon expanding the equations of motion to any order $n$ in the ISD expansion,  we find a very  convenient structure that allows us to disentangle and solve the equations for the various supergravity fields.
To understand this structure, consider the much simpler model of $k$ scalar functions $\varphi_{A}$, $A=1,\ldots k$, of a single variable $r$, obeying a general first-order system of equations.   On general grounds the equations of motion for $n$-th order perturbations $\varphi_{A}^{(n)}$ around some chosen background $\varphi_{A}=\varphi_{A}^{(0)}(r)$ take the form
\begin{equation}
\frac{\rm{d}}{\rm{dr}}\varphi_{A}^{(n)}= N_{A}^{~B} \varphi_{B}^{(n)} + {\cal{S}}_A\,,\label{ODEs}
\end{equation}
where the  matrix $N_{A}^{~B}$ depends on the
% May 31
fields $\varphi_{C}^{(0)}(r)$, and the source term ${\cal{S}}_A$ depends on the fields $\varphi_{C}^{(m)}(r)$, $m<n$, $C=1,\ldots k$.  If the coefficient matrix $N_{A}^{~B}$ were constant, one could readily solve this system by standard techniques, whereas for $N_{A}^{~B}=N_{A}^{~B}(r)$ an analytic solution generally requires that $N_{A}^{~B}$ has some special structure.

In particular, if  $N_{A}^{~B}$ is {\it{triangular}},  i.e.\ if $N_{A}^{~B} =0~\rm{for}~ A<B$, then the equations of motion can be solved iteratively,
% May 31
as we shall explain at length.
For constant $N_{A}^{~B}$, finding a basis in which equation (\ref{ODEs}) is in triangular form is an easy exercise in linear algebra, but
the presence of the derivative operator makes this task highly nontrivial when $N_{A}^{~B}$ is nonconstant.
In fact, the problem of finding a basis in which a
given $N_{A}^{~B}(r)$ takes a triangular form involves solving a system of coupled differential equations that is no easier, in general, than the original system.

A key result of this paper is a simple basis in which the supergravity equations of motion expanded to $n$-th order around an ISD background take a triangular\footnote{In fact, we find that the equations  take a {\it{strictly}} triangular form, analogous to $N_{A}^{~B} =0~\rm{for}~ A\le B$.} form, allowing us to construct an iterative Green's function solution.  In contrast to the toy model above, the fields are not all scalars, and are governed by second-order partial differential equations (i.e., the fields have nontrivial dependence on the angular directions of the cone), but the nature of the simplification is identical.  At each order $n$ in perturbation theory, a privileged field $\varphi_{1}^{(n)}$ at the top of the triangle is sourced by no other fields at order $n$, so that a Green's function solution is straightforward.  The next field $\varphi_{2}^{(n)}$ is sourced only by $\varphi_{1}^{(n)}$, while $\varphi_{3}^{(n)}$ is sourced by $\varphi_{1}^{(n)}$ and $\varphi_{2}^{(n)}$, etc.  Thus, we can solve each successive equation by substituting the solutions from the preceding equations in the triangle.  The same Green's functions apply at every order, so that one need only solve for a single set of Green's functions, one for each field, and then the solutions to the supergravity equations are readily obtained to any desired order in a purely algebraic way.
We
% May 31  remark
stress that the triangular structure that plays a central role in this work appears in the equations of motion expanded around {\it{any}} ISD background, which need not be a warped Calabi-Yau cone
% Liam
(and need not be supersymmetric).  We focus on cones because the explicit metric and  separable structure of the cone  permit direct solution of the equations of motion.

In this work we explain this approach in detail, then determine all necessary Green's functions, so that the enterprising reader can obtain the supergravity solution for a general warped Calabi-Yau cone
attached to a flux compactification, to any desired order.
%\footnote{For other stabilization scenarios in type IIB string theory \cite{BBCQ, Braun}, our results apply given certain assumptions about the scale of supersymmetry breaking, cf. \S?}, to any desired order.
In practice, we give supergravity solutions as functions of the  angular harmonics on the Sasaki-Einstein base of the cone, with radial scalings determined by the corresponding eigenvalues.
For the case of $T^{1,1}$, the necessary eigenvalues and eigenfunctions are available in the literature; to use our method for a more general cone, one would need to compute the angular harmonics on the base.

A related approach was used in \cite{Holographic, LiamsLong} to study the inflationary model of \cite{KKLMMT}, which involves the
attraction of a D3-brane toward an anti-D3-brane in a warped throat.  However, the works \cite{Holographic, LiamsLong} made extensive use of the facts that a D3-brane couples
% May 31 2 PM to only
only to a particular scalar combination of the supergravity fields,
denoted by $\Phi_-$, and that the dominant source for $\Phi_-$ is imaginary anti-self-dual (IASD) flux $G_-$.   Thus, it was possible to restrict attention to the fields $\Phi_-$ and $G_-$, and to truncate at quadratic order.
In this work we fully complete this program for all supergravity fields, to all orders, permitting a much broader range of applications.

% May 31 2 PM
We remark that a similar structure in the equations of motion for global symmetry singlet perturbations linearized around the Klebanov-Strassler background was identified in \cite{BG} and played a role e.g.\ in \cite{BGH, Tolya}.  In contrast to those works, we establish and utilize a triangular structure to all orders, in expansion around a general ISD background.
Our explicit results and separable solutions are not restricted to the singlet sector, but apply only in the approximately-conformal region above the tip of a warped Calabi-Yau cone, whereas
the formulation of \cite{BG} applies throughout the deformed conifold.
% May 31 2 PM

Another useful result of this work is a simple formula for the radial
% May 31 2 PM scalings
scaling (i.e., parametric dependence on $r_{\star}/r_{UV}$) of a general $n$-th order correction.  In a canonical basis, the
$n$-th order corrections at some point in the throat have the same
% May 31  scaling
scalings as the $n$-th order products of the harmonic modes at that point.  In particular, this implies that the `running' sizes of the harmonic modes are faithful expansion parameters.  We anticipate that our formula for the scaling of a general perturbation will be of use in determining the parametric sizes of physical effects mediated through warped geometries.

Although KKLT compactifications provide significant motivation for the geometries described herein, our approach applies more broadly, to type IIB compactifications subject to controllably small violations of the ISD conditions.
In this connection, we remark that one might naively expect that all modes of the supergravity fields have coefficients of order unity at $r = r\uv$, where the throat merges into the bulk.  Then, for a sufficiently long throat, any
% May 31 3 PM
relevant modes will grow exponentially large, and the throat geometry will be destroyed in the IR.  We will find instead that, for a class of throats of broad interest, all relevant modes either violate the ISD conditions or violate the supersymmetry of the background throat geometry.  In particular, we will show that in the concrete example of a Klebanov-Strassler throat in a KKLT compactification, all relevant modes remain perturbatively small all the way to the tip of the throat.  Extending this result to more general throats in more general nearly-ISD compactifications is an interesting direction for the future.

Although we give detailed results for perturbations induced by boundary conditions on the UV brane, corresponding to sources such as D-branes, orientifold planes, fluxes, and quantum effects in the bulk, our methods apply equally well
% May 31  in
to the study of perturbations induced in the infrared.

This paper is organized as follows.  In \S\ref{sec:setup} we explain our expansion scheme in detail, and then expand the equations of motion of type IIB supergravity.  We then present our method: we show that upon obtaining the homogeneous solutions for all supergravity fields, as well as all the associated Green's functions, it is straightforward to write down the inhomogeneous solution for any field of interest, to any desired order.
In \S\ref{sec:harmonic} we summarize the homogeneous solutions for each field, deferring details to Appendix \ref{sec:appendix}.  In \S\ref{sec:Greens} we write down formal Green's function solutions for arbitrary fields. In
%\S\ref{sec:validity} we carefully justify the validity of our expansion scheme.  In
\S\ref{sec:scalings} we obtain the radial scalings of the various contributions to the supergravity fields, allowing efficient identification of the most important fields in a given problem. We conclude in \S\ref{sec:conclusions}.  Appendix \ref{appendix:sources} presents the structure of the source terms in the equations of motion, while Appendix \ref{sec:appendix} contains the details of the homogeneous solutions and Green's functions for the scalar, flux, and metric modes.

\section{Setup and Method}
\label{sec:setup}

%In this section we will characterize the class of compactifications dealt with in this paper and then elaborate on the expansion scheme outlined above. We start
We begin by writing down the equations of motion and describing the ISD background around which we perturb. In \S\ref{sec:expansion} we expand the equations of motion,
and in \S\ref{sec:method} we show that in our chosen basis, the equations of motion for the perturbations take on a triangular form at any order.  Using this structure, we develop an iterative, purely algebraic method for solving the perturbed equations to all orders.
%In \S\ref{sec:consistency} we discuss the validity of our approximation scheme.

\subsection{Equations of motion and background solution}
We consider type IIB compactifications of the form
{\allowdisplaybreaks
\begin{align}
&\dd s^2 = e^{2A(y)} g_{\mu\nu}\dd x^\mu \dd x^\nu +
e^{-2A(y)}g_{mn}\dd y^m\dd y^n\,,\label{compact1}\\
&\tilde{F}_5 =
(1+\star_{10}) \, \dd\alpha(y)\wedge \sqrt{- \det g_{\mu\nu}} \, \dd
x^0\wedge\dd x^1\wedge\dd x^2\wedge\dd x^3\,,\label{compact2}\\
&G_{mnl} = G_{mnl}(y)\,,~~~m,n,l=4,\ldots 9\,,\\
&G_{\mu NL}=0\,,~~~\mu =0,\ldots 3\,, ~~~ N,L=0,\ldots 9\,,\\
&\tau=\tau(y)\label{compact4}\,,
\end{align}
}%enddisplaybreaks
where we are using the conventions and notation of \cite{GKP}, with the modification that $g_{mn}^{\rm here}=\tilde{g}_{mn}^{\rm there}$. We generalize the setup of \cite{GKP} slightly by allowing for a maximally symmetric spacetime $g_{\mu\nu}$.  If we define the quantities
{\allowdisplaybreaks
\begin{eqnarray}
G_{\pm} &\equiv& (\star_6\pm i)G_3\,,\\
\Phi_\pm &\equiv& e^{4A}\pm\alpha\,,\\
\Lambda &\equiv& \Phi_+ G_-+\Phi_- G_+\,,
\end{eqnarray}
}%enddisplaybreaks
then the equations of motion and Bianchi identities take the form
% May 31 2 PM added \,
{\allowdisplaybreaks
\begin{align}
& \nabla^2\Phi_{\pm} = {(\Phi_+ + \Phi_-)^2 \over 96 \, \mathrm{Im} \, \tau } |G_{\pm}|^2 +
\mathcal{R}_4
+ {2 \over \Phi_+ + \Phi_-} |\nabla\Phi_\pm|^2\label{EOM1}\,,\\
&\dd \Lambda + {i\over 2 \, \mathrm{Im}\,\tau}
\dd\tau\wedge(\Lambda+\bar{\Lambda})=0\,,\label{EOM2}\\
&\dd\big( G_3+\tau\, H_3\big) = 0\,,\label{EOM3}\\
&\nabla^2\tau = {\n \tau \cdot \n \tau \over
i \, \mathrm{Im}(\tau)} + {\Phi_+ + \Phi_- \over 48i}\, G_+ \cdot G_-\,,\label{EOM4}\\
&R^6_{mn} =
{\n_{(m} \tau \n_{n)} \bar{\tau} \over 2\, (\mathrm{Im}\,\tau)^2} + {2 \over (\Phi_+ + \Phi_-)^2}
\n_{(m} \Phi_+ \n_{n)} \Phi_- - g_{mn}{\mathcal{R}_4 \over 2 \, (\Phi_++\Phi_-)} \label{EOM5} \\
&\phantom{R^6_{mn} =}-{\Phi_+ + \Phi_- \over 32 \, \mathrm{Im}\,\tau}
\Bigl(G_{+\,(m}^{\phantom{(m}~~pq}\, \bar{G}_{-\,n)\,pq} + G_{-\,(m}^{\phantom{(m}~~pq}\, \bar{G}_{+\,n)\,pq} \Big),\nonumber
\end{align}
}%enddisplaybreaks
where $\mathcal{R}_4$ is the four-dimensional Ricci scalar of $g_{\mu\nu}$, and covariant derivatives $\nabla_m$ and contractions are constructed and performed using $g_{mn}$. We have also dropped all
contributions from localized sources.
We will make use of an equivalent form for the $\Phi_+$ equation of motion:
{\allowdisplaybreaks
\begin{equation}\label{gen_phi}
-\nabla^2\Phi_+^{-1} = {1 \over 96\, \textrm{Im}\,\tau} {(\Phi_++\Phi_-)^2\over
\Phi_+^2}|G_+|^2+{\mathcal{R}_4 \over \Phi_+^2}+{2\over
\Phi_+^2}\Big\{{1\over (\Phi_++\Phi_-)}-{1\over
\Phi_+}\Big\}(\nabla\Phi_+)^2\,.
\end{equation}
}%enddisplaybreaks

In this work we will set $\mathcal{R}_4\rightarrow0$,  $g_{\mu\nu}=\eta_{\mu\nu}$, which is appropriate for modeling late-time physics.  For an example of incorporating curvature corrections in the context of inflation, see \cite{LiamsLong}.
%though we thought it useful to collect the equations in full generality above.  Note that for late time applications, the scale of $\mathcal{R}_4$ will be so small compared to the scale of supersymmetry breaking that there will hardly be any need to keep it.  For an example of incorporating curvature corrections in the context of inflation, see \cite{LiamsLong}.

%\subsection{Background}
%\label{subsec:background}
The background solution of equations (\ref{EOM1})-(\ref{EOM5}) for our analysis will
%is described by the ansatz (\ref{compact1}-\ref{compact4}), and in addition
obey the conditions
\begin{eqnarray}
G_-=0\,,\label{ISD1}\\
\Phi_-=0\,,\label{ISD2}\\
\nabla\tau=0\,.\label{ISD3}
\end{eqnarray}  %We will refer to this as an ISD background.
%
%refer to our background solution, corresponding to the ansatz given by
%equations (\ref{compact1}-\ref{compact4}) and
%(\ref{ISD1},\ref{ISD2}), as an ISD background.
In a slight abuse of language, we will refer to (\ref{ISD1})-(\ref{ISD3}) as the {\it{ISD conditions}}, and to the corresponding background as an ISD solution.  (Properly speaking, (\ref{ISD3}) can be violated in solutions usually described as ISD, e.g.\ in no-scale F-theory compactifications.)
%We will also assume that the dilaton $\tau$ is a constant in the background solution, and in an abuse of language we will subsume this condition in the
As motivation for this starting point, we remark that KKLT compactifications \cite{KKLT} based on conformally Calabi-Yau spaces
% Liam
involve controllably small deviations from ISD backgrounds, as we will explain   in \S\ref{sec:expansion}.

Furthermore, we will assume that the background solution contains a warped throat region.
%Our goal is to provide a framework for analyzing physical systems in this warped throat region.   We will begin with a very simple, computable proxy for the system of interest, then systematically incorporate corrections to this approximation. The computable starting point for our analysis is an {\it{infinite}} throat geometry that is supersymmetric and obeys the ISD conditions (\ref{ISD1}, \ref{ISD2}).
Specifically, we consider a throat for which the internal metric takes the form of a Calabi-Yau cone $\mathcal{C}_6$,
\begin{alignat}{2}
 \dd s^2_{\mathcal{C}_6} & = g_{mn}(y) \dd y^m \dd y^n = \dd r^2 + r^2 \dd s^2_{\mathcal{B}_5} \,, \quad m,n = 4,\ldots 9\,, \label{cone} \\
\intertext{over some Sasaki-Einstein base $\mathcal{B}_5$ with metric $\tilde{g}_{ij}$,}
 \dd s^2_{\mathcal{B}_5} & = \tilde{g}_{ij} (\Psi) \, \dd \Psi^i \dd \Psi^j \,, \quad i,j = 5,\ldots 9\,. \label{angular}
\end{alignat}
%where $\tilde{g}_{ij}$ is the metric on the base space.
(Throughout this paper, we use the letters $i,\, j,\,k,\,l$ to represent angular values for the indices and $m,\,n,\,p,\,q$ for general internal indices.)
%We assume that the dilaton does not run
%in the background solution, and therefore $\mathcal{B}_5$ must be a Sasaki-Einstein space.
We will further assume that the geometry is approximately AdS, so that the background warp factor takes the form
\begin{equation}\label{UVwarp}
e^{-4A} = {C_1+C_2\, \ln r \over r^4}\,,
\end{equation}
where the constants $C_1$ and $C_2$ are determined by the background fluxes $F_5$, $F_3$, and $H_3$.

In many solutions of interest, the throat terminates at a finite radial distance, either smoothly, as in the Klebanov-Strassler solution \cite{KS}, or through the appearance of a horizon or singularity.
In either case, the IR region of the throat,
%Ideally, the throat should terminate smoothly at a finite, minimal value of the warp factor, $e^{A_{\mathrm{min}}}\equiv a_0$.  Then, the IR region of the throat,
below some position $r=r_\text{IR}$, will necessarily deviate from
%
%We do not want the throat to extend indefinitely in the IR direction.  In particular, we would like for the throat to \textit{smoothly}  terminate at a finite, minimal value of the warp factor
%$e^{A_{\mathrm{min}}}\equiv a_0$.  Then in the IR end of the throat, the geometry will necessarily deviate from
the approximately AdS form (\ref{cone}, \ref{angular}, \ref{UVwarp}), and one will need to include corrections arising from the tip in a systematic expansion as well.
%This `tip' geometry will in general be quite complicated to work with, and so in practice one will often have to resort to the approximate form (\ref{cone}, \ref{angular}, \ref{UVwarp}), and include corrections arising from the tip in a systematic expansion as well.
Our approach yields a reliable description of the intermediate regime $r_\text{IR}\ll r \ll r_{UV}$ that is far from the tip and far from the UV brane.

\subsection{Perturbative expansion of the field equations} \label{sec:expansion}

Our strategy is to approximate a highly warped region of a flux compactification in terms of a double expansion around an infinite throat geometry with ISD fluxes.
%supersymmetric,
%ISD, infinite throat geometry.
The system of actual interest deviates in two ways
%There are two types of deviations from this simple model:
from this simple background:
\begin{itemize}
\item{} The throat of interest  has finite length: the UV region is glued into a compact space, with corresponding deviations from the infinite throat solution.
\item{} Effects in the bulk of a stabilized compactification typically violate the ISD conditions (\ref{ISD1}, \ref{ISD2}).
\end{itemize}
Deviations of the first kind will be present even in compact models that everywhere satisfy the ISD conditions, e.g.\ in the warped compactifications of \cite{GKP}.  Moreover, where the throat is glued to the bulk, these deviations  will generally be of order unity, reflecting the transition from the throat to the bulk.  However,  as one moves deeper and deeper into the throat, the bulk geometry has diminishing influence, and use of the infinite throat geometry should hold to better and better approximation.  Thus, we can perform an expansion that is valid at some location $r=r_{\star} \ll r_{UV}$ far below the UV brane, with the infinite throat as the starting point and ${r_{\star}/r\uv}$  controlling corrections.

Deviations of the second kind arise from sources in the bulk. Consider one well-motivated example: to obtain stabilized de Sitter vacua in the scenario of \cite{KKLT}, one incorporates nonperturbative effects on four-cycles, and introduces one or more anti-D3-branes in warped throat regions. These sources lead to controllably small departures from the ISD conditions, and to controllably small breaking of supersymmetry.  The nonperturbative contributions are exponential in the four-cycle volumes, while mass splittings due to a given anti-D3-brane are suppressed by the hierarchy of scales in the corresponding throat, $e^{A_{\mathrm{min}}}\equiv a_0$. Thus, both sorts of corrections are naturally small.  Moreover, the requirement of a de Sitter vacuum links the scale of the nonperturbative effects and the infrared scale of the warped throat, so that all ISD-violating and supersymmetry-violating effects are controlled by the same small parameter, $a_{0}$.  In summary, one has a double expansion in terms of the parameters ${r_{\star}/r\uv}$ and $a_0$.

In practice, we will find it most convenient to use the magnitudes of the  \textit{harmonic} modes evaluated at $r=r_{\star}$ as our expansion parameters.  Specifically, let $\phi$ be any one of the bosonic supergravity fields $\Phi_\pm$, $G_\pm$, $\tau$, $g_{mn}$.  The solution for field $\phi$ about the throat background will be given by a homogeneous piece plus an inhomogeneous piece,
%as is familiar for the solution of a partial differential equation:
\begin{equation}
 \phi  = \phi^{(0)} + \phi_{\mathcal{H}} + \phi_{\mathcal{IH}}\,,
\end{equation}
where  $\phi^{(0)}$ is the background value of the field.  The homogeneous pieces obey simple harmonic equations and have solutions of the form
 \begin{equation}\label{generalHom}
 \phi_{\mathcal{H}} = \sum_I\,\left(c_0^I\,\Big({r\over r_{\star}}\Big)^{\Delta(I)-4}+c_1^I\,\Big({r\over r_{\star}}\Big)^{-\Delta(I)}\right)\, Y^I(\Psi)\,,
\end{equation}
% May 31 3 PM
where $I$ is a multi-index encoding the angular quantum numbers.
The $Y^I(\Psi)$ are angular harmonics that are of order unity at a general point, while the
% May 31 2 PM changes here and below:  $c_i^I$
$c_i^I$, with $i=0,1$,  are numerical coefficients determined by the boundary conditions.  The inhomogeneous piece of a given field then incorporates the effects of source terms in that field's equation of motion.

From (\ref{generalHom}), we see that the $c_i^I$ give the sizes of the harmonic modes at $r=r_{\star}$.   Provided that we work in a region where corrections to the background throat geometry are small, the $c_i^I$ will likewise be small.
%\footnote{We are assuming that the freedom to shift the (unless we have chosen an inefficient separation into homogeneous and inhomogeneous pieces, which we shall avoid).
In practice, we will use the $c_i^I$  as our expansion parameters, i.e.\ we will develop solutions for the inhomogeneous pieces of the fields in terms of a multiple expansion in the $c_i^I$.
%Of course,
Ultimately, the parametric sizes of the $c_i^I$ can be expressed in terms of $a_0$ and $r_{\star}/r\uv$, so that there are only two fundamental expansion parameters.

We now expand the fields around their
% Sohang expectation
values in the ISD background.
For each field $\phi$, we expand as
\begin{align}\label{phiExp}
\phi  = \phi^{(0)}+\phi^{(1)}+\phi^{(2)}+\ldots= \phi^{(0)}+ \phi_{\mathcal{H}}+\phi_{\mathcal{IH}}^{(1)}+\phi_{\mathcal{IH}}^{(2)}+\ldots\,.
\end{align}
where $\phi^{(0)}$ is the background value for the field, $\phi^{(1)}$ represents the sum of corrections to the field linear in the $c^I$, etc.
%  $\phi^{(2)}$ represents the sum of corrections quadratic in the $c_I^\Delta$, etc.
It will also be convenient to use a notation where the homogeneous piece  $\phi_{\mathcal{H}}$ and the inhomogeneous piece  $\phi_{\mathcal{IH}}$ are split.
%\begin{align}
%\phi   &= \phi_{(0)}+ \phi^{\mathcal{H}}+\phi^{\mathcal{IH}}\\
%&= \phi_{(0)}+ \phi^{\mathcal{H}}+\left(\phi^{\mathcal{IH}}\right)^{(1)}+\left(\phi^{\mathcal{IH}}\right)^{(2)}+\ldots\,.\nonumber
%\end{align}
Clearly $\phi_{\mathcal{H}}$  is linear in the $c^I$.  The $\phi_{\mathcal{IH}}^{(n)}$ comprise the inhomogeneous piece of the correction:   $\phi_{\mathcal{IH}}^{(1)}$ represents the sum of inhomogeneous corrections to the field linear in the $c^I$, $\phi_{\mathcal{IH}}^{(2)}$ represents the sum of corrections quadratic in the $c^I$, etc.

%\subsubsection{The $n$-th order equations of motion}

With these preliminaries, we can proceed to expand the supergravity equations (\ref{EOM1}-\ref{EOM5}) around the ISD background.
We will examine the $n$-th order equations of motion, focusing for the moment on terms that involve the $n$-th order corrections, as opposed to products of lower order corrections.
  %since this is all we need to understand the general structure of the equations.
These terms are universal, in the sense that at any order $n$ they take exactly the same form: since we are expanding to order $n$, whenever we take one of the fields in a term of an equation to be at order $n$, all other factors in the term must be taken to be at order zero.
%Also, for the sake of simplicity,  we will ignore all terms coming from the curvature of the four-dimensional space in equations (\ref{EOM1}-\ref{EOM5}), i.e.\ we will set $\mathcal{R}_4 \rightarrow 0$.

The resulting equations for the $n$-th order perturbations around the ISD background following from equations (\ref{EOM1}-\ref{EOM5}) are
{\allowdisplaybreaks
\begin{align}
&\nabla_{(0)}^2\,\Phi_-^{(n)} = \mathrm{Source}_{\Phi_-}(\phi^{(m<n)})\label{nphi-}\,,\\
\nonumber\\
&\dd\left(\Phi_+^{(0)}\,G_-^{(n)}\right)=-\dd\left(\Phi_-^{(n)}\,G_+^{(0)}+\mathrm{Source}_{G_-,\,1}(\phi^{(m<n)})\right) + \mathrm{Source}_{G_-,\,2}(\phi^{(m<n)})\,,\label{nG-}\\
\nonumber\\
&(\star_6^{(0)}+i)\,G_-^{(n)}=  \mathrm{Source}_{G_-,\,3}(\phi^{(m<n)})\,, \label{niasd}\\
\nonumber\\
&\nabla^2_{(0)}\tau_{(n)}= {\Phi_+^{(0)} \over 48i}\,G_+^{(0)}\cdot G_-^{(n)}+\mathrm{Source}_{\tau}(\phi^{(m<n)})\,,\label{ntau}\\
\nonumber\\
-&\hf \Delta_{K}^{(0)}\,g_{mn}^{(n)}= - {\Phi_+^{(0)} \over 32
\mathrm{Im}\,\tau} \left( G^{(0)\phantom{(m}pq}_{+\,(m }\, \bar{G}^{(n)}_{-\,n)\,pq} + G^{(n)\phantom{(m}pq}_{-\,(m }\, \bar{G}^{(0)}_{+\,n)\,pq} \right) \label{nmetric}\\
&\phantom{\hf \Delta_{K}^{(0)}\,g_{mn}^{(n)}= } + 2 (\Phi_+^{-2})^{(0)} \, \n_{(m}\Phi_+^{(0)}\n_{n)}\Phi_-^{(n)}+\mathrm{Source}_g(\phi^{(m<n)})\,,\nonumber\\
\nonumber\\
&\dd \left(G_+^{(n)}\right)= \dd\left(G_-^{(n)} -2i\,\tau_{(n)}\, H_3^{(0)}-\mathrm{Source}_{G_+,\,1}(\phi^{(m<n)})\right)\,,\label{nG+}\\
\nonumber\\
&(\star_6^{(0)}-i)\,G_+^{(n)}=  \mathrm{Source}_{G_+,\,2}(\phi^{(m<n)})\,, \label{nisd}\\
\nonumber\\
-&\nabla^2_{(0)}\oru{(\Phi_+^{-1})}{n}=\nabla^2_{(n)}(\Phi_+^{-1})^{(0)} - {g_s^2\over
96} \mathrm{Im}\,\tau^{(n)}|G_+^{(0)}|^2
\label{nphi+}\\
&\phantom{\nabla^2_{(0)}\oru{(\Phi_+^{-1})}{n}=} + {g_s \over 96} \left(G_+^{(0)}\cdot \bar{G}_+^{(n)}+G_+^{(n)}\cdot
\bar{G}_+^{(0)}+3G_{+~m_1n_1l_1}^{(0)}
\bar{G}_{+~m_2n_2l_2}^{(0)}g_{(0)}^{m_1m_2}g_{(0)}^{n_1n_2}g_{(n)}^{l_1l_2} \right)\nonumber\\
&\phantom{\nabla^2_{(0)}\oru{(\Phi_+^{-1})}{n}=}+ \Big( {g_s\over 48}\,(\Phi_+^{-1})^{(0)}\,|G_+|^2_{(0)}  -  2 (\Phi_+^{-4})^{(0)} \, (\nabla\Phi_+)^2_{(0)} \Big)\,  \Phi_-^{(n)}\nonumber +\mathrm{Source}_{\Phi_+}(\phi^{(m<n)})\,,
\end{align}
}%enddisplaybreaks
where $\E$ denotes the metric kinetic operator
\begin{equation}
\E g^{(n)}_{mn}\equiv \nabla^2\,g^{(n)}_{mn}+\nabla_m\nabla_n g^{(n)} - 2\nabla^p\nabla_{(m} \, g^{(n)}_{n)p}\label{deltaK}\,, \quad g^{(n)} \equiv g_{(0)}^{pq}g^{(n)}_{pq}\,.
\end{equation}
We have used the abbreviation ``$\mathrm{Source}_\varphi(\phi^{(m<n)})$" to stand for all of the source terms in the equation for
% Sohang some
field $\varphi$ involving the fields at previous orders $m < n$. As an illustrative example, we perform the $\tau$ expansion fully in Appendix \ref{appendix:sources}, giving the explicit form of $\mathrm{Source}_\tau(\phi^{(m<n)})$.

\subsection{Method for generating solutions}
\label{sec:method}
We will now outline our algorithmic procedure for generating the solutions to equations (\ref{nphi-}-\ref{nphi+}) to an arbitrary order.

The order in which we arranged equations (\ref{nphi-}-\ref{nphi+})
is of critical significance: it reveals the triangular structure of the $n$-th order equations that will allow us to disentangle and solve the system.
%highlights a triangular structure which the equations take when
% Liam we assume that
%the {\it{background}} satisfies the ISD conditions,
% Sohang and which
%allowing them to be easily disentangled.
Let us emphasize that the equations of motion are triangular (in our chosen basis) whenever the background obeys the ISD conditions
% May 31   (\ref{ISD1})-(\ref{ISD3}),
(\ref{ISD1}-\ref{ISD3}), i.e.\ whenever the background is conformally Calabi-Yau.
%\footnote{The background need not be supersymmetric; moreover, although Calabi-Yau cones are a natural setting for finding explicit solutions to  (\ref{nphi-}-\ref{nphi+}), the triangular structure is present even in compact models.}
In expanding around a background that is not ISD, the perturbed equations of motion will in general be intractably entangled, making an analytic solution impractical even at linear order.
% Liam
% Sohang
%even at linear order impractical.

Assuming that we have solved for the corrections at all orders before $n$, we see that in solving equation (\ref{nphi-}) for $\Phi_-^{(n)}$,  $\mathrm{Source}_{\Phi_-}(\phi^{(m<n)})$ may be taken as given.  Thus we can solve via the scalar Green's function, which we shall denote by $\mathcal{G}_{s}$.  Having the solution for $\Phi_-^{(n)}$, we substitute it into equation (\ref{nG-}) for $G_-^{(n)}$. Then all sources appearing in equations (\ref{nG-}, \ref{niasd}) are given and we can solve for $G_-^{(n)}$ using the flux Green's functions $\mathcal{G}_{G}$.
Continuing in this way, we can generate the $n$-th order solutions for all of the fields.\footnote{A similar method was used in \cite{Ben} to find an all-orders local solution with dynamic SU(2) structure. We thank B.~Heidenreich for helpful discussions of this point.}
%\subsubsection{Homogeneous equations}
The result is an iterative procedure for generating the solutions,  where the results from a lower order are fed into the next higher order.  The seeds for this process are the harmonic modes, which
% referred to in \S\ref{sec:double}.   These harmonic modes may be taken to
obey simple equations without mixing between fields:
{\allowdisplaybreaks
\begin{align}
&\nabla_{(0)}^2\,\Phi_-^{\mathcal{H}} = 0\label{Hom1} \,, \\
&\dd(\Phi_+^{(0)}\,G_-^{\mathcal{H}}) =0\label{hG-} \,, \\
%\nonumber\\
%
&(\star_6^{(0)}+i)\,(\Phi_+^{(0)}\,G_-^{\mathcal{H}}) =  0 \,, \label{hiasd}\\
%\nonumber\\
%
&\dd G_3^{\mathcal{H}} = 0 \label{Gfluxclosed} \,, \\
%\nonumber\\
%
&\nabla^2_{(0)}\tau^{\mathcal{H}} =0 \,, \\
%\nonumber\\ % Liam
%
& \Delta_K^{(0)}\,g_{mn}^{\mathcal{H}}=0 \,, \\
%\nonumber\\
&\nabla^2_{(0)}{(\Phi_+^{-1})}^{\mathcal{H}}=0 \,. \label{Hom7}
\end{align}
}%enddisplaybreaks
Note that when one divides a system of coupled partial differential equations into homogeneous and inhomogeneous pieces, the homogeneous equations are typically coupled.  The fact that we can use the uncoupled system (\ref{Hom1}-\ref{Hom7}) is another fortuitous consequence of the triangular structure.

Let us explain how this works in detail.  At first order, all of the $\mathrm{Source}_\phi(\phi^{(m<n)})=0$.  Then $\Phi_-^{(1)}$ simply obeys the harmonic equation (\ref{Hom1}), and thus
\begin{equation}
\Phi_-^{(1)} = \Phi_-^{\mathcal{H}}\,.
\end{equation}
Substituting these harmonic modes as sources in the $G_-$ equation (\ref{nG-}), we find that, schematically,
%Then $G_-$ is just given by convolving $\Phi_-^{\mathcal{H}} $ with a Green's function and adding $G_-^{\mathcal{H}}$; schematically
\begin{equation}
G_-^{(1)} = \int\, \mathcal{G}_{G} \cdot\Phi_-^{\mathcal{H}}+G_-^{\mathcal{H}}\,.
\end{equation}
Because we are solving equations (\ref{nG-}, \ref{niasd}) with all source terms pre-specified,  $G_-^{\mathcal{H}}$ is given by the uncoupled harmonic equations (\ref{hG-}, \ref{hiasd}).
Working down the triangle in the same fashion, one obtains the solutions for all of the fields as functions of the harmonic solutions.

At order $n>1$,  the  $\mathrm{Source}_\phi(\phi^{(m<n)})\neq 0$.  One needs to carry out the expansion of  the equations of motion to order $n$ to  determine the form of these terms.  One next plugs in the solutions from previous orders for the $\mathrm{Source}_\phi(\phi^{(m<n)})$, and then proceeds down the triangle just as in the linear case.  In this way the solutions for the $n$-th order corrections are determined as functions of the harmonic modes.   Moreover, one can use the same set of Green's functions at all orders, since the structure of the terms involving $n$-th order fields in equations (\ref{nphi-}-\ref{nphi+}) is the same for any $n$.  Note that generally one would expect homogeneous contributions to the solutions at all orders:
\begin{equation}\label{genn}
\phi^{(n)} =  \phi^{(n)}_\mathcal{IH} + \phi^{(n)}_\mathcal{H}\,.
\end{equation}
However, since we are using the coefficients  of the harmonic modes themselves as expansion parameters in our scheme, we have
%\begin{align}
%&\phi_{(1)}_\mathcal{H} \equiv \phi^{\mathcal{H}}\,, \\
%&\phi^{(n)}_\mathcal{H} \equiv 0\,, ~~~~n>1\,,\\
%\end{align}
\begin{equation}
\phi^{(n)}_\mathcal{H} \equiv
\left\{
\begin{array}{cc}
 \phi_\mathcal{H} & \text{for }n=1 \\
 0 & \text{for } n>1
\end{array}\,,
\right.
\end{equation}
where $\phi^{\mathcal{H}}$ is the all-orders resummation of the harmonic modes.

The two key ingredients for our solutions are the seeding harmonic modes and the  Green's functions for equations (\ref{nphi-}-\ref{nphi+}).   We present the harmonic solutions in \S\ref{sec:harmonic} and obtain the Green's functions in \S\ref{sec:Greens}, relegating detailed derivations to Appendix \ref{sec:appendix}.  Our results are presented in terms of the angular harmonics and associated spectroscopy on the base space $\mathcal{B}_5$: we expand all fields (and Green's functions) in these harmonics, separate the equations of motion, and solve the resulting radial equations.   Thus, our solutions require the spectroscopy on $\mathcal{B}_5$ as input.
For the case in which the base space is $\mathcal{B}_5 = T^{1,1}$ (i.e.\  the Klebanov-Strassler throat), all relevant eigenvalues and eigenfunctions are known \cite{CeresoleHarm, CeresoleSCFT, LiamsLong, AntiBrane}.
Moreover, the techniques applied in these works to $T^{1,1}$ can be extended to any homogeneous base space.

A primary goal of this paper is to characterize the effects of perturbations sourced in the bulk, and we have therefore emphasized non-normalizable perturbations  in the discussions below.
A general finite warped throat would involve normalizable perturbations sourced by effects in the IR (including, e.g., the deformation of the conifold, or a supersymmetry-breaking anti-D3-brane), in addition to the non-normalizable perturbations described in the preceding section.  Moreover, boundary conditions at the tip will in general tie together normalizable and non-normalizable modes.  Incorporating normalizable perturbations presents no technical challenge, and one can simply substitute normalizable modes along with non-normalizable modes when generating the Green's function solutions outlined in \S\ref{sec:method}.  Nevertheless, for simplicity of presentation we will restrict our attention to non-normalizable perturbations in this work.

Further details of our perturbative expansion are deferred to \S\ref{sec:scalings}.

\subsection{Matching solutions to boundary values}

The method described so far takes solutions to the uncoupled homogeneous equations (\ref{Hom1}-\ref{Hom7}) as input, with the sizes of the corresponding harmonic modes serving as expansion parameters, and generates an inhomogeneous solution to any desired order.  While this approach efficiently utilizes the triangular structure of the perturbed equations of motion (\ref{nphi-}-\ref{nphi+}), it is not yet adapted to solve a boundary value problem on the cone. We now remedy this.

Suppose that one would like to solve a boundary value problem in which the fields and their derivatives are specified on some slice $r=r_\star$, on which all corrections are small.
To apply the method described above, one needs to extract the values of the
% May 31 2 PM
$c_i^I$ from the boundary data.
%Assuming that both normalizable and non-normalizable modes are turned on, a complete specification of boundary data would for instance be
We first expand the field value and the first radial derivative at $r=r_\star$:
\begin{align}
\delta\phi(r_\star, \Psi)& = \sum_I\,a^I\,Y^I(\Psi)\,,\label{bc1}\\
\partial_r\delta\phi(r_\star, \Psi) &= \sum_I\,\frac{b^I}{r_\star}\,Y^I(\Psi)\,,\label{bc2}
\end{align}
with $\delta\phi = \phi-\phi^{(0)}$, so that the $a^I,\,b^I$ parameterize the boundary conditions.
Expanding $\phi_{\mathcal{IH}}$ in harmonics as
% Liam
\begin{equation}
 \phi_{\mathcal{IH}}(r, \Psi) = \sum_I \, \phi_{\mathcal{IH}}^I(r)\,Y^I(\psi)\,,
 \end{equation}
and using equation (\ref{generalHom}), (\ref{bc1}, \ref{bc2}) give
\begin{align}
&c_0^I+c_1^I+\phi_{\mathcal{IH}}^I(r_\star)=a^I,\label{cbc1}\\
&\left(\Delta(I)-4\right)\,c_0^I-\Delta(I)\,c_1^I+r_\star\,\partial_r\phi_{\mathcal{IH}}^I(r_\star)=b^I\,.\label{cbc2}
\end{align}
We will see in \S\ref{sec:scalings} that  $\phi_{\mathcal{IH}}^I(r_\star)$  and $r_\star\,\partial_r\phi_{\mathcal{IH}}^I(r_\star)$ are both given by power series in the $c^I_i$, with coefficients that are of order unity. Thus, we can obtain the $c^I_i$, which parameterize the homogeneous solutions, as power series in the $a^I,\,b^I$ that parameterize the boundary conditions, by inverting the series (\ref{cbc1}, \ref{cbc2}) to the desired order.

As each of the
% Liam
fields $\phi$ can be expanded in an infinite set of modes, equations (\ref{cbc1}, \ref{cbc2}) represent an infinite system of coupled equations at each order.  However, the triangular structure once again comes to our rescue, so that solving the system is a matter of straightforward\footnote{No boundary value problem of interest will be specified in terms of an infinite number of {\it{independent}} coefficients of harmonics, as such a problem could not even be posed in finite time.  Our approach is applicable when the harmonic expansion truncates, or when the coefficients of higher multipoles are simply related to the coefficients of lower multipoles, e.g. by a closed-form expression for the $a^{I}$, $b^{I}$ for arbitrary $I$.} algebra, as we now explain.  Suppose  for simplicity that all normalizable modes are absent, in which case boundary condition (\ref{cbc1}) is by itself sufficient.  Let us also suppose that there is some small parameter $\epsilon$ controlling the size of the perturbations on the boundary surface, so that we may expand
\begin{align}
c_\phi^I &= (c_\phi^I)^{(1)}+(c_\phi^I)^{(2)}+\ldots\\
a_\phi^I &= (a_\phi^I)^{(1)}+(a_\phi^I)^{(2)}+\ldots
\end{align}
where  $(c_\phi^I)^{(n)}$ and  $(a_\phi^I)^{(n)}$ are the ${\cal{O}}(\epsilon^n)$ parts of the nonnormalizable coefficient and boundary value, respectively, for field $\phi$.

Now begin at first order and at the top level of the triangle.  At this order, $\Phi_-$ is harmonic, so (\ref{cbc1}) becomes
\begin{equation}
(c_{\Phi_-}^I)^{(1)}= (a_{\Phi_-}^I)^{(1)}\,.
\end{equation}
Next, $\Phi_-$ acts as a source for $G_-$.  When we expand the Green's function solution for this source in modes,
\begin{equation}
\left(G_-^{\mathcal{IH}}\right)^{(1)}(r,\Psi) = \int\, \mathcal{G}_{G} \cdot\Phi_-^{\mathcal{H}} = \sum_I\,\left(G_-^{\mathcal{IH}}\right)_I^{(1)}(r)\,Y^I(\Psi)\,,
\end{equation}
we will generically find
% Liam
\begin{equation}
\left(G_-^{\mathcal{IH}}\right)_I^{(1)}
(r=r_\star) = \sum_{J}\,n^{J}_{~I}\,(c_{\Phi_-}^{ J })^{(1)} = \sum_{J}\,n^{J}_{~I}\,(a_{\Phi_-}^{ J })^{(1)} \,,
\end{equation}
where the
% Sohang $n^J$
$n^{J}_{~I}$ are numerical coefficients of order  unity obtained by evaluating the Green's function solutions of \S\ref{sec:Greens} on the boundary surface.  In the final equality we substituted the results from the previous level of the triangle.
Equation (\ref{cbc1}) then gives
% Sohang add I index to n
\begin{equation}
 (c_{G_-}^I)^{(1)}= (a_{G_-}^I)^{(1)}-\sum_{J}\,n^{J}_{~I}\,(a_{\Phi_-}^{ J })^{(1)}\,.
 \end{equation}

One can continue
% Liam
in this way down the triangle.
Then, moving to higher order poses no significant challenge.
% Liam
The contributions of the  $\mathrm{Source}_\phi(\phi^{(n<m)})$ terms to (\ref{cbc1}) are
%numerically
determined by substituting from the previous orders.  For instance, for $\Phi_-$ at second order, we could expand
\begin{equation}
\left(\Phi_-^{\mathcal{IH}}\right)^{(2)}(r,\Psi) = \int\, \mathcal{G}_{s} \cdot\mathrm{Source}_{\Phi_-}(\phi^{(n<2)}) = \sum_I\,\left(\Phi_-^{\mathcal{IH}}\right)_I^{(2)}(r)\,Y^I(\Psi)\,,
\end{equation}
and would generically find
% Liam
\begin{equation}
\left(\Phi_-^{\mathcal{IH}}\right)_I^{(2)}
(r=r_\star) = \sum_{J,\,J',\,\phi,\,\phi'}\,\tilde{n}^{J\,J'}_{~I}\,(c_{\phi}^{ J })^{(1)} (c_{\phi'}^{ J' })^{(1)}\,.
\end{equation}
Equation (\ref{cbc1}) then gives for the second-order $\Phi_-$
 \begin{equation}
 (c_{\Phi_-}^I)^{(2)}= (a_{\Phi_-}^I)^{(2)}-\sum_{J,\,J',\,\phi,\,\phi'}\,\tilde{n}^{J\,J'}_{~I}\,(c_{\phi}^{ J })^{(1)} (c_{\phi'}^{ J' })^{(1)}\,.
 \end{equation}

The reader may inquire why we did not use the $a^I$  as the parameters of our solution from the beginning.  In this case the homogeneous piece of equation (\ref{genn}) would no longer vanish at order $n>1$.  At each order one would have to enforce boundary conditions tying
% Liam these
the new harmonic modes to the inhomogeneous solutions, and the work done in imposing these boundary conditions would effectively amount to the algebraic steps described above.   We find the above approach to be a more systematic way to organize the calculation.

\section{Homogeneous Modes of the Supergravity Fields}
\label{sec:harmonic}

The starting point of our expansion scheme is the set of homogeneous solutions to equations (\ref{Hom1}-\ref{Hom7}). The homogeneous modes are then fed into equations (\ref{nphi-}-\ref{nphi+}), sourcing the inhomogeneous solutions. As seen from equations (\ref{Hom1}-\ref{Hom7}), there are three distinct types of homogeneous equations:

\paragraph{Scalar} The homogeneous modes of the scalar fields $\Phi_-, \Phi^{-1}_+$ and $\tau$ obey the Laplace equation on the cone,
\begin{equation} \label{scalarharm}
 \nabla^2 \Phi^\mathcal{H} = 0 \,,
\end{equation}
where $\nabla^2$ is constructed using the cone metric, equation (\ref{cone}).

\paragraph{Flux} The homogeneous modes of the flux $G_\pm$ obey the system
\begin{align} \label{fluxharm}
&~~~~~~\dd(\Phi_+\,G^\mathcal{H}_-)=0\,,\\
&~~~~~~\dd G_3^\mathcal{H} = 0 \,,\label{fluxBI}
\end{align}
where $\Phi_+$ is given by its background form, equation $(\ref{UVwarp})$.

\paragraph{Metric} The homogeneous modes of the metric perturbations obey
\begin{equation}  \label{metricharm}
\Delta_K \, g^\mathcal{H}_{mn} = 0 \,.
\end{equation}

The solutions below are presented in terms of various harmonics on the angular space $\mathcal{B}_5$.  Details about these harmonics can be found in \S\ref{subsec:harms}. Throughout this section, contractions, covariant derivatives, etc.\ are carried out with respect to the zeroth-order background metric, equations (\ref{cone}), (\ref{angular}). In \S\S\ref{sec:intro},\ref{sec:setup} we denoted the background by $g^{(0)}_{mn}$, but in this section we will drop the superscript for simplicity of notation.  In addition, a tilde above the indices and the derivatives signifies contraction with and construction out of the angular metric $\tilde{g}_{ij}$ on $\mathcal{B}_5$.

\subsection{Homogeneous solutions for the scalars}
\label{sec:homscalar}
Consider first the Laplace equation (\ref{scalarharm}).  Using the cone structure of the background, we can expand $\Phi$ in scalar harmonics $Y^{I_s}(\Psi)$ on $\mathcal{B}_5$,
% the angular base space,
\begin{equation}
\Phi(r,\Psi) = \sum_{I_s} \Phi_{I_s}(r)\, Y^{I_s}(\Psi)\,,
\end{equation}
where the $Y^{I_s}(\Psi)$ diagonalize the angular Laplacian
\begin{equation} \label{Yeigenvalue}
 \tilde{\nabla}^{2} \, Y^{I_s} \equiv \frac{1}{\sqrt{\tilde{g}}} \partial_{i} \Bigl(\sqrt{\tilde{g}} \, \tilde{g}^{ij} \partial_j Y^{I_s}\Bigr) = -\lambda^{I_s}\,Y^{I_s} \,.
\end{equation}
Now using that the Laplacian decomposes,
\begin{equation}
 \nabla^2 = \partial_r^2 + {5\over r} \, \partial_r + {1 \over r^2} \tilde{\nabla}^2\,,
\end{equation}
 the Laplace equation reduces to the following radial equation for the expansion coefficients:
\begin{equation}
\partial_r^2 \, \Phi_{I_s} + {5\over r} \, \partial_r \Phi_{I_s} - {\lambda^{I_s} \over r^2} \Phi_{I_s} = 0\,.
\end{equation}
Thus, the homogeneous solutions for any of the fields $\Phi_-^\mathcal{H}, (\Phi^{-1}_+)^\mathcal{H}, \tau^\mathcal{H}$ take the form
\begin{equation}
 \Phi^\mathcal{H}(r,\Psi) = \sum_{I_s} \left( \Phi_0^{I_s} \, r^{\Delta(I_s)-4} + \Phi_1^{I_s} \, r^{-\Delta(I_s)} \right) \,Y^{I_s}(\Psi)\,,
\end{equation}
where $\Phi_0^{I_s}$ and $\Phi_1^{I_s}$ are constants determined by the boundary conditions, and where we have defined
\begin{equation}
\Delta(I_s) \equiv 2+\sqrt{4+\lambda^{I_s}}\,.\label{scalarScaling}
\end{equation}
By comparison with the standard AdS form, equation (\ref{generalHom}), we see that for a canonically normalized scalar field, $\Delta(I_s)$ corresponds to the dimension of the operator dual to that mode. For the zero mode, $\lambda^{I_s} = 0$, we have $\Delta(I_s) = 4$, but for modes other than the zero mode we have $\lambda^{I_s} \geq 5$ (cf.
%by equation (\ref{lamscond}) of
\S\ref{subsec:harms}), so that generically $\Delta(I_s) \geq 5$.

\subsection{Homogeneous solutions for the fluxes}
\label{sec:homflux}

For the homogeneous perturbations of the three-form fluxes $G_\pm$, we have the system of equations (\ref{fluxharm}), (\ref{fluxBI}).
%\begin{align}
%&\dd(\Phi_+ \,G^\mathcal{H}_-)=0 \,, \label{g-closed} \\
%&\dd G_3^\mathcal{H} = 0 \label{g3closed} \,.
%\end{align}
The solution of this system is a slight generalization of that obtained in \cite{LiamsLong}, now including logarithmic running of $\Phi_+$, equation (\ref{UVwarp}). Here we briefly outline the solution,
%how (\ref{g-closed}, \ref{g3closed}) is solved,
leaving the details to \S\ref{subsubsec:fluxharmappend}.

Because $G^\mathcal{H}_3$ is closed by equation (\ref{fluxBI}), it can  be written locally in terms of a two-form potential $A_2$. We then expand the potential in terms of two-form harmonics and solve equation (\ref{fluxharm}) for the coefficients of the harmonic expansion. The result is, cf.\ equation (\ref{A2solution}),
\begin{align}
G^\mathcal{H}_3 & = \dd A_2 \,,\\
\label{fluxharmpot} A_2 & =
%\sum_{i=1}^{b_2}\,\Big[a^i_- - {a_+^i \over 8\,r^4}\,\Big\{\Big(c_1+{C_2\over 4}\Big)+C_2\,\log\,r\Big\}\Big]\,\omega_2^i\\
%&~~~~~~~~~~+ \sum_{\lambda^{I_2}\neq0}\,
\sum_{I_2} \left( A_-^{I_2} \, r^{-\delta^{I_2}} + A_+^{I_2} \, \left[ (4 - 2 \, \delta^{I_2})(C_1+C_2\,\ln r) + C_2 \right] \,r^{\delta^{I_2}-4}\right)\,Y^{I_2}\,,
\end{align}
where $A_\pm^{I_2}$ are constants of integration and $C_{1,2}$ are the coefficients of the running warp factor $\Phi_+$, cf.\ equation (\ref{UVwarp}).  The $Y_{[ij]}^{I_{2}}(\Psi)$ are the transverse two-form harmonics on $\mathcal{B}_5$ that diagonalize the  Laplace-Beltrami operator
\begin{equation}
{\star_5} \dd Y^{I_{2}} = i \, \delta^{I_{2}} Y^{I_{2}} \,.
\end{equation}
%and satisfy a transversality condition
%\begin{equation}
%{\tilde\star_5} \dd {\tilde\star_5} Y^{I_2} = 0\,.
%\end{equation}
The eigenvalues $\delta^{I_2}$ are real and are symmetric under $\delta^{I_2} \rightarrow -\delta^{I_2}$.  In order for the radial scalings of the modes in equation (\ref{fluxharmpot}) to take on the standard AdS form, equation (\ref{generalHom}), we identify $\Delta(I_2) = \max(\delta^{I_2}, 4-\delta^{I_2})$.  In \S\ref{fluxdimensionappend} we give formulas expressing the resulting scaling dimensions of flux modes in terms of the dimensions
$\Delta(I_s)$ of scalar modes.

\subsection{Homogeneous solutions for the metric}
\label{sec:hommetric}

The homogeneous part of the metric perturbation obeys (\ref{metricharm}). To fully utilize the cone structure of $\mathcal{C}_6$ we decompose $g^\mathcal{H}_{mn}$ into irreducible pieces under general coordinate transformations of the base space $\mathcal{B}_5$. Then $g^\mathcal{H}_{rr}$ transforms as a scalar, $g^\mathcal{H}_{ir}$ transforms as a vector, and the trace, $\tilde{g}^\mathcal{H} \equiv \tilde{g}^{ij} g^\mathcal{H}_{ij}$, and the traceless part, $g^\mathcal{H}_{\{ij\}} \equiv g^\mathcal{H}_{ij} - \frac{1}{5} \tilde{g}_{ij} \tilde{g}^\mathcal{H}$, of $g^\mathcal{H}_{ij}$ transform as a scalar and a symmetric traceless two-tensor, respectively.

In what follows, we will find it convenient to impose a \emph{transverse gauge}, i.e.\ we set
% on both the base space vector $g^\mathcal{H}_{ir}$ and the traceless base space two-tensor $g^\mathcal{H}_{\{ij\}}$, i.e. we use  gauge invariance under  to set
\begin{align}
&\tilde{\nabla}^{\tilde{k}} g^\mathcal{H}_{k\,r}=0 \,, \label{metrictran1}\\
& \tilde{\nabla}^{\tilde{k}} g^\mathcal{H}_{\{k\,i\}} =0 \,. \label{metrictran2}
\end{align}
After imposing the transverse gauge, some residual gauge freedom remains, which we use to impose two additional conditions. First, we impose that the constant mode of the trace, $\tilde{g}^\mathcal{H}$, vanishes. Second, we impose that the Killing vector modes of $g^\mathcal{H}_{ir}$ vanish (cf.\ \S\ref{sec:metricappend} for more details).

Solving the homogeneous equation
% Liam changed equation reference here and below
(\ref{metricharm}) is the subject of \S\ref{sec:metricharmappend}. There it is found that in the transverse gauge specified above, equation (\ref{metricharm}) implies that the only nonvanishing metric component is
%all metric components but the transverse-traceless
$g^\mathcal{H}_{\{ij\}}$, i.e.
\begin{equation}
g^\mathcal{H}_{rr} = g^\mathcal{H}_{ir} = \tilde{g}^\mathcal{H} = 0 \,.
\end{equation}
Furthermore, when we expand $g^\mathcal{H}_{\{ij\}}$ in transverse-traceless two-tensor harmonics, equation (\ref{metricharm}) is reduced to a radial equation for the coefficients with the solution (cf.\ equation (\ref{hommetricsol})),
\begin{equation}
 g^\mathcal{H}_{\{ij\}} = \sum_{I_t} \Bigl( g_0^{I_t} \, r^{\Delta(I_t)-2} + g_1^{I_t} \, r^{-\Delta(I_t) +2}\Bigr)\,Y^{I_t}_{\{ij\}}(\Psi) \,,
\end{equation}
where $g_0^{I_t}$ and $g_1^{I_t}$ are integration constants determined by the boundary conditions, and where we have defined
\begin{equation} \label{metricdimension}
 \Delta (I_t) \equiv 2 + \sqrt{\lambda^{I_t}-4}\,.
\end{equation}
The $Y^{I_t}_{\{ij\}}$ are the transverse-traceless symmetric two-tensor harmonics on  $\mathcal{B}_5$,
%the angular base space
\begin{equation}
\tilde{\nabla}^{\tilde{k}}Y^{I_t}_{\{kj\}}=0\,, \quad \tilde{g}^{ij}Y^{I_t}_{\{ij\}} = 0\,,
\end{equation}
that diagonalize the angular Lichnerowicz operator
\begin{equation} \label{Yteigenvalues}
\tilde{\nabla}^2 Y^{I_t}
_{\{ij\}}-2\tilde{\nabla}^{\tilde{k}}\tilde{\nabla}_{(i} Y^{I_t}_{\{j)k\}} =  - \lambda^{I_t} Y^{I_t}_{\{ij\}}\,.
\end{equation}

\subsection{Summary: radial scalings of the homogeneous modes}
\label{sec:homScalings}

In this subsection we summarize the radial scalings of all supergravity fields $\phi$ and the dimensions $\Delta(\phi)$ of the dual operators, as obtained in \S\S\ref{sec:homscalar}, \ref{sec:homflux}, \ref{sec:hommetric}. The results are presented in Table \ref{homNon}, which we now explain.

\begin{table}[h!!]
\begin{center}
\begin{tabular}{lll}
\multicolumn{3}{c}{\bf \emph{Homogeneous Scalings of the Non-Normalizable Modes}} \\
\addlinespace[3pt]
\toprule[1pt]
\rowcolor[gray]{0.85} \textbf{Field} & \textbf{Scaling} & \textbf{Dimension} \\ \addlinespace[2pt]
\toprule[1pt]
\addlinespace[3pt]
$r^{-4} \, \Phi_-^{\mathcal{H}}$ & $r^{\Delta({\Phi_-})-4}$ &  $\Delta({\Phi_-}) = \Delta(I_s) - 4\,, ~\lambda^{I_s}\neq 0$ \\ \addlinespace[3pt]
$G_-^{\mathcal{H}}$ & $r^{\Delta({G_-})-4}$ &
$\Delta({G_-}) = \Delta(\delta^{I_2} \geq 2)$
%$\Delta({G_-}) =
%\left\{\begin{array}{ll}
% -1+\Delta(I_s) \\
% -2+\Delta(I_s)\,,~\lambda^{I_s}\neq 0 \\
% -3+\Delta(I_s)\,,~\lambda^{I_s}\neq 0 \\
%\end{array}\right.$
\\ \addlinespace[3pt]
$\tau^{\mathcal{H}}$ & $r^{\Delta({\tau})-4}$ & $\Delta({\tau}) = \Delta(I_s)\,,~ \lambda^{I_s} \neq 0$ \\ \addlinespace[3pt]
$r^{-2} g_{\{ij\}}^{\mathcal{H}}$ & $r^{\Delta({g})-4}$ &   $\Delta({g}) = \Delta(I_t)$ \\ \addlinespace[3pt]
$G_+^{\mathcal{H}}$ & $r^{\Delta({G_+})-4}$ &
$\Delta({G_+}) = \Delta(\delta^{I_2} \geq 2), \, \Delta(\delta^{I_2} \leq -2)$
%$\Delta({G_+}) = \Delta({G_-}) ,\,4+\Delta({G_-}) $
\\ \addlinespace[3pt]
$r^{4} (\Phi_+^{-1})^{\mathcal{H}}$ & $r^{\Delta({\Phi^{-1}_+})-4}$ & $\Delta({\Phi^{-1}_+}) = \Delta(I_s) + 4$ \\ \addlinespace[6pt]

\multicolumn{3}{c}{\bf \emph{Homogeneous Scalings of the Normalizable Modes}} \\
\addlinespace[3pt]
\toprule[1pt]
\rowcolor[gray]{0.85} \textbf{Field} & \textbf{Scaling} & \textbf{Dimension} \\ \addlinespace[2pt]
\toprule[1pt]
\addlinespace[3pt]
$r^{-4} \, \Phi_-^{\mathcal{H}}$ & $r^{-\Delta({\Phi_-})}$ &  $\Delta({\Phi_-}) = \Delta(I_s) + 4$ \\ \addlinespace[3pt]
$G_-^{\mathcal{H}}$ & $r^{-\Delta({G_-})}$ &
$\Delta({G_-}) = \Delta(\delta^{I_2} \leq -2),\,\Delta(b_2)$
%$\Delta({G_-})
%=\left\{\begin{array}{ll}
% -1+\Delta(I_s) \\
% -2+\Delta(I_s)\,,~\lambda^{I_s}\neq 0 \\
% -3+\Delta(I_s)\,,~\lambda^{I_s}\neq 0 \\
% 4\,,~b_2\neq 0 \\
%\end{array}\right.$
\\ \addlinespace[3pt]
$\tau^{\mathcal{H}}$ & $r^{-\Delta({\tau})}$ & $\Delta({\tau}) = \Delta(I_s)$ \\ \addlinespace[3pt]
$r^{-2} g_{\{ij\}}^{\mathcal{H}}$ & $r^{-\Delta({g})}$ &   $\Delta({g}) = \Delta(I_t)$ \\ \addlinespace[3pt]
$G_+^{\mathcal{H}}$ & $r^{-\Delta({G_+})}$ &
$\Delta({G_+}) = \Delta(\delta^{I_2} \geq 2),\, \Delta(\delta^{I_2} \leq -2),\, \Delta(b_2)$
%$\Delta({G_+}) =\Delta({G_-}) ,\, 4 + \Delta({G_-})$
\\ \addlinespace[3pt]
$r^{4} (\Phi_+^{-1})^{\mathcal{H}}$ & $r^{-\Delta({\Phi^{-1}_+})}$ & $\Delta({\Phi^{-1}_+}) = \Delta(I_s) - 4\,,~\lambda^{I_s}\neq 0$
\end{tabular}
\end{center}
\caption{The radial scalings of the homogeneous modes of the supergravity fields. Here $\Delta(I_s) = 2 + \sqrt{4+\lambda^{I_s}}$, where the $\lambda^{I_s}$ are the eigenvalues of the angular scalar Laplacian, cf.\ equation (\ref{Yeigenvalue}). Furthermore, $\Delta(I_t) = 2 + \sqrt{\lambda^{I_t} - 4}$, where the $\lambda^{I_t}$ are the eigenvalues of the angular Lichnerowicz operator, cf.\ equation (\ref{Yteigenvalues}).
The expressions
% May 31
% May 31 3 PM
$\Delta(\delta^{I_2} \geq 2)$, $\Delta(\delta^{I_2} \leq -2)$, and $\Delta(b_2)$ appearing in the flux dimensions  can be found in equations (\ref{fluxdimensions1}, \ref{fluxdimensions0}, \ref{fluxdimensions2}).
Although we have not explicitly displayed this in the tables, the modes of $G_\pm$ can have additional logarithmic running of the form $r^{\Delta_G-4}\,\ln r$ and $r^{-\Delta_G}\,\ln r$ for the non-normalizable and normalizable modes,
respectively; cf.\ equations (\ref{IASD}, \ref{ISD}).
%For the non-normalizable modes the zero modes of $\Phi^\mathcal{H}_-$ and $\tau^\mathcal{H}$ are excluded since for $\Phi^\mathcal{H}_{-}$ it can be gauged away while for $\tau^\mathcal{H}$ it can be absorbed into the background. For the normalizable modes the zero mode of $(\Phi^{-1}_+)^\mathcal{H}$ is excluded since it can also be absorbed into the background.
}\label{homNon}
\end{table}

For canonically normalized fields $\phi$, the radial scalings of the modes and the dimensions of the operators of the dual field theory are related via the standard AdS formula (\ref{generalHom}).
To start with, the scalar field $\tau$ is canonically normalized, so the dimension of the operator dual to $\tau$ is given by
\begin{equation}
\Delta(\tau) = \Delta(I_s) = 2 + \sqrt{4 + \lambda^{I_s}} \,.
\end{equation}
The same is true for the potential $A_2$, and the dimensions $\Delta(G_\pm)$ can be read off from (\ref{fluxdimensions1}, \ref{fluxdimensions0}, \ref{fluxdimensions2}), taking into account the discussion at the end of \S\ref{fluxdimensionappend}.
%at the end of \S\ref{sec:homflux}.
For the ISD flux $G_+$ both $A_+$ and $A_-$ modes can be turned on, so that all modes are present except for non-normalizable Betti modes:
\begin{alignat}{2}
 \textit{Non-normalizable:} \quad \Delta(G_+) & = \Delta(\delta^{I_2} \geq 2),\, \Delta(\delta^{I_2} \leq -2) \,, \\
 \textit{Normalizable:} \quad \Delta(G_+) & = \Delta(\delta^{I_2} \geq 2),\, \Delta(\delta^{I_2} \leq -2),\, \Delta(b_2) \,,
\intertext{while for the IASD flux $G_-$ only $A_+$ can be turned on, and we have}
 \textit{Non-normalizable:} \quad \Delta(G_-) & = \Delta(\delta^{I_2} \geq 2) \,, \\
 \textit{Normalizable:} \quad \Delta(G_-) & = \Delta(\delta^{I_2} \leq -2),\, \Delta(b_2) \,,
\end{alignat}
where the expressions for $\Delta(\delta^{I_2} \geq 2)$, $\Delta(\delta^{I_2} \leq -2)$, and $\Delta(b_2)$ are given in equations (\ref{fluxdimensions1}), (\ref{fluxdimensions2}), and (\ref{fluxdimensions0}), respectively.

Next, it is the warped internal metric $e^{-2A}\,g_{\{ij\}}\sim r^{-2}\,g_{\{ij\}}$ that is the canonical field \cite{CeresoleHarm, CeresoleSCFT}, corresponding to a dual operator with dimension
\begin{equation}
 \Delta(g) = \Delta(I_t) = 2 + \sqrt{\lambda^{I_t} - 4} \,,
\end{equation}
as anticipated by the notation.  Finally, $\Phi_-$ and $\Phi^{-1}_+$ are not canonical fields, but as explained in \cite{Holographic}, the combinations $r^{-4} \Phi_-$ and $r^4 \Phi^{-1}_+$ exhibit the same radial scaling as do the corresponding canonical variables.   Now comparing the non-normalizable and normalizable modes of $r^{-4} \Phi_-$ with equation (\ref{generalHom}) one can identify the operator dimensions
\begin{alignat}{2}
 \textit{Non-normalizable:} \quad \Delta(\Phi_-) & = \Delta(I_s) - 4 \,, \\
 \textit{Normalizable:} \quad \Delta(\Phi_-) & = \Delta(I_s) + 4 \,.
\intertext{Similarly, by comparing the non-normalizable and normalizable modes of $r^{4} \Phi^{-1}_+$ with (\ref{generalHom}) one can identify the operator dimensions}
 \textit{Non-normalizable:} \quad \Delta(\Phi^{-1}_+) & = \Delta(I_s) + 4 \,, \\
 \textit{Normalizable:} \quad \Delta(\Phi^{-1}_+) & = \Delta(I_s) - 4 \,.
\end{alignat}
Notice that $\Delta(\Phi_-)$ and $\Delta(\Phi^{-1}_+)$ exchange roles in going from the normalizable modes to the non-normalizable modes.

In
Table \ref{homNon} we have excluded the zero modes of both $\tau$ and $\Phi_-$ for the non-normalizable modes
(scaling like $r^{0}$) while for the normalizable modes we have excluded that of $\Phi_+^{-1}$ (scaling like $r^{-4}$). For $\tau$,
the non-normalizable zero mode corresponds to a constant shift of the axion $\mathrm{Re} \, \tau \equiv C_0$ and the dilaton $\mathrm{Im} \, \tau \equiv e^{-\phi}$.
A constant shift of the dilaton can be absorbed in the background value of $g^{-1}_s \equiv \mathrm{Im} \, \tau^{(0)}$, while the axion $C_0$ is shift-symmetric. The non-normalizable zero mode of $\Phi_-$ can be gauged away using a constant shift of $\alpha$, thus preserving the background $\Phi^{(0)}_-=0$. The normalizable zero mode of $\Phi_+^{-1}$ corresponds to a shift of  the
% Sohang background value of the warp factor,
constant $C_1$ in the warp factor (\ref{UVwarp}),
which we will also absorb into the background value.
%$C_1$ in the warp factor and let us into the background.

\section{Inhomogeneous Modes:  Green's Function Solutions}
\label{sec:Greens}
The final ingredient of our expansion scheme is the set of inhomogeneous solutions to equations (\ref{nphi-}-\ref{nphi+}). In this section we will write down the Green's function solutions for the inhomogeneous scalar, flux and metric modes, again relegating detailed derivations to the appendix. As discussed in \S\ref{sec:setup}, the structure of the equations is the same at every order. Thus, we only need to write down one set of scalar, flux and metric Green's functions, $\mathcal{G}_s$, $\mathcal{G}_{G_\pm}$, and $\mathcal{G}_{g}$, which are used at all orders.

The initial seeds for the inhomogeneous pieces are the homogeneous solutions obtained in \S\ref{sec:harmonic}. The homogeneous modes are given by angular harmonics multiplying radial powers $r^\alpha$ (possibly including logarithmic running $(\ln r)^m$, in the case of flux). Thus, the source terms are of a non-localized nature, and the standard Green's functions for localized sources give divergences at the origin and at infinity when convoluted with the non-localized sources. One could introduce regulated Green's function with cutoffs at $r_\text{IR}$ and $r_\text{UV}$, but these introduce large counterterms, and in what follows we will take a more direct route by solving the equations explicitly.
%Again, we split up the calculation in terms of scalars, fluxes and metric.

\subsection{Inhomogeneous  solutions for the scalars}
From equations (\ref{nphi-}, \ref{ntau}, \ref{nphi+}) we see that $n$-th order perturbations of the scalar fields $\Phi_-, \Phi_+^{-1}$ and $\tau$ obey Poisson's equation on the cone
\begin{equation} \label{poisson}
 \nabla^2_{(0)} \Phi^{(n)} = \mathcal{S}^{(n)}_{\Phi} \,,
\end{equation}
where $\nabla^2_{(0)}$ is constructed from the background metric of the cone, equation (\ref{cone}).
%\begin{align}
%\nabla^{2}_{(0)} \Phi_-^{(n)} & = \mathcal{S}^{(n)}_{\Phi_-} \,, \\
%\nabla^{2}_{(0)} \tau^{(n)} & = \mathcal{S}^{(n)}_{\tau} \,, \\
%\nabla^{2}_{(0)} \big(\Phi^{-1}_+\big)^{(n)} & = \mathcal{S}^{(n)}_{\Phi_+^{-1}} \,.
%\end{align}
The source dependence on the
% Sohang the
fields at order $n$ can be read off explicitly from equations (\ref{nphi-}, \ref{ntau}, \ref{nphi+}), while the dependence on the fields at order $m<n$ is left implicit:
%LM Correction below
\begin{align}
 \mathcal{S}^{(n)}_{\Phi_-} & = \mathrm{Source}_{\Phi_-}(\phi^{m<n}) \,, \\
 \label{phi+sourcen} \mathcal{S}^{(n)}_{\Phi_+^{-1}} & = {g_s \over 96} \left( G_+^{(0)}\cdot \bar{G}_+^{(n)}+G_+^{(n)}\cdot
\bar{G}_+^{(0)} +3G_{+~m_1n_1l_1}^{(0)}
\bar{G}_{+~m_2n_2l_2}^{(0)}g_{(0)}^{m_1m_2}g_{(0)}^{n_1n_2}g_{(n)}^{l_1l_2} \right) \\
&- {g_s^2\over
96} \mathrm{Im}\,\tau^{(t)}|G_+^{(0)}|^2\
+ \Big[ {g_s\over 48}\,(\Phi_+^{-1})^{(0)}\,|G_+|^2_{(0)}  -  2 (\Phi_+^{-4})^{(0)} \, (\nabla\Phi_+)^2_{(0)} \Big]\,  \Phi_-^{(n)} + \mathrm{Source}_{\Phi_+}(\phi^{m<n}) \,,\nonumber \\
\mathcal{S}^{(n)}_{\tau} & = \mathrm{Source}_{\tau}(\phi^{m<n}) -i \,\Phi_+^{(0)}\,G_+^{(0)}\cdot G_-^{(n)} \,. \label{taungreen}
\end{align}
%The source terms depend on the background fields, fields at lower order in perturbation theory as well as fields at the same order. Thus, as discussed in \S\ref{}, we have to solve the equations according to order dictated by the tower-like structure, equations (\ref{nphi-}-\ref{nphi+}), with the homogeneous modes as initial seeds. The homogeneous modes are given by angular harmonics $Y(\Psi)$ together with various radial powers $r^\alpha (\ln r)^m$.

We start by expanding the fields and the sources in terms of angular harmonics
\begin{align}
 \Phi^{(n)}(r,\Psi) & = \sum_{I_s} \Phi^{(n)}_{I_s}(r) \, Y^{I_s}(\Psi)\,, \\
 \mathcal{S}^{(n)}_{\Phi}(r,\Psi) & = \sum_{I_s} \mathcal{S}^{(n)}_{I_s}(r) \, Y^{I_s}(\Psi) \,,
\end{align}
so that Poisson's equation (\ref{poisson}) reduces to an equation for the radial coefficients
\begin{equation} \label{radialpoisson}
 \left( \partial_r^2 + \frac{5}{r} \partial_r - \frac{\lambda^{I_s}}{r^2} \right) \Phi^{(n)}_{I_s}(r) = \mathcal{S}^{(n)}_{I_s}(r) \,.
\end{equation}
As discussed above, the source $\mathcal{S}_{I_s}$ will involve a sum of various radial scalings due to the homogeneous modes\begin{equation}
 \mathcal{S}_{I_s}(r) = \sum_{\alpha,m} \mathcal{S}^{(n)}_{I_s}(\alpha,m) \, r^{\alpha} \, (\ln r)^{m} \,,
\end{equation}
and the inhomogeneous solution to Poisson's equation (\ref{poisson}) is
\begin{equation}
 \Phi^{(n)}_\mathcal{IH}(r) = \sum_{I_s} \sum_{\alpha,m} \Phi^{(n)}_{I_s}(r;\alpha,m) \, Y^{I_s}(\Psi) \,,
\end{equation}
where $\Phi^{(n)}_{I_s}(r;\alpha,m)$ is given in equations (\ref{singlesol1}, \ref{singlesol2}). The solution for $\Phi^{(n)}_{I_s}(r;\alpha,m) $ depends on the value of $\alpha$:

\paragraph{Case: $\alpha + 2 \neq -2 \pm(\Delta(I_s)-2)$.}
The solution to equation (\ref{radialpoisson}) is given by
\begin{equation} \label{singlesol1}
 \Phi^{(n)}_{I_s}{(r;\alpha,m}) = \mathcal{S}^{(n)}_{I_s}(\alpha,m) \, r^{\alpha+2} \, \big( a_0 + a_1 \ln(r) + \ldots + a_m \, (\ln r)^m \big) \,,
% \left( \sum_{k=0}^{m} a_k^{(m)} \, (\ln r)^k \,,
\end{equation}
where the coefficients $a_k$ are given by
\begin{equation}
 a_k = (-1)^{k+m+1} \, \frac{m!/k!}{2 \, \Delta(I_s)-4} \left[ (\alpha+2+\Delta(I_s))^{k-1-m} - (\alpha+2-\Delta(I_s)+4)^{k-1-m} \right] \,.
\end{equation}

\paragraph{Case: $\alpha + 2 = -2 \pm (\Delta(I_s)-2)$.} The  solution to equation (\ref{radialpoisson}) is given by
\begin{equation} \label{singlesol2}
 \Phi^{(n)}_{I_s}{(r;\alpha,m}) = \mathcal{S}^{(n)}_{I_s}(\alpha,m) \, r^{\alpha+2} \, \big( b_0 + b_1 \ln(r) + \ldots + b_{m+1} \, (\ln r)^{m+1} \big) \,,
% \left( \sum_{k=0}^{m} a_k^{(m)} \, (\ln r)^k \,,
\end{equation}
where the coefficients $b_k$ are given by
\begin{equation}
 b_k = (-1)^{k+m+1} \, \frac{m!}{k!} \, (\pm2\Delta(I_s)\mp4)^{k-2-m} \,, \quad \alpha +2 = -2 \pm(\Delta(I_s)-2) \,.
\end{equation}

\subsection{Inhomogeneous  solutions for the fluxes}

We now find the inhomogeneous modes for $G_\pm$ solving equations (\ref{nG-}, \ref{niasd}, \ref{nG+}, \ref{nisd}). The equations of motion for the $n$-th order perturbation of $G_-$ take the form
\begin{align}
 \dd \left( \Phi_+^{(0)} G_-^{(n)} + \mathcal{S}^{(n)}_{G_-,1} \right) & = \mathcal{S}^{(n)}_{G_-,3} \,, \\
 (\star_6^{(0)} + i) \, \Phi_+^{(0)} G_-^{(n)} & = \mathcal{S}^{(n)}_{G_-,2} \,.
\end{align}
Here the sources $\mathcal{S}^{(n)}_{G_-,1}, \mathcal{S}^{(n)}_{G_-,2}$ are three-forms and $\mathcal{S}^{(n)}_{G_-,3}$ is a four-form. The expressions for the sources in terms of the $n$-th order fields can be read off from equations (\ref{nG-}, \ref{niasd}), where again the dependence on the fields at lower order is left implicit,
\begin{align}
 \mathcal{S}^{(n)}_{G_-,1} & = \Phi_-^{(n)}\,G_+^{(0)} + \mathrm{Source}_{G_-,1}(\phi^{m<n}) \,, \\
 \mathcal{S}^{(n)}_{G_-,2} & = \mathrm{Source}_{G_-,2}(\phi^{m<n}) \,, \\
 \mathcal{S}^{(n)}_{G_-,3} & = \mathrm{Source}_{G_-,3}(\phi^{m<n}) \,.
\end{align}
The equations of motion for the $n$-th order perturbation of $G_+$ are similar to those of  $G_-$:
\begin{align}
 \dd \left( G_+^{(n)} + \mathcal{S}^{(n)}_{G_+,1} \right) & = 0 \,, \\
 (\star_6^{(0)} - i) \, G_+^{(n)} & = \mathcal{S}^{(n)}_{G_+,2} \,,
\end{align}
where the three-form sources $\mathcal{S}^{(n)}_{G_+,1}, \mathcal{S}^{(n)}_{G_+,2}$ can be read off from equations (\ref{nG+}, \ref{nisd})
\begin{align}
 \mathcal{S}^{(n)}_{G_+,1} & = - G_-^{(n)} + 2 i \tau^{(n)} \, H_3^{(0)} + \mathrm{Source}_{G_+,1} (\phi^{m<n}) \,, \\
 \mathcal{S}^{(n)}_{G_+,2} &=  \mathrm{Source}_{G_-,2}(\phi^{m<n}) \,.
\end{align}
Both systems are of the form
\begin{align}
 \dd \left( \Sigma_\pm + \mathcal{S}_1 \right) &= \mathcal{S}_3 \,, \\
 (\star_6^{(0)} \mp i) \, \Sigma_\pm & = \mathcal{S}_2 \,,
\end{align}
with $\Sigma_- = \Phi^{(0)}_+ G_-^{(n)}$ and $\Sigma_+ = G^{(n)}_+$.
%and.
We first solve the two simpler systems
%\begin{align}
% \dd \left( \Sigma_\pm^\mathrm{I} + \mathcal{S}_1 \right) &= 0 & \dd \Sigma_\pm^\mathrm{II} &= \mathcal{S}_3 \\
% (\star_6^{(0)} \mp i) \, \Sigma^\mathrm{I}_\pm & = \mathcal{S}_2 & (\star_6^{(0)} \mp i) \, \Sigma^\mathrm{II}_\pm & = 0
%\end{align}
\begin{equation}
\mathrm{I}: \quad \left.
\begin{array}{l}
 \dd \left( \Sigma_\pm^{(\mathrm{I})} + \mathcal{S}_1 \right) = 0 \\
 (\star_6^{(0)} \mp i) \, \Sigma^{(\mathrm{I})}_\pm = \mathcal{S}_2
\end{array}
\right. \,,
~~~~~~~~~~~~~~~
\mathrm{II}: \quad
\left.
\begin{array}{l}
 \dd \Sigma_\pm^{(\mathrm{II})} = \mathcal{S}_3 \\
 (\star_6^{(0)} \mp i) \, \Sigma^{(\mathrm{II})}_\pm = 0
\end{array}
\right. \,.
\end{equation}
By linearity the full solution is $\Sigma_\pm = \Sigma^{(\mathrm{I})}_\pm + \Sigma^{(\mathrm{II})}_\pm$. The solving of I and II is the subject of \S\ref{sec:fluxgreensappend} and here we only present the results.

\paragraph{Flux Green's function I:}
From the first equation we see that $\Sigma^{(\mathrm{I})}_\pm + \mathcal{S}_1$ is closed and can locally be expressed as $\dd \chi_\pm$ for some two-form $\chi_\pm$.  The solution in terms of this potential is
\begin{align}
 \Sigma^{(\mathrm{I})}_\pm & = - \mathcal{S}_1 + \dd \chi_\pm \,, \\
 \chi_\pm(y) & = \int_{\mathcal{C}_6} \mathcal{G}^{(\mathrm{I})}_{G}(y,y') \wedge \left( \mathcal{S}_2 + \big(\star_6^{(0)} \mp i \big) \mathcal{S}_1 \right)(y') \,,
\end{align}
where the explicit form of $\mathcal{G}^{(\mathrm{I})}$ is given in equation (\ref{fluxgreens1}). The  indices of the above equation should be interpreted in the following way: the Green's function $(\mathcal{G}^{(\mathrm{I})})_{mn,p'q's'}$ has two legs in the $y$ coordinate system and three legs in the $y'$ coordinate system. When we wedge $\mathcal{G}_G^{(\mathrm{I})}$ with the three-form source $\mathcal{S}_2 + (\star_6^{(0)} \mp i ) \mathcal{S}_1$ we produce a six-form in the $y'$ coordinates which is integrated over the whole manifold $\mathcal{C}_6$, resulting in a two-form $\chi_\pm(y)$ in the $y$ coordinate system.
\paragraph{Flux Green's function II:} In a similar way the solution to system II is given by
\begin{equation}
 \Sigma^{(\mathrm{II})}_\pm = \int_{\mathcal{C}_6} \mathcal{G}^{(\mathrm{II})}_{G}(y,y') \wedge \mathcal{S}_3(y') \,,
\end{equation}
where the explicit form of $\mathcal{G}^{(\mathrm{II})}$ is given in equation (\ref{fluxgreens2}). Here $\mathcal{S}_3$ is a four-form and $(\mathcal{G}^{(\mathrm{II})})_{mnp,q's'}$ is a $(3+2')$-form producing a three-form $\Sigma^{(\mathrm{II})}_\pm$.

\subsection{Inhomogeneous  solutions for the metric}
The $n$-th order perturbations of the metric $g_{mn}$ obey
\begin{equation}
 \Delta_K^{(0)} g_{mn}^{(n)} = (\mathcal{S}^{(n)}_g)_{mn} \,,
\end{equation}
where the source can be read off from equation (\ref{nmetric}),
\begin{align}
 (\mathcal{S}^{(n)}_{g})_{mn} & = {\Phi_+^{(0)} \over 16
\mathrm{Im}\,\tau} \left( G^{(0)\phantom{(m}pq}_{+\,(m }\, \bar{G}^{(n)}_{-\,n)\,pq} + G^{(n)\phantom{(m}pq}_{-\,(m }\, \bar{G}^{(0)}_{+\,n)\,pq} \right) \\
& - 4 \big(\Phi_+^{(0)}\big)^{-2} \, \n^{(0)}_{(m}\Phi_+^{(0)}\n^{(0)}_{n)}\Phi_-^{(n)}+\mathrm{Source}_g(\phi^{m<n}) \,. \nonumber
\end{align}

As in the homogeneous case, we utilize the cone structure and decompose the metric perturbations into irreducible pieces under general coordinate transformations of the base space. We continue to impose a \emph{transverse} gauge on the irreducible vector and tensor at each order in perturbation theory, i.e.\ we set
\begin{align}
&\tilde{\nabla}^{\tilde{k}} g^{(n)}_{k\,r}=0 \,, \label{inhommetrictran1}\\
& \tilde{\nabla}^{\tilde{k}} g^{(n)}_{\{k\,i\}} =0 \,, \label{inhommetrictran2}
\end{align}
together with the additional constraint on the constant mode and Killing vector modes as discussed in \S\ref{sec:hommetric}. We end up with a Green's function solution of the form
\begin{equation}
 (g_{mn}^{(n)})^\mathcal{IH}(y) = \int_{\mathcal{M}'} \dd^6 y' \sqrt{g'} \,\,(\mathcal{G}_{g})_{mn}^{\phantom{mn}m'n'} (y,y') \, (\mathcal{S}^{(n)}_{g})_{m'n'}(y') \,.
\end{equation}
The Green's function $(\mathcal{G}_{g})_{mn}^{\phantom{mn}m'n'}(y,y')$  is valid only in the gauge specified above, cf.\ equations (\ref{metricgauge1}-\ref{metricgauge4}) in \S\ref{sec:metricgreensappend}. Note that all components not listed in (\ref{MetricGreens1}-\ref{MetricGreens6}) vanish in this gauge.

\section{Radial Scalings of Corrections}
\label{sec:scalings}

The results described above depend implicitly and explicitly on the angular harmonics, and corresponding eigenvalues, associated with the scalar, flux, and metric perturbations.  Thus, although one can use our results to obtain an explicit solution to any desired order on a cone whose angular harmonics are known (e.g., the conifold), this is little consolation when one is faced with computing the eigenfunctions in a more general example.
%%In particular,
%can be challenging, particularly for base spaces $\mathcal{B}_5$ that are not homogeneous.
Fortunately, for many questions of physical interest\footnote{For example, one might want to  estimate the scale of the mass term induced for some object, such as an anti-D3-brane \cite{AAB} or a D3-brane \cite{LiamsLong}, or determine the soft masses in a toy visible sector \cite{Dym,Paul,Seq}.} it suffices to determine how corrections scale with $r$, obviating the full Green's function solution.  In this section we present results adapted to extracting radial scalings without obtaining  the full angular dependence of the corresponding solutions.

The main result of this section is equation (\ref{genscaling}), which qualitatively states that the  $n$-th order correction $\phi^{(n)}$ of a field $\phi$   scales like a sum of products of $n$ harmonic modes
\begin{equation} \label{harmonicproduct}
 \hat\phi^{(n)} \sim \sum_{i_1, \ldots, i_n} \hat\phi^\mathcal{H}_{i_1} \cdots \hat\phi^\mathcal{H}_{i_n} \,,
\end{equation}
where the sum runs over subsets of the fields $\{\hat{\Phi}_+,\, \hat{G}_+,\, \hat{\tau},\, \hat{g}_{\{ij\}},\, \hat{G}_-,\, \hat{\Phi}_+^{-1} \}$, and the hatted variables are defined in equation (\ref{canvar}). Throughout this section we will use $\sim$ to signify that two objects have the same radial scaling, but may differ by order-unity angular functions, e.g.\ we will write
$r^\alpha \chi_1(\Psi)  \sim r^\alpha \chi_2(\Psi)$, for angular functions $\chi_{1,2}(\Psi)$ that are of order unity at generic points.

One complication in equation (\ref{harmonicproduct}) is that not every possible product of harmonic modes contributes in the sum, and one must trace through the expanded equations (\ref{nphi-}-\ref{nphi+}) to see which combinations appear for a given field.
%This is quite clear at first order: e.g.
For example, from equation (\ref{nphi-}) for $\Phi_-$, one sees that none of the harmonic modes apart from $\Phi_-$ itself contributes to the correction at first order.
The results from the first and second order calculations are presented in Tables \ref{inhomlin} and \ref{inhomsec}, respectively.  We expect that at higher order in the expansion, all possible products will contribute, as the
% Sohang equations become more richly interwoven.
number of ways a particular combination can propagate
% Liam up
through the equations of motion becomes large.

When checking which products of harmonics appear, we will not rule out the possibility that contractions of indices or convolutions of angular harmonics with Green's functions result in a vanishing contribution. If a particular mode is critical to an analysis, the associated product would need to be examined in detail  by tracing through the equations of motion.

\subsection{First-order and second-order scalings}  % Liam
\label{sec:linscalings}

We begin by determining the radial scalings of the inhomogeneous modes at first order, in terms of the first-order homogeneous modes obtained in \S\ref{sec:homScalings}.
%We work down the tower like structure and show that $\hat{\phi}^{(n)} \sim \sum_i \hat\phi_i^\mathcal{H}$, including the correct range of the sum for each field, conveniently summarized in Table \ref{inhomlin}.
%Then, using the results of \S\ref{sec:homScalings}, where the  radial scalings of the harmonic modes were found to be $\hat\phi^\mathcal{H} \sim \sum_{\Delta(\phi)} (r^{\Delta(\phi)-4}+r^{-\Delta(\phi)})$, we infer the radial scalings of the first-order perturbations.
To make full use of the triangular structure of the equations of motion, we begin at the top of the triangle, with the scalar field $\Phi^{(1)}_-$, and work our way downward.

\paragraph{First level $\Phi^{(1)}_-$:} At first order, equation (\ref{nphi-}) for $\Phi_-^{(1)}$ reads $\nabla^2_{(0)} \Phi^{(1)}_- = 0$, so that $\Phi^{(1)}_-$ is solely determined by its harmonic mode,
\begin{equation} \label{phi1}
 \Phi_-^{(1)} = \Phi_-^{\mathcal{H}} \,.
\end{equation}

\paragraph{Second level $G^{(1)}_-$:} From equation (\ref{nG-}) we get at first order $\dd ( \Phi_+^{(0)} G^{(1)}_-) = - \dd (\Phi^{(1)}_- G^{(0)}_+)$, so that $G_-^{(1)}$ is sourced by $\Phi^{(1)}_-$. Using equation (\ref{phi1}) for $\Phi^{(1)}_-$ together with the radial scalings of the background fields, $\Phi_+^{(0)} \sim r^{-4}$ and $G^{(0)}_+ \sim r^0$, we infer that
\begin{equation}
 G_-^{(1)} \sim  r^{-4}\Phi_-^{\mathcal{H}} + G_-^{\mathcal{H}}\,, \label{1G-Scaling}
\end{equation}
where we  also include the homogeneous contribution $G_-^\mathcal{H}$ in the first-order solution.

\paragraph{Third level $\tau^{(1)}$:} Equation (\ref{ntau}) for $\tau^{(1)}$ reads at first order $\nabla^2_{(0)} \tau^{(1)} = \Phi_+^{(0)} /(48i) \, G_+^{(0)} \cdot G^{(1)}_{-}$.
To find the radial scaling for $\tau^{(1)}$ we substitute the radial scaling for $G^{(1)}_-$, equation (\ref{1G-Scaling}), and the radial scalings for the background fields, yielding
\begin{equation}
 \tau^{(1)} \sim r^{-4} \Phi_-^{\mathcal{H}} + G_-^{\mathcal{H}} + \tau^{\mathcal{H}}\,.
\end{equation}
Thus, $\tau^{(1)}$ inherits a dependence on $r^{-4} \Phi_-^\mathcal{H}$ through the solution for $G^{(1)}_-$.
%As we work down the tower, the fields at the lower levels inherit the scalings of the fields at the higher levels.

\paragraph{Higher levels $g^{(1)}_{\{ij\}}, G^{(1)}_+, (\Phi^{-1}_+)^{(1)}$:} We continue in a similar manner, solving for the radial scalings of all the fields.  The result is most efficiently presented in terms of new fields $\hat\phi$, which are defined
such that they scale with $r$ in the same way as the corresponding canonical degrees of freedom:
\begin{equation} \label{canvar}
 \hat{\Phi}_- \equiv r^{-4} \Phi_- \,, \quad
 \hat{G}_- \equiv G_- \,, \quad
 \hat{\tau} \equiv \tau \,, \quad
 \hat{g}_{mn} \equiv r^{-2} g_{mn}\,, \quad
 \hat{G}_+ \equiv G_+ \,, \quad
 \hat{\Phi}^{-1}_+ \equiv r^{4} \Phi^{-1}_+ \,.
\end{equation}
Then, the radial scalings at linear order are very simple:
{\allowdisplaybreaks
\begin{alignat}{13}
{\hat{\Phi}_{-}}^{(1)} & \sim{} & {\hat{\Phi}_{-}}^{\mathcal{H}} & \,, \label{1phi-scale} \\
{\hat{G}_-}^{(1)} & \sim{} & {\hat{\Phi}_{-}}^{\mathcal{H}} &+{}& {\hat{G}_-}^{\mathcal{H}} & \,, \label{1G-scale} \\
\hat{\tau}^{(1)} & \sim{} & {\hat{\Phi}_{-}}^{\mathcal{H}} & +{} & {\hat{G}_-}^{\mathcal{H}} &+{}&\hat{\tau}^{\mathcal{H}} & \,, \label{1tauscale}\\
\hat{g}_{ij}^{(1)} & \sim{} & {\hat{\Phi}_{-}}^{\mathcal{H}} &+{}& {\hat{G}_-}^{\mathcal{H}} &{} +{}&{}&+{}&  \hat{g}_{\{ij\}}^{\mathcal{H}} & \,, \label{1metricscale}\\
\hat{G}_+^{(1)} & \sim{} & {\hat{\Phi}_{-}}^{\mathcal{H}} &+{}& {\hat{G}_-}^{\mathcal{H}} &+{}& \hat{\tau}^{\mathcal{H}} & +{} &{}&+{}& \hat{G}_+^{\mathcal{H}} &\,, \label{1G+scale}\\
({\hat{\Phi}^{-1}_+})^{(1)} & \sim{} & {\hat{\Phi}_{-}}^{\mathcal{H}} &+{}& {\hat{G}_-}^{\mathcal{H}}  &+{}& \hat{\tau}^{\mathcal{H}} &+{}& \hat{g}_{\{ij\}}^{\mathcal{H}} &{}+& \hat{G}_+^{\mathcal{H}}  &+{}& ({\hat{\Phi}^{-1}_+})^{\mathcal{H}} \label{1phi+scale}\,.
\end{alignat}
}%endallowdisplaybreaks
Notice that in terms of the fields $\hat\phi$, the first-order perturbation takes the simple form $\hat{\phi}^{(1)} \sim \sum_i \hat\phi_i^\mathcal{H}$.
The content of equations (\ref{1phi-scale}-\ref{1phi+scale}) is also summarized in Table \ref{inhomlin}.

It is now easy to obtain the radial scaling for the first-order fields, using the results for the harmonic scalings obtained in \S\ref{sec:homScalings}.
%, where we showed that the canonically normalized fields take the form
%\begin{align}\label{homogeneouscn}
%\hat{\phi}^{\mathcal{H}}(r,\Psi) & = \sum_{\Delta(\phi)} \, \left[ c_0^{\Delta(\phi)}\,\left({r\over r_{\star}}\right)^{\Delta(\phi)-4} Y_0^{\Delta(\phi)}(\Psi) + c_1^{\Delta(\phi)}\,\left({r\over r_{\star}}\right)^{-\Delta(\phi)} Y_1^{\Delta(\phi)}(\Psi) \, \right] \,.
%\end{align}
%Here we define the integration constants $c^{\Delta(\phi)}_{0}$ and $c^{\Delta(\phi)}_{1}$ such that they determine the sizes of the non-normalizable and normalizable homogeneous modes, at the scale $r=r_{\star}$, using that the angular harmonics $Y_{0,1}^{\Delta(\phi)}(\Psi)$ are of order unity at generic points of the angular space.
Restricting attention henceforth to the non-normalizable modes, we find that the radial scalings and the sizes of the modes at first order are
\begin{equation} \label{firstorderscaling}
 \hat{\phi}^{(1)}(r,\Psi) = \sum_{\phi} \sum_{\Delta(\phi)} c_0^{\Delta(\phi)}\,\left({r\over r_{\star}}\right)^{\Delta(\phi)-4} h^{\Delta(\phi)}_0(\Psi) \,.
\end{equation}
where the first sum runs over contributing fields, and the explicit form of the angular functions $h^{\Delta(\phi)}_{0}(\Psi)$ can  be obtained from the full Green's function analysis.

As an example, Table \ref{inhomlin} together with equation (\ref{firstorderscaling}) dictates that the solution for the first-order perturbation $\hat{G}_-^{(1)}(r,\Psi)$ takes the form
\begin{align}
 \hat{G}_-^{(1)}(r,\Psi) & = \sum_{\Delta(\Phi_-)} \left( c_0^{\Delta(\Phi_-)} \, \left({r\over r_{\star}}\right)^{\Delta(\Phi_-)-4} h_0^{\Delta(\Phi_-)}(\Psi)  \right)
 & + \sum_{\Delta(G_-)} \left( c_0^{\Delta(G_-)}\,\left({r\over r_{\star}}\right)^{\Delta(G_-)-4} h_0^{\Delta(G_-)}(\Psi)  \right) \,,\nonumber
\end{align}
for some order-unity angular functions $h_{0}^{\Delta(\Phi_-)}(\Psi)$, $h_{0}^{\Delta(G_-)}(\Psi)$.

\begin{table}[h!!]
\begin{center}
\begin{tabular}{l@{~~~\,}c@{~~~\,}c@{~~~\,}c@{~~~\,}c@{~~~\,}c@{~~~\,}c@{~~~\,}c}
%\multicolumn{7}{c}{\bf \emph{Inhomogeneous Scalings for the}} \\
\multicolumn{7}{c}{\bf \emph{Radial Scalings at 1${}^\text{st}$ Order}} \\
\addlinespace[3pt]
\toprule[1pt]
\addlinespace[3pt]
%\rowcolor[gray]{0.85} \textbf{Field} & \multicolumn{5}{>{\columncolor[gray]{0.85}}l}{\textbf{Amplitude and scaling}} \\ \addlinespace[2pt]
%\toprule[1pt]
%\addlinespace[3pt]
\rowcolor[gray]{0.85}
\multicolumn{1}{r}{} & $\Phi_-$ & $G_-$ & $\tau$ & $g$ & $G_+$ & $\hspace{0pt} \Phi_+^{-1} \hspace{0pt} $ \\
\multicolumn{1}{r}{${\Phi_-^{(1)}}$} & \checkmark & & & & & \\
\multicolumn{1}{r}{${G_-^{(1)}}$} & \checkmark & \checkmark & & & & \\ \addlinespace[3pt]
\multicolumn{1}{r}{${\tau^{(1)}}$} & \checkmark & \checkmark & \checkmark & & & \\
\multicolumn{1}{r}{${g^{(1)}}$} & \checkmark & \checkmark & & \checkmark & &\\ \addlinespace[3pt]
\multicolumn{1}{r}{${G_+^{(1)}}$} & \checkmark & \checkmark & \checkmark & & \checkmark & \\
\multicolumn{1}{r}{${(\Phi_+^{-1})^{(1)}}$} & \checkmark & \checkmark & \checkmark & \checkmark & \checkmark & \checkmark \\
\addlinespace[3pt]
\toprule[1pt]
\end{tabular}
\end{center}
\caption{In this table we summarize the contents of equations (\ref{1phi-scale}-\ref{1phi+scale}).
The fields in the leftmost column label the first-order modes in equations (\ref{1phi-scale}-\ref{1phi+scale}), while the fields in the shaded top row label the homogeneous modes.
A checkmark ($\checkmark$) indicates that the first-order mode receives a contribution with the corresponding homogeneous scaling, while an empty space indicates that no such scaling is present.}\label{inhomlin}
%For example, the row $\boldsymbol{\Phi_-^{(1)}}$ contains a check mark under $\Phi_-$ thus $\hat{\Phi}^{(1)}_- \sim \hat{\Phi}_-^\mathcal{H}$, while the row $\boldsymbol{G_-^{(1)}}$ contains check marks under $\Phi_-$ and $G_-$ thus $\hat{G}^{(1)}_- \sim \hat{\Phi}_-^\mathcal{H} + \hat{G}_-^\mathcal{H}$
\end{table}

One can perform a similar exercise for the second-order perturbations.  We omit the derivation and present the results in Table \ref{inhomsec}.

% Liam removed "for Modes" from title to match 1st order table
\begin{table}[h!!]
\begin{center}
\begin{tabular}{>{\columncolor[gray]{0.85}[5pt]} c c@{~~~\,}c@{~~~\,}c@{~~~\,}c@{~~~\,}c@{~~~\,}c >{\columncolor[gray]{0.85}[3pt]} c c@{~~~\,}c@{~~~\,}c@{~~~\,}c@{~~~\,}c@{~~~\,}c}
\multicolumn{14}{c}{\bf \emph{Radial Scalings at $2^\text{nd}$ Order}} \\
\addlinespace[3pt]
\toprule[1pt]
\addlinespace[3pt]
%\rowcolor[gray]{0.85} \textbf{Field} & \multicolumn{5}{>{\columncolor[gray]{0.85}}l}{\textbf{Amplitude and scaling}} \\ \addlinespace[2pt]
%\toprule[1pt]
%\addlinespace[3pt]
\rowcolor[gray]{0.85}
\multicolumn{1}{c}{${\Phi_-^{(2)}}$} & $\Phi_-$ & $G_-$ & $\tau$ & $g$ & $G_+$ & $\hspace{0pt} \Phi_+^{-1} \hspace{0pt} $ &
\multicolumn{1}{c}{${G_-^{(2)}}$} & $\Phi_-$ & $G_-$ & $\tau$ & $g$ & $G_+$ & $\hspace{0pt} \Phi_+^{-1} \hspace{0pt}$ \\ \addlinespace[3pt]
$\Phi_-$ & \checkmark & \checkmark & & \checkmark & & &
$\Phi_-$ & \checkmark & \checkmark & \checkmark & \checkmark & \checkmark & \checkmark \\ \addlinespace[3pt]
$G_-$ & \checkmark & \checkmark &&&& &
$G_-$ & \checkmark & \checkmark & \checkmark & \checkmark & \checkmark & \checkmark \\ \addlinespace[3pt]
$\tau$ &&&&&& &
$\tau$ & \checkmark & \checkmark &&&& \\ \addlinespace[3pt]
$g$ & \checkmark &&&&& &
$g$ & \checkmark & \checkmark &&&& \\ \addlinespace[3pt]
$G_+$ &&&&&& &
$G_+$ & \checkmark & \checkmark &&&& \\ \addlinespace[3pt]
$\hspace{0pt} \Phi_+^{-1} \hspace{0pt}$ &&&&&& &
$\hspace{0pt} \Phi_+^{-1} \hspace{0pt}$ & \checkmark & \checkmark &&&& \\ \addlinespace[3pt]

\rowcolor[gray]{0.85}
\multicolumn{1}{c}{${\tau^{(2)}}$} & $\Phi_-$ & $G_-$ & $\tau$ & $g$ & $G_+$ & $\hspace{0pt} \Phi_+^{-1} \hspace{0pt} $ &
\multicolumn{1}{c}{${g^{(2)}}$} & $\Phi_-$ & $G_-$ & $\tau$ & $g$ & $G_+$ & $\hspace{0pt} \Phi_+^{-1} \hspace{0pt}$ \\ \addlinespace[3pt]
$\Phi_-$ & \checkmark & \checkmark & \checkmark & \checkmark & \checkmark & \checkmark &
$\Phi_-$ & \checkmark & \checkmark & \checkmark & \checkmark & \checkmark & \checkmark \\ \addlinespace[3pt]
$G_-$ & \checkmark & \checkmark & \checkmark & \checkmark & \checkmark & \checkmark &
$G_-$ & \checkmark & \checkmark & \checkmark & \checkmark & \checkmark & \checkmark \\ \addlinespace[3pt]
$\tau$ & \checkmark & \checkmark & \checkmark & \checkmark & & &
$\tau$ & \checkmark & \checkmark & \checkmark & & & \\ \addlinespace[3pt]
$g$ & \checkmark & \checkmark & \checkmark &&& &
$g$ & \checkmark & \checkmark & & \checkmark && \\ \addlinespace[3pt]
$G_+$ & \checkmark & \checkmark & &&& &
$G_+$ & \checkmark & \checkmark & &&& \\ \addlinespace[3pt]
$\hspace{0pt} \Phi_+^{-1} \hspace{0pt}$ & \checkmark & \checkmark &&&& &
$\hspace{0pt} \Phi_+^{-1} \hspace{0pt}$ & \checkmark & \checkmark &&&& \\ \addlinespace[3pt]

\rowcolor[gray]{0.85}
\multicolumn{1}{c}{${G_+^{(2)}}$} & $\Phi_-$ & $G_-$ & $\tau$ & $g$ & $G_+$ & $\hspace{0pt} \Phi_+^{-1} \hspace{0pt}$ &
\multicolumn{1}{c}{$\hspace{-4pt}{(\Phi^{-1}_+)^{(2)}}\hspace{-4pt}$} & $\Phi_-$ & $G_-$ & $\tau$ & $g$ & $G_+$ & $\hspace{0pt} \Phi_+^{-1} \hspace{0pt}$ \\ \addlinespace[3pt]
$\Phi_-$ & \checkmark & \checkmark & \checkmark & \checkmark & \checkmark & \checkmark &
$\Phi_-$ & \checkmark & \checkmark & \checkmark & \checkmark & \checkmark & \checkmark \\ \addlinespace[3pt]
$G_-$ & \checkmark & \checkmark & \checkmark & \checkmark & \checkmark & \checkmark  &
$G_-$ & \checkmark & \checkmark & \checkmark & \checkmark & \checkmark & \checkmark \\ \addlinespace[3pt]
$\tau$ & \checkmark & \checkmark & \checkmark & \checkmark & \checkmark & &
$\tau$ & \checkmark & \checkmark & \checkmark & \checkmark & \checkmark & \checkmark \\ \addlinespace[3pt]
$g$ & \checkmark & \checkmark & \checkmark & && &
$g$ & \checkmark & \checkmark & \checkmark & \checkmark & \checkmark & \checkmark \\ \addlinespace[3pt]
$G_+$ & \checkmark & \checkmark & \checkmark &&& &
$G_+$ & \checkmark & \checkmark & \checkmark & \checkmark & \checkmark & \checkmark \\ \addlinespace[3pt]
$\hspace{0pt} \Phi_+^{-1} \hspace{0pt}$ & \checkmark & \checkmark &&&& &
$\hspace{0pt} \Phi_+^{-1} \hspace{0pt}$ & \checkmark & \checkmark & \checkmark & \checkmark & \checkmark & \checkmark \\
\addlinespace[3pt]
\toprule[1pt]

\end{tabular}
\end{center}
\caption{At second order, the perturbation $\hat{\phi}^{(2)}$ of a canonically normalized field $\hat{\phi}$ has the radial scaling of a sum of products of two canonically normalized homogeneous modes, i.e. $\hat\phi^{(2)} \sim \sum_{ij} \hat{\phi}_i^\mathcal{H} \hat{\phi}_j^\mathcal{H}$, where the range of $ij$ is read off from the above table. The shaded rows label $\hat{\phi_i}$,
the shaded columns label $\hat{\phi}_j$, and the fields inside white spaces label $\hat{\phi}$. For intersections indicated by a check mark ($\checkmark$), the corresponding term is present in the sum, while for an empty space, no such term is present.}\label{inhomsec}
%For example, using the content of the top left $6 \times 6$ block, labelled by $\Phi^{(2)}_-$, one can reproduce the scaling of $(\Phi_-^{(2)})_\mathcal{IH}$ in equation equation \ref{qwerty}. We emphasize that whether or not a coefficient $C^{\,\phi}_{ij}$ with a check mark is truely non-zero depends on the full Green's function solution, though generically it will be non-zero, while a coefficient without a check mark is always zero.
\end{table}

\subsection{Higher-order scalings}

We now go on to prove that the $n$-th order perturbation scales as a sum of products of $n$ harmonic modes, as in (\ref{harmonicproduct}).
To see this, we introduce new coordinates $\hat{y}^{\hat m} = (\hat r, \hat \Psi^i)$ related to the coordinates $y^m=(r,\Psi^i)$ through
\begin{equation}
 \hat r = \ln r \,, \quad \hat\Psi^i = \Psi^i \,.
\end{equation}
These coordinates are convenient because when taking derivatives with respect to them we do not change the scaling with $r$, i.e.
$\partial_{\hat{m}} \, \phi \sim \phi$.
This is obvious for angular derivatives, while for radial derivatives it follows from $\frac{\partial}{\partial \hat{r}} = r \, \frac{\partial}{\partial r}$.
%\begin{equation}
% \frac{\partial}{\partial \hat{r}} = \frac{\partial}{\partial \ln r} = r \, \frac{\partial}{\partial r}
%\end{equation}
When a tensor is expressed in this basis, the radial components and the angular components scale in the same way since $\dd \hat{r} = \frac{\dd r}{r}$, e.g.
$
 (\hat{G}_\pm)_{\hat r jk} \sim (\hat{G}_\pm)_{ijk} \sim \hat{G}_\pm \,.
$
Furthermore, we observe that for all non-zero background fields $\phi^{(0)}$, the corresponding hatted variables $\hat{\phi}^{(0)}$ are of order unity in the entire background throat solution:
%$r_\text{IR} \lesssim r \lesssim r_\text{UV}$, that is
\begin{equation}
 \hat\phi^{(0)} \sim r^0 \,.
\end{equation}

The equations of motion (\ref{EOM1}-\ref{EOM5}) now take the form
{\allowdisplaybreaks
\begin{align}
\label{scEOM1}
&	r^{-4}\,(\hat{\nabla}^2+5\,\hat{g}^{\hat r \hat{m}} \, \partial_{\hat{m}}) (r^4\,\hat{\Phi}_{\pm})
	= {(\hat{\Phi}_++\hat{\Phi}_-)^2  \over 96 \, \mathrm{Im}\,\hat{\tau} } |\hat{G}_{\pm}|^{\hat{2}} +
	{2\over (\hat{\Phi}_++\hat{\Phi}_-)} |\hat{\nabla}\hat{\Phi}_\pm|^{\hat{2}} \,,\\
&	\dd \hat{\Lambda} + {i\over 2 \, \mathrm{Im}\,\hat{\tau}}	\dd\hat{\tau}\wedge(\hat{\Lambda}+\bar{\hat{\Lambda}})=0\,,\label{scEOM2}\\
&	\dd \hat{G}_3=-\dd\hat{\tau}\wedge H_3\,,\label{scEOM3}\\
%\,
&	\,(\hat{\nabla}^2+5\,\hat{g}^{\hat{r}\hat{m}} \, \partial_{\hat{m}})\,\hat{\tau} = { \hat{\nabla} \hat{\tau} \, \hat{\cdot} \, \hat{\nabla} \hat{\tau}  \over  i\mathrm{Im}(\hat{\tau})} + {\hat{\Phi}_+ +\hat{\Phi}_- \over 48i} \, \hat{G}_+ \, \hat{\cdot} \, \hat{G}_-   \label{scEOM4}\,,\\
\label{scEOM5}
&\hat{R}^6_{\hat{m}\hat{n}} +\hat{\Xi}_{\hat{m}\hat{n}} =
 {\hat{\nabla}_{(\hat{m}} \hat{\tau} \, \hat{\nabla}_{\hat{n})} \bar{\hat{\tau}}
\over 2 \, (\mathrm{Im} \, \hat{\tau})^2} + {2 \over (\hat{\Phi}_+ + \hat{\Phi}_-)^2}
\hat{\nabla}_{(\hat{m}} \hat{\Phi}_+ \hat{\nabla}_{\hat{n})} \hat{\Phi}_- \\
& \phantom{\hat{R}^6_{\alpha\beta} +\hat{\Xi}_{\alpha\beta} =\{}-{\hat{\Phi}_+ + \hat{\Phi}_- \over 32 \,
\mathrm{Im}\,\hat{\tau}} \left( \hat{G}_{+\,(\hat{m}}^{\phantom{(\alpha}~~\hat{p}\hat{q}}\, \bar{\hat{G}}_{-\,\hat{n})\,\hat{p}\hat{q}} + \hat{G}_{-\,(\hat{m}}^{\phantom{(\alpha}~~\hat{p}\hat{q}}\, \bar{\hat{G}}_{+\,\hat{n})\,\hat{p}\hat{q}}  \right) \,. \nonumber
\end{align}
}%enddisplaybreaks
In the above equations, a hat over a contraction, a modulus-squared, or a raised index indicates use of the metric $\hat{g}_{\hat{m}\hat{n}}$. Moreover, the  Ricci tensor $\hat{R}_{\hat{m}\hat{n}}^6$ and all  derivative operators $\hat{\nabla}_{\hat{m}}$ are constructed using the  metric $\hat{g}_{\hat{m}\hat{n}}$. Furthermore, $\hat{\Xi}_{\hat{m}\hat{n}}$  represents the term generated by performing the conformal transformation from $R^6_{mn}$ to $\hat{R}_{\hat{m}\hat{n}}^6$, which involves derivatives of the coordinate $\hat{r}$:
\begin{align}
 \hat{\Xi}_{\hat{m}\hat{n}} & \equiv -4\, \hat{\nabla}_{\hat{m}} \hat{\nabla}_{\hat{n}} \, \hat{r} - \hat{g}_{\hat{m}\hat{n}} \, \hat{\nabla}^2 \, \hat{r} + 4 \hat{\nabla}_{\hat{m}} \hat{\nabla}_{\hat{n}} \, \hat{r} - 4 \, \hat{g}_{\hat{m}\hat{n}} \, \hat{g}^{\hat{p}\hat{q}} \, \hat{\nabla}_{\hat{p}} \, \hat{r} \hat{\nabla}_{\hat{q}} \,  \hat{r} \\
& =  4 \, \hat{\Gamma}^{\hat{r}}_{\hat{m}\hat{n}} + \hat{g}_{\hat{m}\hat{n}} \, \hat{g}^{\hat{p}\hat{q}} \, \hat{\Gamma}^{\hat{r}}_{\hat{p}\hat{q}} + 4 \, \delta^{\hat{r}}_{\hat{m}} \, \delta^{\hat{r}}_{\hat{n}} + 4 \, \hat{g}_{\hat{m}\hat{n}} \, \hat{g}^{\hat{r}\hat{r}} \,,
\end{align}
where $\hat{\Gamma}^{\gamma}_{\hat{m}\hat{n}}$ is the Christoffel connection constructed from $\hat{g}_{\hat{m}\hat{n}}$. Finally, we have also defined
\begin{equation}
\hat{\Lambda} = \hat{\Phi}_- \hat{G}_+ + \hat{\Phi}_+ \hat{G}_- \,.
\end{equation}

From the form (\ref{scEOM1}-\ref{scEOM5}) of the supergravity equations in terms of the hatted fields and coordinates, we can deduce the desired result (\ref{harmonicproduct}).  Because
all background fields $\hat{\phi}^{(0)}$ scale as $r^0$, all derivatives $\hat{\nabla}$ are logarithmic, and no coefficient in the equations depends on $r$, the $n$-th order perturbation $\hat{\phi}^{(n)}$
will inherit its radial scaling exclusively from the other perturbations.  That is, if one were to expand any of the equations (\ref{scEOM1}-\ref{scEOM5}) to $n$-th order, then matching the radial scalings on either side of the equation one would find a relation of the form
%That is, since we have defined all the hatted fields so that they all scale as $r^{0}$ at zeroth order, $\hat{\phi}^{(0)} \sim r^0$, and since all coefficients of the various terms in (\ref{scEOM1}-\ref{scEOM5}) are order 1---all powers of $r$ cancel when re-expressing the equations in terms of the hatted fields---then when we expand these equations to order $n$ the resulting equation for each field $\hat{\phi}_i^{(n)}$ implies that the solution takes the form
\begin{equation}
 \hat{\phi}^{(n)} \sim \sum_i \hat\phi^{(n)}_i + \sum_{p=1}^n \sum_{i,j} \hat\phi_i^{(p)} \hat\phi_j^{(n-p)} + \sum_{p,q=1}^{n} \sum_{i,j,k} \hat\phi_i^{(p)} \hat\phi_j^{(q)} \hat\phi_j^{(n-p-q)} + \ldots \label{scRelation}
\end{equation}
where the sums run over whichever fields appear in the equation under consideration.
%\begin{equation}
% \hat{\phi}^{(n)} \sim \sum_{i_1} \hat\phi^{(n)}_{i_1} + \sum_{m_1=1}^n \sum_{i_1,i_2} \hat\phi_{i_1}^{(m_1)} \hat\phi_{i_2}^{(n-m_1)} + \sum_{m_1,m_2=1}^{n} \sum_{i_1,i_2,i_3} \hat\phi_{i_1}^{(m_1)} \hat\phi_{i_2}^{(m_2)} \hat\phi_{i_3}^{(n-m_1-m_2)} + \ldots + \sum_{i_1 \ldots i_n} \hat{\phi}^{(1)}_{i_1} \cdots \hat{\phi}^{(1)}_{i_n}
%\end{equation}
%The above form is what we would expect from perturbation theory (up to radial scalings): a perturbation at a given order depends on the other %fields at that order and on all the fields at previous orders. In the variables we are employing, no additional radial scalings are introduced. We can %then use the tower like structure to solve for the $n$-th order perturbations in terms of $m$-th order perturbations with $m<n$.
We have seen that the scalings of all the fields at linear order are given by the scalings of the homogeneous modes.  Therefore, by
% Sohang repeatedly
iteratively applying equation (\ref{scRelation}), we deduce that $n$-th order perturbations scale as
\begin{align}
\hat{\phi}^{(n)}(r,\Psi) & = \sum_{i_1, \ldots, i_{n}}
c_{0}^{\Delta(i_1)} \cdots c_{0}^{\Delta(i_n)} \,\, \left(\frac{r}{r_{\star}}\right)^{\Delta(i_1)+\ldots+\Delta(i_n)-4n} \times h^{\Delta(i_1) \ldots \Delta(i_n)}(\Psi) \,,  \label{genscaling} \\
&\equiv \sum_{i_1, \ldots, i_{n}}\hat\phi_{0}^{\Delta(i_1)}(r) \cdots \hat\phi_{0}^{\Delta(i_n)}(r)\times h^{\Delta(i_1) \ldots \Delta(i_n)}(\Psi) \,, \nonumber
\end{align}
%\begin{align}
where the $h^{\Delta(i_1) \ldots \Delta(i_n)}(\Psi)$ are angular functions that are of order unity at generic points in the angular space, and we have defined the running couplings
\begin{equation}
\hat{\phi}_0^{\Delta(\phi)}(r)\equiv c_0^{\Delta(\phi)} \, \Big({r\over r_{\star}}\Big)^{\Delta(\phi)-4}\,.
\end{equation}
%we can rewrite the non-normalizable part of the homogeneous solution (\ref{homogeneouscn}) as
%\begin{equation}
%\hat{\phi}^{\mathcal{H}}(r,\Psi) = \sum_{\Delta(\phi)}  \hat{\phi}_0^{\Delta(\phi)} (r) Y_0^{\Delta(\phi)}(\Psi)  \,.
%\end{equation}

The formula (\ref{genscaling}) is one of our main results.  It states that in the basis specified in (\ref{canvar}), the size of the $n$-th order perturbation of any field $\hat\phi$ can be read off in terms of the sizes  $c_{0}^{\Delta}$ of all the homogeneous modes at $r=r_\star$, and the dimensions $\Delta$ characterizing the spectrum of Kaluza-Klein masses.  That corrections at order $n$ are proportional to degree $n$ products of the perturbation parameters is of course not surprising.  However, equation (\ref{genscaling}) says more than this: it shows that in solving the equations of motion, no addition radial scaling is introduced that would affect the sizes of the corrections: the sizes of the $n$-th order inhomogeneous corrections at some point in the throat are immediately determined by $n$-th order products of harmonic modes \textit{at that point}.  It follows that throat perturbation theory is naturally organized as an expansion in the
% Sohang
running sizes of the harmonic modes, and the expansion is convergent as long as the seeding harmonic modes are small.

\subsection{Conditions for consistency}
\label{sec:consistency}

We now turn to explaining why our perturbative expansion can consistently describe a warped throat, despite the presence of relevant perturbations.
On general grounds, one might expect the boundary conditions on the UV brane to activate all possible modes, with coefficients that are not much smaller than unity.  In particular, any relevant modes will grow toward the infrared, and, given  enough range of renormalization group evolution, would ultimately become large and destroy the IR region of the throat.  This is a critical issue not just for our perturbation scheme, but for the existence of metastable vacua in which antibranes break supersymmetry.  If effects in the bulk induce corrections to the throat geometry that grow precipitously large in the IR, then the vacuum energy of an antibrane at the tip of the throat is poorly approximated by the antibrane action in the uncorrected background,
\begin{equation}
V_{\overline{\mathrm{D}3}}^{(0)} = T_3\,\Phi_+^{(0)} \,,
\end{equation} and the vacuum energy will in general not remain small in string units, so that the compactification will be destabilized.  This fundamental requirement that effects in the bulk do not destabilize the throat, and with it the entire compactification, therefore implies the existence of a perturbative expansion around a background throat geometry.  Our task is to assess whether this requirement can be met without undue fine tuning.
% Liam

In a finite warped throat, the hierarchy of scales is finite, so that if every relevant mode has a sufficiently small coefficient in the UV, all perturbations will remain small throughout the throat.
If effects in the bulk source some relevant mode
\begin{equation}
\phi^{\mathcal{H}} = c_{\mathrm{UV}}^\Delta\,\left({r_{\star}\over r\uv}\right)^{\Delta-4}\,
\end{equation}
with $\Delta < 4$,
% (we have taken $r_{\star} =r\uv$ in the above),
then this mode will become dangerously large at the tip of the throat, $r= r_\text{IR}$, if
\begin{equation}
c_{\mathrm{UV}}^\Delta\,\left({ r_\text{IR}\over r\uv}\right)^{\Delta-4} \gtrsim 1\,.
\end{equation}
Thus, using $\frac{ r_\text{IR}}{r\uv} \sim e^{A_{\mathrm{min}}} \equiv a_0$, we see that the size of the mode in the UV must be
\begin{equation}\label{UVcond}
c_{\mathrm{UV}}^\Delta \lesssim a_0^{4-\Delta}
\end{equation}
in order for the entire throat to be stable against corrections from this mode.

Let us now discuss the circumstances in which (\ref{UVcond}) can hold for all relevant modes.
%For concreteness, we initially focus on a Klebanov-Strassler throat, then indicate how our results can be generalized.
One obviously sufficient condition arises when there are no relevant modes (i.e.\ modes with $\Delta <4$) that
are sourced in the bulk.  This can occur if an unbroken symmetry, such as supersymmetry, forbids all relevant modes.\footnote{See \cite{KST} for a construction utilizing discrete symmetries to protect a non-supersymmetric throat.}  In fact, a Klebanov-Strassler throat
attached to a supersymmetric, ISD flux compactification is stable against compactification effects, because every relevant mode either violates the ISD conditions or violates the supersymmetry\footnote{To be precise, there are relevant perturbations that are consistent with four-dimensional ${\cal{N}}=1$ supersymmetry, but the supercharges preserved are different from those preserved by the background.} of the background throat geometry.
Thus, in a supersymmetric, ISD compactification, the existence of a Klebanov-Strassler throat does not require any unnatural fine-tuning of relevant perturbations.

However, in the same example there exist relevant modes that are incompatible with the supersymmetry of the background throat, but could be
sourced by supersymmetry-breaking effects, e.g.\ by distant antibranes,
fluxes, or nonperturbative effects.  Thus, one should ask whether supersymmetry breaking in the compact space tends to induce
perturbations that destroy the IR region of the throat.

Before proceeding, we emphasize that, by construction,
% Liam assumption,
in any stabilized vacuum in which an anti-D3-brane in a warped throat makes a
dominant contribution to supersymmetry breaking, the scale of the moduli potential and of any bulk sources of supersymmetry breaking must obey
\begin{equation}\label{Modulicond}
V_{bulk} \lesssim 2\,a_0^{4}\,T_3
\end{equation} lest the supersymmetry-breaking energy drive decompactification.
Crucially, this relationship links the scale of supersymmetry-breaking bulk perturbations to the IR scale of the throat.
Arranging this near-equality between disparate contributions -- e.g., anti-D3-brane supersymmetry breaking and gaugino condensation on D7-branes -- obviously requires a degree of fine-tuning.  We are asking whether {\it{further}} fine-tuning is required to subdue instabilities associated with relevant perturbations of the throat that are sourced in the bulk.\footnote{To differentiate these issues, imagine two warped throat backgrounds A, B with identical IR scales, with A admitting a large number of relevant modes, and B having no relevant modes whatsoever.  Arranging for (\ref{Modulicond}) to hold requires fine-tuning in either case, but throat A is vulnerable to large corrections from relevant modes sourced in the bulk, while B is not.}

If the scale of bulk supersymmetry breaking obeys (\ref{Modulicond}), then every supersymmetry-breaking perturbation has a small coefficient, which by (\ref{Modulicond}) can be expressed in terms of the IR scale $a_0$ of the throat.  The particular power of $a_0$ multiplying a given mode,
\begin{equation}
\phi \propto a_{0}^{Q}
\end{equation}
can be obtained by a spurion analysis, as in \cite{LiamsLong}.

The dangerous modes in a general throat can be extracted by examining the homogeneous solutions presented in \S\ref{sec:harmonic} (cf.\ Table \ref{homNon}).
%. and equations (\ref{delta+bound}, \ref{delta-bound}):
We easily see that the fields $\Phi_-$, $G_3$, and $g_{\{ij\}}$ can all possess relevant (i.e.\ $\Delta \leq 4$) modes, while all modes of the remaining supergravity fields are  irrelevant. Evidently, a throat is robust if
\begin{equation} \label{robust}
Q > 4-\Delta
\end{equation} for all modes of $\Phi_-$, $G_3$, and $g_{\{ij\}}$.

Let us now verify that the Klebanov-Strassler throat obeys (\ref{robust}), using the spectroscopic data for $T^{1,1}$ obtained in  \cite{CeresoleHarm, CeresoleSCFT, LiamsLong, AntiBrane}.
First, as explained in \cite{LiamsLong}, the harmonic modes of $\Phi_-$ have $Q=4$, while $G_{3}$ perturbations that are not purely ISD have $Q=2$, corresponding to double and single insertions, respectively, of the supersymmetry-breaking spurion $F_{X} \propto a_{0}^2$.   As the lowest-dimension mode  of flux has $\Delta =5/2 >2$, perturbations of $\Phi_-$  and $G_{3}$ are harmless.  Finally, the two relevant modes of $g_{\{ij\}}$ with $\Delta = 2,\,3$ are the bottom components of supermultiplets, and hence have $Q=4$, completing the proof.  Extending this argument to more general throats would be straightforward given the necessary spectroscopic data, but is beyond the scope of this work.

The arguments above refer only to harmonic modes.  One might have worried that even if all harmonic modes remain small down to the tip, the solutions for the inhomogeneous modes could have scalings that are even more relevant than those of the harmonic modes.  In fact, this is not a problem: our result (\ref{genscaling}) makes it evident that whenever the harmonic modes are small, the expansion is convergent. As we have just presented a spurion argument that shows that the harmonic modes remain small in a Klebanov-Strassler throat attached to a compactification with weakly broken supersymmetry, it follows that a consistent perturbation expansion exists in such a throat.

\subsection{Truncation of the expansion: a worked example}
\label{sec:truncation}

The preceding sections have provided a perturbative solution near some location of interest, $r_{\star}$, in a {\it{double}} expansion in terms of $a_0$ and $r_{\star}/r_{UV}$.  (In particular, the
% Liam
parametric sizes of the $c_I^\Delta$ can be expressed in terms of $a_0$ and $r_{\star}/r_{UV}$.)  To make use of such a solution, we must consistently truncate the double expansion to some desired accuracy.  The simplest way to accomplish this is to specify the relative sizes of the two expansion parameters,
\begin{equation}
\frac{r_{\star}}{r_{UV}} \sim a_{0}^{P}\,,
\end{equation} for some
% Sohang
$P \in (0,1]$,
so that in practice there is a single expansion parameter, taken to be $a_{0}$ in the above.
Then, if the size of some mode in the UV is
\begin{equation}
\phi^{\mathrm{UV}}\sim a_0^{Q_i}\,,
\end{equation}
 the size of the mode at $r=r_{\star}$ is
 \begin{equation}
c_0^\Delta\equiv \phi(r_{\star})\sim a_0^{Q_i}\,\left(\frac{r_{\star}}{r\uv}\right)^{\Delta-4} \sim a_0^{Q_i+(\Delta-4)\cdot P}\,.
 \end{equation}
Truncation is then straightforward.

We will illustrate the necessary steps in the concrete example of the region near the tip of a Klebanov-Strassler throat, where $\frac{r_{\star}}{r\uv}\sim a_0$, so that $P=1$.\footnote{For simplicity we will neglect perturbations generated in the IR, even when studying the tip region.  This is consistent, for example, if we are investigating the potential along a direction corresponding to an isometry preserved by the deformation of the tip, as in \cite{AAB}.}
Suppose that we are interested in going up to an accuracy $\sim a_0^{1.5}$.  The most relevant scalings of each field are \cite{CeresoleHarm, CeresoleSCFT, LiamsLong, AntiBrane}
{\allowdisplaybreaks
\begin{align}
\Phi_-&:~\Delta_{\Phi_-} = 1.5,\,\ldots\\
G_-&:~\Delta_{G_-}= 2.5,\,3,\,3.5,\,\ldots\\
\tau&:~\Delta_{\tau}=4+\Delta_{\Phi_-}\\
g_{mn}&:~\Delta_g=2,\,3,\,5.29,\,\ldots\\
G_+&:~\Delta_{G_+}=\Delta_{G_-},\,\ldots\\
\Phi_+^{-1}&:~\Delta_{\Phi_+}=8,\,\ldots
\end{align}}%enddisplaybreaks

Now we need the sizes of the modes in the UV. We have already seen that $\Phi_-$ scales as $a_0^4$ in the UV, while $G_-$ scales as $a_0^2$. The relevant modes of $G_+$ come paired with modes of $G_-$ and thus inherit the $a_0^2$ scaling.  We have already shown that the two relevant modes of the metric scale like $a_0^4$ in the UV.  The $\Delta_g=5.29$ mode of the metric, the leading mode of $\tau$, and  the $\Delta_{\Phi_+}=8$ mode of $\Phi_+$ are all allowed by supersymmetry and the ISD conditions, and are therefore of order unity in the UV.

With the above data, we can estimate the sizes of the modes at the tip in terms of $a_0$.  We find that the leading homogeneous modes of each field have scalings\footnote{One must be careful to compare the scaling of the hatted fields, as these modes are the proper perturbation variables.}
{\allowdisplaybreaks
\begin{align}
 \hat{\Phi}_- &\sim a_0^{1.5}\,, \\
\hat{G}_- &\sim a_0^{0.5}, a_0^{1.0}, a_0^{1.5}\,,\\
  \hat{\tau}~ &\sim a_0^{1.5}\,, \\
\hat{g}_{mn} &\sim a_0^{1.29}\,, \\
 \hat{G}_+ &\sim a_0^{0.5}, a_0^{1.0}, a_0^{1.5}\,,\\
 \hat{\Phi}_+^{-1} &\sim a_0^{8}\,.
\end{align}
}%enddisplaybreaks
Notice that there is a hierarchy between the various modes and therefore it would be inconsistent to truncate at the same order in each.  To reach the desired accuracy of $a_0^{1.5}$, one considers combinations of the above modes whose net
% Liam scaling
size is at least $a_0^{1.5}$, taking into account the restrictions presented in
Tables \ref{inhomlin} and \ref{inhomsec}.  For example, the mode of $\hat{G}_-$ scaling as $a_0^{0.5}$ and the mode of $\hat{G}_+$ scaling as $a_0^{1.0}$ present a possible contribution.  Consulting Table \ref{inhomsec}, we find that this combination of homogeneous modes can source second-order perturbations of all fields except $\Phi_-$.

\section{Conclusions}
\label{sec:conclusions}
We have developed a method that yields local solutions of type IIB supergravity to any desired order in an expansion around a warped Calabi-Yau cone.  Our approach relies on the observation that the equations of motion expanded to any order in perturbations around a background with ISD fluxes
are easily disentangled.  Specifically, we identified a basis of fields in which the
equations for the $n$-th order perturbations take a triangular form.  As a result, one can write down a Green's function solution to any desired order in a purely algebraic way.  This is a striking simplification, as in expansion around a general background the equations of motion are typically intractably coupled.

Next, we obtained all necessary Green's functions, as functions of the angular harmonics on the Sasaki-Einstein base of the cone.  For cones with known harmonics, such as the conifold, it is straightforward to obtain explicit solutions using the tools presented herein.  We also presented a simple expression for the radial scaling of a general $n$-th order perturbation, so that the size of any desired perturbation is readily estimated.  Our result demonstrates that the sizes of the harmonic modes at a given point in the throat serve as faithful expansion parameters.  For the case of a Klebanov-Strassler throat attached to a KKLT compactification, we showed that our expansion is convergent above the tip, and we provided a spectroscopic criterion for assessing convergence in a more general throat.

We anticipate that our results will have applications to local model-building in flux compactifications of type IIB string theory.  Our tools
simplify the task of characterizing the effective action of a sector of fields localized on D-branes in a throat region, which is a common problem in the study of local models of particle physics and of inflation.  In addition, the methods presented here could be
useful in the study of the long-distance supergravity solutions induced by supersymmetry breaking on anti-D3-branes.  Previous attempts  in each direction have required considerable ingenuity in the choice of ansatz and the basis of fields, and in most cases it has not been evident whether one could in practice proceed to higher order.  Our purely algebraic approach yields a solution to any desired order in terms of a single set of Green's functions.

A second application is to the construction of non-supersymmetric AdS/CFT dual pairs.  Taking a supersymmetric warped Calabi-Yau cone as the background, one can construct families of non-supersymmetric solutions to any desired order in the supersymmetry breaking parameter, as functions of the harmonics on the base.  This provides the prospect of
exploring new aspects of non-supersymmetric, strongly coupled, approximately conformal field theories.\footnote{Solutions making use of the expansion in $r_{\star}/r_{UV}$ would be dual to effective conformal field theories, in the spirit of \cite{Fitzpatrick:2010zm}.}

The principal limitation of our approach is that the Green's functions and separable solutions that we have provided apply only in the approximately-AdS region of a warped Calabi-Yau cone.  The triangular structure of the equations of motion, however, is far more general, applying in expansion around any conformally Calabi-Yau flux compactification.  Extending our methods to more general supergravity backgrounds is a very interesting question for the future.

\subsection*{Acknowledgements}
We are grateful to Thomas Bachlechner, Daniel Baumann, Marcus Berg, Tarun Chitra, A.~Liam Fitzpatrick, Raphael Flauger, Ben Heidenreich, David Marsh, Paul McGuirk, Enrico Pajer, and Gang Xu for helpful discussions, and we thank Anatoly Dymarsky, Shamit Kachru, and David Marsh for comments on a draft.  The research of L.M. was supported by the Alfred P. Sloan Foundation and by the NSF under grant PHY-0757868.  The research of S.G.\ was supported by an NSF Graduate Research Fellowship.  We gratefully acknowledge support for this work by the Swedish Foundation for International Cooperation in Research and Higher Education.

\appendix

\section{Structure of the Source Terms} \label{appendix:sources}

In \S\ref{sec:expansion} we left the source terms in the equations of motion implicit.  In this appendix we will work out the source term for the dilaton as an example.
Expanding the kinetic term, we find
\begin{eqnarray}
(\nabla^2\,\tau)_{(n)} = \sum_{l=0}^n\,\nabla^2_{(l)}\,\tau_{(n-l)}= \nabla^2_{(0)}\,\tau_{(n)}+\sum_{l=1}^{n-1}\,\nabla^2_{(l)}\,\tau_{(n-l)}\,.
\end{eqnarray}
For the first term on the right-hand side of equation (\ref{EOM4})  we have
\begin{equation}
\left( {\n \tau \cdot \n \tau \over
i\mathrm{Im}(\tau)}\right)_{(n)} =\sum_{l=0}^{n-2}\,\sum_{q=1}^{l-1}\,{(-)^{l}\,l!\,g_s\over i}\,\mathrm{Im}\,\tau_{(l)}\, \partial_m\tau_{(q)}\,\partial^m\tau_{(n-l-q)},
 \end{equation}
 using the fact that $\partial_m\,\tau_{(0)} = 0$.
 For the second term on the right-hand side we get
 \begin{equation}
 \left({\Phi_+ + \Phi_- \over 48i}\, G_+ \cdot G_-\right)_{(n)}=- {2i\,e^{4A_{(0)}}}\,G_+^{(0)} \cdot G_-^{(n)}-2i  \sum_{l=0}^{n-1}\,\sum_{q=0}^{l-1}\, (\Phi_-^{(l)}+\Phi_+^{(l)})\,G_+^{(q)} \cdot G_-^{(n-l-q)}\,,
 \end{equation}
 using the fact that $G_-^{(0)}=0$.  The $n$-th order equation of motion for $\tau$ is then
 \begin{align}
 \nabla^2_{(0)}\tau_{(n)}= &{\Phi_+^{(0)} \over 48i}\,G_+^{(0)}\cdot G_-^{(n)}-\sum_{l=1}^{n-1}\,\nabla^2_{(l)}\,\tau_{(n-l)}+ \sum_{l=0}^{n-2}\,\sum_{q=1}^{l-1}\,{(-)^{l}\,l!\,g_s\over i}\,\mathrm{Im}\,\tau_{(l)}\, \partial_m\tau_{(q)}\,\partial^m\tau_{(n-l-q)} \\
 &-2i  \sum_{l=0}^{n-1}\,\sum_{q=0}^{l-1}\, (\Phi_-^{(l)}+\Phi_+^{(l)})\,G_+^{(q)} \cdot G_-^{(n-l-q)}\,.\nonumber
 \end{align}
This is then of the form (\ref{ntau}), with
 \begin{align}
 \mathrm{Source}_{\tau}(\phi^{(m<n)}) = &-\sum_{l=1}^{n-1}\,\nabla^2_{(l)}\,\tau_{(n-l)}+ \sum_{l=0}^{n-2}\,\sum_{q=1}^{l-1}\,{(-)^{l}\,l!\,g_s\over i}\,\mathrm{Im}\,\tau_{(l)}\, \partial_m\tau_{(q)}\,\partial^m\tau_{(n-l-q)} \\
 &-2i  \sum_{l=0}^{n-1}\,\sum_{q=0}^{l-1}\, (\Phi_-^{(l)}+\Phi_+^{(l)})\,G_+^{(q)} \cdot G_-^{(n-l-q)}\,.\nonumber
 \end{align}
The remaining $\mathrm{Source}_\varphi(\phi^{(m<n)})$ can be obtained in like fashion.

\section{Harmonic Solutions and Green's Functions}

\label{sec:appendix}
In this appendix we derive the harmonic solutions and Green's functions that are needed in the main text. We separate the equations of motion for scalar, flux and metric perturbations on a Calabi-Yau cone into radial and angular equations, and then solve the resulting radial equations.  This yields the homogeneous solutions and Green's functions on the cone, given the harmonics on the Sasaki-Einstein base as well as the associated spectrum of Hodge-de Rham eigenvalues (see \cite{Salam, Kim} for seminal related work). In the case that the base is $T^{1,1}$, the spectroscopy is well understood \cite{CeresoleSCFT, CeresoleHarm} (see also \cite{KM, LiamsLong}), and is conveniently presented in \cite{AntiBrane}.

In the main body of the text we have considered a six-dimensional cone, but many of the results of this appendix hold for any $(n+1)$-dimensional cone.
However, in our treatment of fluxes in \S\ref{sec:fluxappend}, we specialize to $n=5$.

\subsection{Angular harmonics on an Einstein manifold}
\label{subsec:harms}
We will begin by defining the angular harmonics
%that we use throughout this appendix
and establishing their relevant properties.
Some of the properties below are specific to $n=5$, and we indicate this where applicable.

Consider a general $(n+1)$-dimensional Calabi-Yau cone $\mathcal{C}_{n+1}$:
\begin{eqnarray}
\dd s_{\mathcal{C}_{n+1}}^2 = g_{mn}\dd y^m \dd y^n &=& \dd r^2 + r^2 \dd s_{\mathcal{B}_n}^2\\
&=& \dd r^2 + r^2 \tilde{g}_{ij}\dd \Psi^i \dd\Psi^j\,,
\end{eqnarray}
where we use $i,j,k,l$ for indices which lie in the angular space
only, and $m,n,p,q$ for indices which run over both $r$ and the
angular directions.  Here $\tilde{g}_{ij}$ is the metric on the base space $\mathcal{B}_n$, which
%The space $\mathcal{B}_n$
must be a Sasaki-Einstein manifold, with
\begin{equation}\label{sasein}
\tilde{R}_{ij} = (n-1)\tilde{g}_{ij}\,,
\end{equation}
where $\tilde{R}_{ij}$ is the Ricci tensor built from
$\tilde{g}_{ij}$.  In the following, we will use a tilde above indices (derivative operators) to denote contraction with (construction from) the metric $\tilde{g}_{ij}$.

We now discuss the various tensor harmonics on the angular space $\mathcal{B}_n$. A complete basis for scalar functions on $\mathcal{B}_n$  are the scalar harmonics
\begin{equation}
Y^{I_s}(\Psi)\,.
\end{equation}
A complete basis for one-forms on $\mathcal{B}_n$ are the transverse and longitudinal harmonics
\begin{equation}
Y_{i}^{I_v}(\Psi)\,, ~~ \tilde{\nabla}_iY^{I_s}(\Psi)\,.
\end{equation}
A complete basis for two-forms on $\mathcal{B}_n$ are the transverse and longitudinal harmonics
\begin{equation}
Y_{[ij]}^{I_{2}}(\Psi)\,, ~~ \tilde{\nabla}_{[i}Y^{I_v}_{j]}(\Psi)\,,
\end{equation}
where square brackets denote antisymmetrization.
A complete basis for symmetric, two-index tensors on $\mathcal{B}_n$ are the transverse and longitudinal harmonics
\begin{equation}
Y^{I_t}_{\{ij\}}(\Psi)\,,~~\tilde{\nabla}_{\{i}Y_{j\}}^{I_v}(\Psi)\,,~~\tilde{\nabla}_{\{i}\tilde{\nabla}_{j\}}Y^{I_s}(\Psi)\,,~~\tilde{g}_{ij}Y^{I_s}(\Psi)\,,
\end{equation}
where curly  brackets around indices denote the symmetric traceless part:
\begin{equation}
A_{\{ij\}}= \hf \left( A_{ij}+A_{ji}\right) - {\tilde{g}_{ij} \over n}  A^{\tilde{k}}_{~k} \,.
\end{equation}
The transverse harmonics obey
{\allowdisplaybreaks
\begin{eqnarray}
\tilde{\nabla}^{\tilde{k}}Y^{I_v}_{k} & = & 0\,,\\
\tilde\n^{\tilde k} Y^{I_2}_{[ki]} & = & 0\,, \label{2formtrans} \\
\tilde{\nabla}^{\tilde{k}}Y^{I_t}_{\{ki\}} & = & 0\,.
\end{eqnarray}
}%enddisplaybreaks

\subsubsection{Eigenvalue properties}

The zero-, one- and two-form harmonics $Y^{I_s}$, $Y^{I_v}_{i}$ and $Y^{I_2}_{[ij]}$ are eigenfunctions of the Hodge-de Rham operator $\tilde\Delta = \tilde\delta \dd + \dd \tilde\delta$, where $\dd$ denotes the exterior derivative and $\tilde\delta = (-1)^{n(k+1)+1} \enn \dd \enn$ denotes its adjoint acting on $k$-forms on $\mathcal{B}_n$. The symmetric two-index tensor harmonic $Y^{I_t}_{\{ij\}}$ is an eigenfunction of the Lichnerowicz operator $\tilde\Delta_L$ (cf.\ e.g.\ \cite{Duff}).  These equations are efficiently expressed as
{\allowdisplaybreaks
\begin{eqnarray}
 \tilde\Delta_0 Y^{I_s} & = & \lambda^{I_s} Y^{I_s}\,, \\
 \tilde\Delta_1 Y^{I_v}_i & = & \lambda^{I_v} Y^{I_v}_i\,, \\
 \tilde\Delta_2 Y^{I_2}_{ij} & = & \lambda^{I_2} Y^{I_2}_{ij}\,, \\
 \tilde\Delta_L Y^{I_t}_{ij} & = & \lambda^{I_t} Y^{I_t}_{ij}\,.
\end{eqnarray}
}%enddisplaybreaks
Using the relationships
{\allowdisplaybreaks
\begin{eqnarray}
 \tilde\delta \dd Y^{I_s} & = & - \n^2 Y^{I_s} \,, \\
 (\tilde\delta \dd Y^{I_v})_i & = & - 2 \n^k \n_{[k} Y^{I_v}_{i]} \,, \\
 (\tilde\delta \dd Y^{I_2})_{ij} & = & - 3 \n^k \n_{[i} Y^{I_2}_{jk]} \,,
\end{eqnarray}
together with
\begin{eqnarray}
 \tilde\delta Y^{I_s} & = & 0\,, \\
 \tilde\delta Y^{I_v} & = & - \tilde\n^{\tilde k} Y^{I_v}_k\,, \\
 (\tilde\delta Y^{I_2})_j & = & - \tilde\n^{\tilde k} Y^{I_2}_{kj} \,,
\end{eqnarray}
one can derive the explicit form of the Hodge-de Rham and Lichnerowicz operators:
{\allowdisplaybreaks
\begin{eqnarray}
 \tilde\Delta_0 Y^{I_s} & = & - \tilde{\nabla}^{2} Y^{I_s} \,, \\
 \tilde\Delta_1 Y^{I_v}_i & = & - \tilde{\nabla}^{\tilde{2}} Y^{I_v}_i + \tilde{R}_i^{\phantom{i}\tilde{j}} Y^{I_v}_j \,, \\
 \tilde\Delta_2 Y^{I_2}_{ij} & = & - \tilde{\nabla}^{2} Y^{I_2}_{ij} + 2 \tilde{R}_{\phantom{k}ij}^{\tilde{k}\phantom{ij}\tilde{l}} Y^{I_2}_{kl} - 2 \tilde{R}_{[i}^{\phantom{[i}\tilde{k}} Y^{I_2}_{j]k} \,, \\
 \tilde\Delta_L Y^{I_t}_{ij} & = & - \tilde \n^{\tilde 2} Y^{I_t}_{ij} + 2 \tilde{R}_{\phantom{k}ij}^{\tilde{k}\phantom{ij}\tilde{l}} Y^{I_t}_{kl} + 2 \tilde{R}_{(i}^{\phantom{(i}\tilde{k}} Y^{I_t}_{j)k}\,.
\end{eqnarray}
}%enddisplaybreaks
Notice that the transversality of the one- and two-form harmonics $Y^{I_v}$ and $Y^{I_2}$ is simply the statement that they are co-closed, $\tilde \delta Y^{I_v} = \tilde \delta Y^{I_2} = 0$. Using the transversality of the harmonics, the above eigenvalue equations can also be written as
\begin{eqnarray}
 \tilde{\nabla}^2 Y^{I_s} & = & - \lambda^{I_s}Y^{I_s}\,, \\
 2 \tilde{\nabla}^{\tilde{k}} \tilde{\nabla}_{[k} Y^{I_v}_{i]} & = & - \lambda^{I_v} Y^{I_v}_{i}\,, \\
 3 \tilde{\nabla}^{\tilde{k}} \tilde{\nabla}_{[i} Y^{I_2}_{jk]} & = & - \lambda^{I_2} Y^{I_2}_{[ij]}\,, \\
 \tilde{\nabla}^2 Y^{I_t}_{\{ij\}} - 2 \tilde{\nabla}^{\tilde{k}} \tilde{\nabla}_{(i} Y^{I_t}_{\{j)k\}} & = & - \lambda^{I_t} Y^{I_t}_{\{ij\}}\,.
\end{eqnarray}

We also note that when $n$ is odd, the Hodge-de Rham operator for a tranverse $\left(\frac{n-1}{2}\right)$-form can be expressed in terms of the square of the first-order operator $\enn \dd$. In the case of interest for us, $n=5$, the two-form $Y_{[ij]}^{I_{2}}$ is an eigenfunction of $\star_5 \dd$,
\begin{eqnarray}
\star_5 \dd Y^{I_2} = i\, \delta^{I_2} \,Y^{I_2}\label{LapBelt}\,, \quad \delta^{I_2} \in \mathbb{R}
\end{eqnarray}
such that $\tilde \delta \dd Y^{I_2} = - (\star_5 \dd)^2Y^{I_2} = + (\delta^{I_2})^2 Y^{I_2}$, i.e.\
\begin{equation}
 \tilde{\Delta}_2 \, Y^{I_2} = \lambda^{I_2} \, Y^{I_2} \,, \quad \lambda^{I_2} \equiv (\delta^{I_2})^2.
\end{equation}

\subsubsection{Orthogonality properties} \label{orthprop}

We normalize the harmonics such that
\begin{eqnarray}
%\langle\, Y_{I_s}, \, Y^{I'_s} \, \rangle & \equiv &
\int \dd^n \Psi \sqrt{\tilde{g}} \, \bar{Y}_{I_s}\,Y^{I_s'} & = &\delta_{I_s}^{I_s'}\,,\\
%\langle\,Y_{I_v},\,Y^{I'_v}\,\rangle & \equiv &
\int \dd^n \Psi \sqrt{\tilde{g}} \, \bar{Y}_{I_v}^{\tilde{k}} \,Y_k^{I_v'} & = & \delta_{I_v}^{I_v'}\,,\\
%\langle\,Y_{I_2},\,Y^{I'_2}\,\rangle & \equiv &
\int \dd^n \Psi \sqrt{\tilde{g}} \, \bar{Y}_{I_2}^{[\tilde{k}\tilde{l}]}\,Y^{I_2'}_{[kl]} & = &\delta_{I_2}^{I_2'}\,,\\
%\langle\,Y_{I_t},\,Y^{I'_t} \, \rangle & \equiv &
\int \dd^n \Psi \sqrt{\tilde{g}} \, \bar{Y}_{I_t}^{\{\tilde{i}\tilde{j}\}}\, Y^{I_t'}_{\{ij\}} & = & \delta_{I_t}^{I_t'}\,.
\end{eqnarray}
Here we use a bar to denote complex conjugation, $\bar{Y} \equiv Y^*$. From the above orthonormality properties and  equation (\ref{sasein}) one can derive the remaining set of orthonormality conditions:
\begin{eqnarray}
 \int \dd^n\Psi
\sqrt{\tilde{g}} \, \tilde{\nabla}^{\tilde{k}} \bar{Y}_{I_s} \, \tilde{\nabla}_k Y^{I_s'} & = & \lambda^{I_s} \, \delta^{I_s}_{I_s'}\,, \label{vnorms}\\
 \int \dd^n\Psi
\sqrt{\tilde{g}} \, (\tilde{g}^{\tilde{k}\tilde{l}} \bar{Y}_{I_s}) \, (\tilde{g}_{kl} Y^{I_s'}) & = & n \, \delta_{I_s}^{I_s'}\,,\label{tnormtr}\\
 \int \dd^n\Psi
\sqrt{\tilde{g}} \, \tilde{\nabla}^{[\tilde{k}} \bar{Y}_{I_v}^{\tilde{l}]} \, \tilde{\nabla}_{[k} Y^{I_v'}_{l]} & = & \hf \lambda^{I_v} \, \delta_{I_v}^{I_v'}\,, \label{2normv}\\
 \int \dd^n\Psi
\sqrt{\tilde{g}} \, \tilde{\nabla}^{\{\tilde{k}} \bar{Y}_{I_v}^{\tilde{l}\}} \, \tilde{\nabla}_{\{k} Y^{I_v'}_{l\}} & = & \hf\Bigl(\lambda^{I_v}-2(n-1)\Bigr) \, \delta_{I_v}^{I_v'}\,, \label{tnormv}\\
\int \dd^n\Psi
\sqrt{\tilde{g}} \, \tilde{\nabla}^{\{\tilde{k}} \tilde{\nabla}^{\tilde{l}\}} \bar{Y}^{I_s} \,
\tilde{\nabla}_{\{k} \tilde{\nabla}_{l\}}Y^{I_s'} & = & {(n-1)\over n} \lambda^{I_s} (\lambda^{I_s}-n) \, \delta_{I_s}^{I_s'} \label{tnorms}\,.
\end{eqnarray}
All remaining inner products---those between transverse and longitudinal harmonics, or between longitudinal harmonics with different numbers of derivatives---vanish.

One can learn much from equations (\ref{vnorms}--\ref{tnorms}).  Since the inner products must be positive definite, we see from equation (\ref{vnorms}) that $\tilde{\nabla}_iY^{I_s}$ vanishes if and only if $\lambda^{I_s} = 0$.  It is known (see \cite{Duff}) that compact Einstein spaces always support exactly one zero mode---the constant mode $Y^{I_s}(\Psi)=\text{const.}$
From equations (\ref{tnormv}) and (\ref{tnorms}) one deduces that $\lambda^{I_s} \geq n  ~~ \mathrm{or}~~ \lambda^{I_s}=0$, while $\lambda^{I_v} \geq 2(n-1)$.
Both of these conditions are known to hold for an Einstein space (with scaling as in equation (\ref{sasein})), see \cite{Duff}. The value $\lambda^{I_s} = n$ occurs only for the trivial case of the sphere,
$\mathcal{B}_n = \mathcal{S}^n$ \cite{Yano}.  This corresponds to the $(n+1)$-dimensional ``cone" being merely flat Euclidean space.
Next, the condition $\tilde{\nabla}_{\{i}Y^{I_v}_{j\}}=0$ is just the condition that $Y^{I_v}_i$ is a Killing vector, and so there is one harmonic with $\lambda^{I_v} = 2(n-1)$
%saturating the inequality in equation (\ref{lamv})
for each continuous isometry of $\mathcal{B}_n$.

For the two-form harmonics there is no lower bound on the eigenvalues $\delta^{I_2}$.  Indeed,  by conjugation of equation (\ref{LapBelt}) one sees that the spectrum is symmetric  under $\delta^{I_2} \rightarrow - \delta^{I_2}$.  Modes with $\delta^{I_2} = 0$ have a special significance: when $\delta^{I_2}=0$, $\dd Y^{I_2} = 0$.
%.  This would imply
%\begin{equation}
%\dd Y^{I_2} = 0,~~~~\mathrm{if}~\delta^{I_2}=0\,.
%\end{equation}
Combining this with the transversality condition (\ref{2formtrans}), we see that such a $Y^{I_2}$ must be harmonic, and is therefore a Betti form.
We will denote these Betti two-forms as
\begin{equation}
\omega_2^i,~~~~i=1,2,\ldots b_2\,,
\end{equation}
where $b_2$ is the  second Betti number of $\mathcal{B}_5$.

\subsection{Flux solutions and Green's functions}
\label{sec:fluxappend}
The harmonic three-form flux solutions were obtained in \cite{LiamsLong}. In \S\ref{subsubsec:fluxharmappend} we present a slight generalization of those solutions that allows for logarithmic running of the warp factor. Then, in \S\ref{sec:fluxgreensappend} we derive the Green's functions for the three-form flux. In this section we specialize to the case of $n=5$.

\subsubsection{Homogeneous flux solutions}
\label{subsubsec:fluxharmappend}
We wish to obtain the solution to the system of differential equations
%(\ref{fluxharm}), (\ref{hiasd}),
(\ref{hG-}, \ref{Gfluxclosed}),
%\begin{eqnarray}
% \dd (\Phi_+^{(0)}\,G_-) & = &0 \label{fluxeq1} \\
% \dd G_3 & = & 0\,, \label{fluxeq2}
%\end{eqnarray}
where the IASD part of the flux is given by $G_- = (\star_6 - i) G_3$, % Liam
and the expression for $\Phi_+^{(0)}$ in an ISD background is given in terms of the warp factor (cf.\ equation (\ref{UVwarp})),
\begin{equation}
   {2 \over \Phi_+^{(0)} } = e^{- 4 A^{(0)}} = {C_1+C_2 \ln r \over r^4}\,.
\end{equation}
Because $G_3$ is closed,
% by equation (\ref{fluxeq2}),
it can  be written locally in terms of a two-form potential $A_2$ as
%\begin{equation}
$G_3 = \dd A_2$.
%\end{equation}
Generically, $A_2$ will have a harmonic expansion
\begin{align}
A_2 = & \sum_{I_2} a^{I_2}(r)\,Y^{I_2}(\Psi)+ \sum_{I_v} a^{I_v}(r)\,\dd\,Y^{I_v}(\Psi)\\
&~~~~+\sum_{I_v} b^{I_v}(r)\,{\dd r\over r}\wedge Y^{I_v}(\Psi)+\sum_{I_s} b^{I_s}(r)\,{\dd r \over r}\wedge \dd\,Y^{I_s}(\Psi)\nonumber\,.
\end{align}
We have the obvious gauge symmetry $A_2 \rightarrow A_2+\dd \chi_1$, for a one-form gauge parameter $\chi_1$, and by expanding  $\chi_1$ in harmonics, we can set $b^{I_v}=b^{I_s} = 0$:
\begin{equation} \label{gaugefixedA}
A_2 =  \sum_{I_2} a^{I_2}(r)\,Y^{I_2}(\Psi)+ \sum_{I_v} a^{I_v}(r)\,\dd\,Y^{I_v}(\Psi),~~~~\mathrm{gauge}~\mathrm{fixed}\,.
\end{equation}
Now we insert this form of $A_2$ into equation (\ref{hG-}). Since the equations are linear we can consider a single mode at a time, and we have two cases: non-exact and exact modes.

\paragraph{Non-exact modes:} Consider the non-exact mode
\begin{equation}
A_2 = A^{I_2}(r)\,Y^{I_2}(\Psi)\,.
\end{equation}
Using the identities
\begin{align}
 & \star_6 \left({\dd r\over r}\wedge \Omega_2\right) = \star_5 \Omega_2 \,, \label{hodgeid1}\\
 & \star_6\Omega_3 = -\left({\dd r\over r}\wedge \star_5 \Omega_3\right)\,, \label{hodgeid2}
\end{align}
for arbitrary two- and three-forms $\Omega_2$ and $\Omega_3$ on $\mathcal{B}_5$, together with $\star_5 \dd Y^{I_2} = i \delta^{I_2} Y^{I_2}$, we get for the flux
\begin{equation} \label{Gpm}
 G_\pm = \pm i \left( r \partial_r A^{I_2} \mp \delta^{I_2} A^{I_2} \right) \left( \frac{\dd r}{r} \wedge Y^{I_2} \mp i \star_5 Y^{I_2} \right).
\end{equation}
Inserting the above expression for $G_-$ into equation (\ref{hG-}) yields
\begin{equation}
r \partial_r f^{I_2}(r) - \delta^{I_2} \, f^{I_2}(r) = 0 \,,
\end{equation}
where $f^{I_2} (r) \equiv \Phi_+^{(0)} (r) \, \left(r \partial_r A^{I_2}(r) + \delta^{I_2} A^{I_2}(r) \right)$. Solving the above equation we find
\begin{equation} \label{asol}
A^{I_2}(r) = A_-^{I_2} r^{-\delta^{I_2}} + A_+^{I_2} r^{\delta^{I_2} - 4} \left[ (4 - 2 \delta^{I_2})(C_1+C_2 \log r) + C_2 \right]\,,
\end{equation}
where $A_\pm^{I_2}$ are integration constants. The IASD/ISD components of this solution are
\begin{align}
 G_- & = + i \, (2\delta^{I_2} - 4)^2 \, A^{I_2}_+ \, r^{\delta^{I_2} - 4} \, \left( C_1+C_2\,\ln r \right) \left( {\dd r \over r} \wedge Y^{I_2} + i \star_5 Y^{I_2} \right)\,, \label{IASD} \\
 G_+ & = - i \left( 2 \delta^{I_2} A^{I_2}_- r^{-\delta^{I_2}} + 2A^{I_2}_+ r^{\delta^{I_2} - 4} \left[ (8 - 4 \delta^{I_2})\left(C_1 + C_2 \ln r \right) + \delta^{I_2} C_2 \right] \right) \left( {\dd r \over r} \wedge Y^{I_2} - i \star_5 Y^{I_2} \right) \label{ISD}\,.
\end{align}
Notice that the mode $A^{I_2}_-$ does not contribute to $G_-$.

For the Betti modes with $\delta^{I_2} = 0$ the above solutions reduce to
\begin{equation}
A^{I_2}(r) = A_-^{I_2} + A_+^{I_2} r^{-4} \Bigl( 4(C_1+C_2 \log r) + C_2 \Bigr)\,,
\end{equation}
with IASD/ISD flux components
\begin{align} \label{bettiIASD/ISD}
 G_\pm = \mp \, 32 \, i \, \frac{A_+^{I_2}}{\Phi_+^{(0)}} \left( {\dd r \over r} \wedge Y^{I_2} \mp i \star_5 Y^{I_2} \right) \,.
\end{align}

\paragraph{Exact modes:} Consider the exact mode
\begin{equation}
A_2 = A^{I_v}(r) \,\dd Y^{I_v}\,.
\end{equation}
The flux is
\begin{equation}
G_\pm = \pm i r \partial_r A^{I_v} \left({\dd r \over r} \wedge \dd Y^{I_v} \mp i \star_5 \dd Y^{I_v} \right) \,.
\end{equation}
Plugging this expression into equation (\ref{hG-})
%(\ref{fluxeq1})
and using $\star_5 \dd \star_5 \dd Y^{I_v} = \tilde\delta \dd Y^{I_v} = \lambda^{I_v} Y^{I_v}$,
we get
\begin{equation}\label{exact}
 \dd \left( \Phi_+^{(0)} \, G_- \right)= r \partial_r \left( \Phi_+^{(0)} r \partial_r A^{I_v} \right) {\dd r \over r} \wedge \star_5 \dd Y^{I_v} + \lambda^{I_v} \left( \Phi_+^{(0)} r \partial_r A^{I_v} \right) \star_5 Y^{I_v} = 0.
\end{equation}
From the discussion in \S\ref{orthprop}, we know that $\lambda^{I_v} \geq 8$,
% by equation (\ref{lamv}),
so the second term on the right in equation (\ref{exact}) can only vanish if
$A^{I_v}(r) = \mathrm{const}$.
%\end{equation}
Thus, for this mode the flux vanishes:
\begin{equation}
G_3 \propto \dd \big(\dd Y^{I_v}\big) = 0\,.
\end{equation}
Moreover, the mode is topologically trivial. Thus, the exact modes are unphysical.

\paragraph{Total solution:} To summarize, our solution is
\begin{align}
G_3 & = \dd A_2\\
A_2 & = \sum_{I_2} \left\{ A_-^{I_2} r^{-\delta^{I_2}} + A_+^{I_2} r^{\delta^{I_2} - 4} \left[ (4 - 2 \delta^{I_2}) (C_1+C_2 \ln r) + C_2 \right] \right\} Y^{I_2}\,, \label{A2solution}
\end{align}
where the sum over $I_2$ runs over all non-exact modes, including the Betti modes with $\delta^{I_2} = 0$.

\subsubsection{Scaling dimensions for modes of flux}
\label{fluxdimensionappend}
In \cite{LiamsLong} explicit expressions for all possible closed IASD three-forms on a cone were given in terms of the scalar harmonic functions of the cone, the K\"ahler potential $k$, the K\"ahler form $J$, and the holomorphic three-form $\Omega$.   This in particular allows one to determine the set of radial scalings of flux modes in terms of the radial scalings of the scalar modes.  One finds that the allowed Laplace-Beltrami eigenvalues are
\begin{equation}\label{fluxEig}
\delta^{I_2} = \pm \left\{\begin{array}{ll}
		-1+\Delta(I_s) \\
		-2+\Delta(I_s)\,,&~~~~\lambda^{I_s} \neq 0\\
		-3+\Delta(I_s)\,,&~~~~\lambda^{I_s} \neq 0\\
		0\,,&~~~~b_2\neq 0
		\end{array}\right.\,,
\end{equation}
 Now, since $-3 + \Delta(I_{s}) \geq 2$ for $\lambda^{I_s} \neq 0$  (see paragraph below equation (\ref{scalarScaling})), we find that $|\delta^{I_2}| \geq 2$, apart from the Betti modes, that is
\begin{equation}
 \delta^{I_2} \geq 2,\quad\mathrm{or}\quad \delta^{I_2} =0,\quad\mathrm{or}\quad \delta^{I_2} \leq -2 \,.
\end{equation}

In order for the radial scalings of the modes in equation (\ref{fluxharmpot}) to take on the standard AdS form, equation (\ref{generalHom}), we identify $\Delta(I_2) = \max(\delta^{I_2}, 4-\delta^{I_2})$.  Thus, the operator dimensions corresponding to  modes with $\delta^{I_2}\geq 2$ are given by $\Delta(\delta^{I_2} \geq 2) = |\delta^{I_s}|$, i.e.\
\begin{equation} \label{fluxdimensions1}
\Delta(\delta^{I_2} \geq 2) = \left\{\begin{array}{ll}
		-1+\Delta(I_s) \\
		-2+\Delta(I_s)\,,&~~~~\lambda^{I_s} \neq 0\\
		-3+\Delta(I_s)\,,&~~~~\lambda^{I_s} \neq 0
		\end{array}\right.\,,
\end{equation}
The dimensions of the Betti modes with $\delta^{I_2} = 0$ are given by
\begin{equation} \label{fluxdimensions0}
 \Delta(b_2) = 4\,,
\end{equation}
while the modes with $\delta^{I_2} \leq -2$ have $\Delta(\delta^{I_2} \leq -2) = 4 +  |\delta^{I_s}|$, i.e.\
\begin{equation} \label{fluxdimensions2}
\Delta(\delta^{I_2} \leq -2) = 4 + \left\{\begin{array}{ll}
		-1+\Delta(I_s) \\
		-2+\Delta(I_s)\,,&~~~~\lambda^{I_s} \neq 0\\
		-3+\Delta(I_s)\,,&~~~~\lambda^{I_s} \neq 0\\
		\end{array}\right.\,.
\end{equation}

The ISD/IASD parts $G^\mathcal{H}_\pm$ of the flux solutions are presented in equations (\ref{ISD}, \ref{IASD}) and in equation (\ref{bettiIASD/ISD}) for the Betti modes.  From these expressions one can see that $G^\mathcal{H}_-$ always vanishes for the $A_-^{I_2}$ mode, which scales like $r^{-\delta^{I_2}}$.  Whether this mode corresponds to the normalizable mode $r^{-\Delta(I_2)}$ or the non-normalizable mode $r^{\Delta(I_2)-4}$ depends on the value of $\delta^{I_2}$.  For $\delta^{I_2}\geq 2$ we have $r^{-\delta^{I_2}}=r^{-\Delta(I_2)}$ and this is the normalizable mode. For $\delta^{I_2} < 2$ we have $r^{-\delta^{I_2}} = r^{\Delta(I_2)-4}$ and this is the non-normalizable mode. For the Betti modes we see from equation (\ref{bettiIASD/ISD}) that both $G_+$ and $G_-$ vanish for the non-normalizable mode, scaling like $r^{0}$. These modes are still physical, and they correspond to nontrivial topological configurations.  So, to summarize,
\begin{itemize}
 \item For $\delta^{I_2} \geq 2$, the IASD flux $G_-$ vanishes in the normalizable mode.
 \item For $\delta^{I_2} \leq -2$, the IASD flux $G_-$ vanishes in the non-normalizable mode.
 \item For $\delta^{I_2}=0$, the total flux vanishes in the non-normalizable mode.
\end{itemize}

\subsubsection{Flux Green's functions}
\label{sec:fluxgreensappend}

We want to solve the system of equations
\begin{eqnarray}
 \dd \left( \Sigma_\pm + \mathcal{S}_1 \right) & = & \mathcal{S}_3 \,,\label{fluxinhom1}\\
 \left( \star_6 \mp i \right) \, \Sigma_\pm & = & \mathcal{S}_2 \label{fluxinhom2}\,,
\end{eqnarray}
for two three-form sources $\mathcal{S}_1$ and $\mathcal{S}_2$ and a
four-form source $\mathcal{S}_3$. We will do so in two steps.
\paragraph{System I:} First, we will solve the system of equations with $\mathcal{S}_3 = 0$, % Liam
\begin{eqnarray}
 \dd \left( \Sigma_{\pm}^{(\mathrm{I})}+\mathcal{S}_1 \right) & = & 0\,, \label{exactinhom1}\\
 \left( \s_6 \mp i \right) \Sigma_\pm^{(\mathrm{I})} & = & \mathcal{S}_2\,, \label{exactinhom2}
\end{eqnarray}
\paragraph{System II:} Second, we will solve the system of equations with $\mathcal{S}_1, \mathcal{S}_2 = 0$, % Liam
\begin{eqnarray}
 \dd \Sigma_\pm^{(\mathrm{II})} & = & \mathcal{S}_3 \,, \label{nonexinhom1}\\
 \left(\s_6 \mp i \right) \, \Sigma_\pm^{(\mathrm{II})} & = & 0\,. \label{nonexinhom2}
\end{eqnarray}
The solution to the original system (\ref{fluxinhom1}), (\ref{fluxinhom2}) is then obtained by adding the two solutions above,
\begin{equation}
\Sigma_\pm = \Sigma_\pm^{(\mathrm{I})}+\Sigma_\pm^{(\mathrm{II})}\,.
\end{equation}

\paragraph{Solution to system I:} We first note that equation (\ref{exactinhom1}) implies that the combination $\Theta_\pm \equiv \Sigma_\pm^{(\mathrm{I})}+\mathcal{S}_1$ is closed, so that we can locally solve equation (\ref{exactinhom1}) in terms of a two-form potential
\begin{equation}
\Theta_\pm = \dd \chi_\pm\,,
\end{equation}
where $\chi_\pm$ is defined only in one coordinate patch. In terms of $\Theta_\pm$, equation (\ref{exactinhom2}) becomes
\begin{equation}\label{newexact}
\left(\s_6 \mp i \right) \Theta_\pm =  \mathcal{S}_2 + \left( \s_6 \mp i \right) \mathcal{S}_1 \equiv \mathcal{S}_\pm \,,
\end{equation}
where we defined the three-form $\mathcal{S}_\pm$ in the last line. To solve this equation we expand $\chi_\pm$ and $\mathcal{S}_\pm$ in harmonics and then equate the coefficients of the independent modes. Note that a three-form on $\mathcal{B}_5$ can always be dualized to give a two-form on $\mathcal{B}_5$.  Thus we have
\begin{equation}
 \mathcal{S}_\pm = \dd r \wedge \mathcal{T}_\pm + \star_5 \tilde{\mathcal{T}}_\pm,
\end{equation}
for  $\mathcal{T}_\pm$ and $\tilde{\mathcal{T}}_\pm$ two-forms on $\mathcal{B}_5$. Now from the definition of $\mathcal{S}_\pm$, equation (\ref{newexact}), we find that
\begin{equation} \label{esspm}
 \left( \s_6 \pm i \right) \mathcal{S}_\pm = 0\,,
\end{equation}
which gives $\tilde{\mathcal{T}}_\pm = \pm \, i r \mathcal{T}_\pm$, so that we get
\begin{equation}
 \mathcal{S}_\pm = \dd r \wedge \mathcal{T} \pm i \, r \star_5 \mathcal{T}_\pm.
\end{equation}
Thus, $\mathcal{S}_\pm$ has the harmonic expansion
\begin{equation} \label{Sexp}
 \mathcal{S}_\pm =
 \sum_{I_2} r \, \mathcal{S}^{I_2}_\pm \left( \frac{\dd r}{r} \wedge Y^{I_2} \pm i \star_5 Y^{I_2} \right) +
 \sum_{I_v} r \, \mathcal{S}^{I_v}_\pm \left( \frac{\dd r}{r} \wedge \dd Y^{I_v} \pm i \star_5 \dd Y^{I_v} \right).
\end{equation}
Just as for the potential $A_2$ of the previous subsection, we can choose a gauge in which $\chi_\pm$ has an expansion
\begin{equation}
 \chi_\pm = \sum_{I_2} \, \mathcal{\chi}^{I_2}_\pm (r) \, Y^{I_2} + \sum_{I_v} \, \mathcal{\chi}^{I_v}_\pm (r) \, \dd Y^{I_v}\,.
\end{equation}
Therefore
\begin{eqnarray}
 (\s_6 \mp i) \dd \chi_\pm
 & = & \mp i \Bigg\{ \sum_{I_2} (r \partial_r \chi_\pm^{I_2} \pm \lambda^{I_2} \chi_\pm^{I_2}) \left( \frac{\dd r}{r} \wedge Y^{I_2} \pm i \star_5 Y^{I_2} \right) \nonumber \\
 & & + \sum_{I_v} r \partial_r \chi_\pm^{I_v} \left( \frac{\dd r}{r} \wedge \dd Y^{I_v} \pm i \star_5 \dd Y^{I_v} \right) \Bigg\} \,.
\end{eqnarray}
Inserting this into equation (\ref{newexact}) we find the differential equations
\begin{eqnarray}
 \partial_r \mathcal{\chi}^{I_2}_\pm \pm \frac{\lambda^{I_2}}{r} \mathcal{\chi}^{I_2}_\pm & = & \pm \, i \mathcal{S}^{I_2}_\pm \,,\\
 \partial_r \mathcal{\chi}^{I_v}_\pm & = & \pm \, i \mathcal{S}^{I_v}_\pm \,,
\end{eqnarray}
with solutions
\begin{eqnarray} \label{chi2}
 \mathcal{\chi}^{I_2}_\pm (r) & = & \pm i \int_0^{\infty} dr' \, \vartheta(r-r') \, \left( {r \over r'} \right)^{\pm\lambda^{I_2}} \, \mathcal{S}^{I_2}_\pm (r') \,,\\
 \mathcal{\chi}^{I_v}_\pm (r) & = & \pm i \int_0^{\infty} dr' \, \vartheta(r-r') \, \mathcal{S}^{I_v}_\pm (r')\,.\label{chiv}
\end{eqnarray}
In writing down the above solutions we have introduced a modified step function $\vartheta$ suitable for non-localized sources $\mathcal{S}$ that takes care of the boundary behavior of the integrand in the IR and the UV:
\begin{equation} \label{modtheta}
 \vartheta(r-r') = \left\{ \begin{array}{cl} \theta(r-r') & \text{for integrands that go to zero at zero,} \\ - \theta(r'-r) & \text{for integrands that go to zero at infinity.} \end{array} \right. \,
\end{equation}
The orthonormality relations of \S\ref{orthprop} imply
\begin{eqnarray}
 {\pm} \, i \mathcal{S}_\pm^{I_2} \, \dd r & = & \int_{\mathcal{B}_5} \frac{\dd r}{r} \wedge 2 \, \bar{Y}_{I_2} \wedge \mathcal{S}_\pm\,, \label{coeff2} \\
 {\pm} \, i \mathcal{S}_\pm^{I_v} \, \dd r & = & \int_{\mathcal{B}_5} \frac{\dd r}{r} \wedge \lambda_{I_v}^{-1} \, \dd \bar{Y}_{I_v} \wedge \mathcal{S}_\pm \,. \label{coeffv}
\end{eqnarray}
Using this together with the solutions (\ref{chi2}) and (\ref{chiv}), we can write down the Green's function solution for $\Sigma_\pm^{(\mathrm{I})}$ in terms of the sources $\mathcal{S}_1$ and $\mathcal{S}_2$:
\begin{align}
 \Sigma^{(\mathrm{I})}_\pm & = \dd \chi_\pm - \mathcal{S}_1\,, \\
 \chi_\pm(y) & = \int_{\mathcal{C}_6} \mathcal{G}^{(\mathrm{I})}_\pm(y, y') \wedge \mathcal{S}_\pm(y')\,, \label{chi2solution} \\
 \mathcal{S}_\pm & =  \mathcal{S}_2 + \left( \star_6 \mp i \right) \mathcal{S}_1\,, \\
 \mathcal{G}^{(\mathrm{I})}_\pm(y,y') & =
 \sum_{I_2} Y^{I_2}(\Psi) \left[ \vartheta(r-r') \left( \frac{r'}{r} \right)^{\pm \lambda^{I_2}} \frac{\dd r'}{r'} \wedge 2 \, \bar{Y}^{I_2}(\Psi') \right] \nonumber \label{fluxgreens1} \\
  &\phantom{=}+
 \sum_{I_v} \dd Y^{I_v}(\Psi) \left[ \vartheta(r-r') \frac{\dd r'}{r'} \wedge \lambda_{I_v}^{-1} \, \dd \bar{Y}^{I_v}(\Psi') \right] \,.
 \end{align}
The index structures of the above equations are as follows:
\begin{align}
 \left( \chi_\pm(y) \right)_{ij} & = \frac{1}{3!} \int \dd^6 y' \sqrt{g'} \, \big( \mathcal{G}^{(\mathrm{I})}_\pm(y,y') \big)_{ij}^{\phantom{ij}mnp} \left( \star_6^{-1} \mathcal{S}_\pm(y') \right)_{mnp} \,, \\
 \big( \mathcal{G}^{(\mathrm{I})}_\pm(y,y') \big)_{ij,rkl} & =
 \sum_{I_2} Y^{I_2}_{ij}(\Psi) \vartheta(r-r') \left( \frac{r'}{r} \right)^{\pm \lambda^{I_2}} \frac{1}{r'} \, 2 \, \bar{Y}^{I_2}_{kl}(\Psi') \nonumber \\
 &\phantom{=}+
 \sum_{I_v} 2 \tilde{\nabla}_{[i} Y^{I_v}_{j]} (\Psi) \vartheta(r-r') \frac{1}{r'} \, \lambda_{I_v}^{-1} \, 2\tilde{\nabla}_{[k}\bar{Y}^{I_v}_{l]} (\Psi') \,,
\end{align}
where the full metric $g_{mn}$ is used to raise and lower the indices, and the modified theta function $\vartheta$ was introduced in equation (\ref{modtheta}).

\paragraph{Solution to system II:} We now solve the system (\ref{nonexinhom1}), (\ref{nonexinhom2}).  Equation (\ref{nonexinhom2}) tells us that
\begin{equation}
(\star_6\mp i)\Sigma_\pm^{(\mathrm{I})} = 0\,.
\end{equation}
The general solution to this equation is of the form of $S_\mp$ in equation (\ref{Sexp}), i.e.
\begin{equation}
 \Sigma_\pm^{(\mathrm{I})} = \sum_{I_2} \sigma^{I_2}_\pm (r) \left( \frac{\dd r}{r} \wedge Y^{I_2} \mp i \star_5 Y^{I_2} \right) + \sum_{I_v} \sigma^{I_v}_\pm (r) \left( \frac{\dd r}{r} \wedge \dd Y^{I_v} \mp i \star_5 \dd Y^{I_v} \right) \,. ~~~~
\end{equation}
A general four-form $\mathcal{S}_3$ can be expanded
\begin{eqnarray}
\mathcal{S}_3 & = & \sum_{I_2} \mathcal{S}_3^{I_2}(r)\, \dd r \wedge \star_5 Y^{I_2}+ \sum_{I_2} \mathcal{S}_3^{I_v} (r) \dd r\wedge \star_5 \dd Y^{I_v} \\
& \phantom{=} &+  \sum_{I_v} \tilde{\mathcal{S}}_3^{I_v}(r)\, \star_5 Y^{I_v}+ \sum_{I_s} \mathcal{S}_3^{I_s}(r)\, \star_5 \dd Y^{I_s}\,.\nonumber
\end{eqnarray}
Equation (\ref{nonexinhom1}) implies that $\mathcal{S}_3$ is closed. Upon imposing this, we find the constraints
\begin{align}
&\mathcal{S}_3^{I_s}=0\,,\\
&\mathcal{S}_3^{I_v} = {1\over \lambda^{I_v}}\,\partial_r\tilde{\mathcal{S}}_3^{I_v}\,.
\end{align}

Substituting these expansions into equation (\ref{nonexinhom1}) and collecting the coefficients of the independent harmonics, we find the radial equations
\begin{eqnarray}
 \partial_r \sigma_\pm^{I_2} \pm {\lambda^{I_2} \over r} \, \sigma_\pm^{I_2} & = & \pm \, i \mathcal{S}_3^{I_2} \,, \\
 \lambda^{I_v} \sigma_\pm^{I_v} & = & \pm \, i \tilde{\mathcal{S}}_3^{I_v}\,,
\end{eqnarray}
with solutions
\begin{eqnarray}
 \sigma_\pm^{I_2}(r) & = & \pm i \int_0^\infty \dd r'  \vartheta(r-r') \, \left({r\over r'}\right)^{\pm \lambda^{I_2}} \, \mathcal{S}_3^{I_2}(r')\,, \label{sigma2} \\
 \sigma_\pm^{I_v}(r) & = & \pm i \int_0^\infty \dd r' \delta(r-r') \, \lambda_{I_v}^{-1} \, \tilde{\mathcal{S}}_3^{I_v}(r') \label{sigmav} \,.
\end{eqnarray}
%where the modified step function $\vartheta(r-r')$ was introduced in equation (\ref{modtheta}).
Using the orthonormality properties in \S\ref{orthprop},
\begin{eqnarray}
 \int_{\mathcal{B}_5} 2 \, \bar{Y}_{I_2} \wedge \mathcal{S}_3 & = & \mathcal{S}_3^{I_2} \, \dd r \,, \label{coeff2second}\\
 \int_{\mathcal{B}_5} \dd r \wedge \bar{Y}_{I_v} \wedge \mathcal{S}_3 & = & \tilde{\mathcal{S}}^{I_v}_3 \, \dd r \,. \label{coeffvsecond}
\end{eqnarray}
We can now use the solutions (\ref{sigma2}) and (\ref{sigmav}) to write down the Green's function solution for $\Sigma_\pm^{(\mathrm{II})}$ in terms of the source $\mathcal{S}_3$:
\begin{align}
 \Sigma^{(\mathrm{II})}_\pm(y) & = \int_{\mathcal{C}_6} \mathcal{G}^{(\mathrm{II})}_\pm(y, y') \wedge \mathcal{S}_3(y')\,, \\
 \mathcal{G}^{(\mathrm{II})}_\pm(y, y') & =
 \sum_{I_2}  \left( \frac{\dd r}{r} \mp i \star_5 \right) \wedge Y^{I_2}(\Psi) \left[ \pm i \vartheta(r-r') \left( \frac{r'}{r} \right)^{\pm \lambda^{I_2}} 2 \, \bar{Y}_{I_2}(\Psi') \right] \nonumber \label{fluxgreens2}   \\
 + &
 \sum_{I_v} \left( \frac{\dd r}{r} \mp i \star_5 \right) \wedge \dd Y^{I_v}(\Psi) \Big[ {\pm} i \lambda_{I_v}^{-1} \delta(r-r') \, \dd r' \wedge \bar{Y}_{I_v}(\Psi') \Big] \,.
 \end{align}
%with $\vartheta$ given in equation (\ref{modtheta}).

\paragraph{Total Solution:}

% Liam
The total solution to the system (\ref{fluxinhom1}, \ref{fluxinhom2}) is just the sum of the pieces from each of the two steps:
%For convenience we collect the full Green's function solution here.
\begin{align}
 \Sigma_\pm & = \Sigma_\pm^{(\mathrm{I})} + \Sigma_\pm^{(\mathrm{II})} = \dd \chi_\pm - \mathcal{S}_1 + \int_{\mathcal{C}_6} \mathcal{G}^{(\mathrm{II})}_\pm \wedge\mathcal{S}_3 \,, \\
 \chi_\pm(y) & = \int_{\mathcal{C}_6} \mathcal{G}^{(\mathrm{I})}_\pm(y,y') \wedge \mathcal{S}_\pm(y') \,, \\
 \mathcal{S}_\pm & = \left( \star_6 \mp i \right) \mathcal{S}_1 + \mathcal{S}_2 \,,
\end{align} where the Green's functions $\mathcal{G}^{(\mathrm{I})}_\pm(y,y')$,  $\mathcal{G}^{(\mathrm{I})}_\pm(y,y')$ are given in equations (\ref{fluxgreens1}) and (\ref{fluxgreens2}), respectively.

\subsection{Metric solutions and Green's functions}
\label{sec:metricappend}
Now we wish to solve the equations % Liam
of motion for the metric perturbations $\delta g_{mn} \equiv h_{mn}$ on a general $(n+1)$-dimensional Calabi-Yau cone $\mathcal{C}_{n+1}$. The linearized Einstein equations take the form
\begin{equation}
\Delta_K h_{mn} = \mathcal{S}_{mn},
\end{equation}
where $\mathcal{S}_{mn}$ denotes source terms, and the kinetic operator $\Delta_K$ defined in (\ref{deltaK}) is constructed using the background metric $g_{mn}$.
%
%
% with action on $h_{mn}$ given by
%\begin{equation}
%\Delta_K h_{mn}\equiv \nabla^2h_{mn}+\nabla_m\nabla_n h_{p}^{~p} - 2
%\nabla^p\nabla_{(m}h_{n)p} \,.
%\end{equation}
The general solution takes the form
\begin{equation}
h_{mn}(y) = h^{\mathcal{H}}_{mn}(y) + \int \dd^{6}y' \sqrt{g} \, (\mathcal{G}_g)_{mn}^{\phantom{mn}m'n'}(y,y') \, \mathcal{S}_{m'n'}(y'),
\end{equation}
where $h^{\mathcal{H}}_{mn}$ is a homogenous solution (i.e., $\Delta_K h^{\mathcal{H}}_{mn} = 0$), and where $(\mathcal{G}_g)_{mn}^{\phantom{mn}m'n'}(y,y')$ denotes the metric Green's function. In \S\ref{sec:metricharmappend} we will solve for the homogeneous perturbations in terms of angular harmonics on $\mathcal{B}_n$.  In \S\ref{sec:metricgreensappend} we obtain the metric Green's function. To this end we separate the radial and angular variables in the operator $\Delta_K$:
\begin{align}
\label{ij}
 \Delta_K h_{ij} & = \left( \partial_r^2 + {n-4 \over r} \partial_r + {4\over r^2} \right) h_{ij}
 + {1\over r^2} \left(\tilde{\nabla}^{2} h_{ij} - 2\tilde{\nabla}^{\tilde{k}} \tilde{\nabla}_{(i}h_{j)k} \right) \nonumber \\
 &\phantom{=} -2 \left( \partial_r - {2-n \over r} \right) \tilde{\nabla}_{(i}h_{j)r} - {2 \over r} \tilde{g}_{ij} \tilde{\nabla}^{\tilde{k}} h_{kr} \nonumber \\
 &\phantom{=} + \left[ {1\over r^2} \tilde{\nabla}_i \tilde{\nabla}_j + \tilde{g}_{ij} {1 \over r} \left( \partial_r - {2 \over r} \right) \right] h_k^{\tilde{k}}
 + \left[ \tilde{\nabla}_i \tilde{\nabla}_j - \tilde{g}_{ij} r \left( \partial_r - {2-2n \over r} \right) \right] h_{rr} \, , \\ \nonumber \\
\label{ir}
 \Delta_K h_{ir} & = {2n-2 \over r^2} h_{ir} + {1 \over r^2} \left( \tilde{\nabla}^{2} h_{ir} - \tilde{\nabla}^{\tilde{k}}\tilde{\nabla}_{i} h_{kr} \right) \nonumber \\
 &\phantom{=} - {1 \over r^2} \left( \partial_r - {2 \over r} \right) \tilde{\nabla}^{\tilde{k}} h_{ik} + {1 \over r^2} \left( \partial_r - {2\over r} \right) \tilde{\nabla}_i h_k^{\tilde{k}} + {1-n \over r} \tilde{\nabla}_i h_{rr} \, , \\ \nonumber \\
 \Delta_K h_{rr} & = {1\over r^2} \left( \partial_r^2 - {2 \over r} \partial_r + {2 \over r^2} \right) h_k^{\tilde{k}} - \left( {n \over r} \partial_r - {1 \over r^2} \tilde{\nabla}^{2} \right) h_{rr} - {2 \over r^2} \partial_r \tilde{\nabla}^{\tilde{k}} h_{kr} \, . \label{rr}
\end{align}

Throughout this work we impose a \emph{transverse} gauge on the metric perturbations:
\begin{eqnarray}
\tilde{\nabla}^{\tilde{k}}h_{\{ik\}} &=& 0\label{gc1}\,,\\
\tilde{\nabla}^{\tilde{k}}h_{kr}&=&0\label{gc2}\,.
\end{eqnarray}
This gauge condition projects out the longitudinal harmonics
$\tilde{\nabla}_{\{i}\tilde{\nabla}_{j\}}Y^{I_s}$,
$\tilde{\nabla}_{\{i}Y^{I_v}_{j\}}$, and $\tilde{\nabla}_iY^{I_s}$,
and we get the following harmonic expansions
\begin{eqnarray}
h_{\{ij\}} & = & \sum_{I_t}\,\phi^{I_t}(r)Y^{I_t}_{\{ij\}}(\Psi)\label{texp}\,,\\
h_{ir} & = & \sum_{I_v}\,b^{I_v}(r)Y^{I_v}_{i}(\Psi)\,,\\
h_k^{\tilde{k}} & = & \sum_{I_s}\,\pi^{I_s}(r)Y^{I_s}(\Psi)\,,\\
h_{rr} & = & \sum_{I_s}\,\mathfrak{r}^{I_s}(r)Y^{I_s}(\Psi)\label{rexp}\,.
\end{eqnarray}
By expanding the gauge parameter $\xi_m$ in angular harmonics, one can easily show that there always exists $\xi_m$ such that the gauge (\ref{gc1}, \ref{gc2}) is attainable via
\begin{equation}\label{gtransf}
h_{mn} \longrightarrow h_{mn} + 2\nabla_{(m}\xi_{n)}\,.
\end{equation}
There is, however, a residual gauge freedom.  The gauge conditions (\ref{gc1}, \ref{gc2}) are preserved under (\ref{gtransf})
if
\begin{eqnarray}
\tilde{\nabla}^{\tilde{k}}\nabla_{\{k}\xi_{i\}} = 0\,,\\
\tilde{\nabla}^{\tilde{k}}\nabla_{(k}\xi_{r)}=0\,.
\end{eqnarray}
The most general form for $\xi$ is then
\begin{eqnarray}
\xi_i & = & \sum_{K_v} \Lambda^{K_v}(r) \,Y_i^{K_v}(\Psi)\,,\\
\xi_r&=&\epsilon(r)\,,
\end{eqnarray}
%\begin{eqnarray}
%\xi_i &=& \sum_{I_v,~\mathrm{Killing}}\Lambda^{I_v,~\mathrm{Killing}}(r)\,Y_i^{I_v,~\mathrm{Killing}}(\Psi)\,,\\
%\xi_r&=&\epsilon(r)\,,
%\end{eqnarray}
where the $Y_i^{K_v}(\Psi)$ are the Killing vectors on $\mathcal{B}_n$ with $\lambda^{K_v} = 2 (n-1)$.  The radial fields then transform as
\begin{eqnarray}
 \phi^{I_t} & \longrightarrow & \phi^{I_t} \, , \\
 b^{K_v} & \longrightarrow & b^{K_v} + \left( \partial_r - {2 \over r} \right) \Lambda^{K_v} \, , \\
 \pi_0 & \longrightarrow & \pi_0 + n r \epsilon \, , \\
 \mathfrak{r}_0 & \longrightarrow & \mathfrak{r}_0 + \partial_r \epsilon \, ,
\end{eqnarray}
where $\pi_0$, $\mathfrak{r}_0$ are zero modes, i.e.\ correspond to harmonics with $\lambda^{I_s} = 0$.
%Note  that once we compactify on $\mathcal{B}_5$ we have only the radial direction as a noncompact coordinate and are left with 1-dimensional General Relativity, $\mathfrak{r}_0$ being the metric, and 1-dimensional Yang-Mills, $b^{K_v}$ being the gauge field.  Then $\epsilon$ and $\Lambda^{K_v}$ parameterize the usual diffeomorphisms and gauge transformations.  Note further that 1-dimension General Relativity and Yang-Mills are trivial---they are pure gauge.
We will find it convenient to use the residual gauge symmetry to impose $\pi_0 = 0$ and $b^{K_v} = 0$, i.e.\ we set
\begin{eqnarray}
\int\dd^{n}\Psi\,\sqrt{\tilde{g}}\, \bar{Y}^{\tilde{k}}_{K_v}(\Psi)\,h_{kr}(r,\Psi)& = & 0\,, \\
\int\dd^{n}\Psi\,\sqrt{\tilde{g}}\,h^{\tilde{k}}_k(r,\Psi) &=& 0\label{metricgauge4}\,.
\end{eqnarray}

\subsubsection{Homogeneous metric perturbations}
\label{sec:metricharmappend}
%Now for the harmonic solution.
Using the expansions (\ref{texp}--\ref{rexp}) and the separation (\ref{ij}), and then collecting the coefficients of independent harmonics, the homogeneous equation
\begin{equation}
 \Delta_K h_{ij} =0
\end{equation}
gives the radial equations
\begin{eqnarray}
 \left(\partial_r^2 + {n - 4 \over r} \partial_r + {4 - \lambda^{I_t} \over r^2} \right) \phi^{I_t} \, Y^{I_t}_{\{ij\}}
 & = & 0 \,, \label{ijt} \\ \nonumber \\
 -2\left( \partial_r + {2-n \over r} \right) b^{I_v} \, \tilde{\nabla}_{\{i} Y^{I_v}_{j\}}
 & = & 0 \,, ~~~~ \lambda^{I_v} \neq \lambda^{K_v} \label{ijv} \\ \nonumber \\
 \left(\frac{1}{r^2} {n-2 \over n} \pi^{I_s}+ \mathfrak{r}^{I_s} \right) \, \tilde{\nabla}_{\{i} \tilde{\nabla}_{j\}} Y^{I_s}
 & = & 0 \,, ~~~~ \lambda^{I_s} \neq 0, n \label{ijs} \\ \nonumber \\
 \Bigg[ \frac{1}{n} \left( \partial^2_r + {2n-4 \over r} \partial_r - {2n-4 \over r^2} - \frac{\lambda^{I_s}}{r^2} {2n-2 \over n} \right) \pi^{I_s}
 & - & \left( r\partial_r - (2-2n) + \frac{\lambda^{I_s}}{n} \right) \mathfrak{r}^{I_s} \Bigg] \, \tilde{g}_{ij} Y^{I_s} \nonumber \\
 & = & 0 \,. \label{ijtr}
\end{eqnarray}
Note that equations (\ref{ijv}) and (\ref{ijs}) should not be applied for values of the quantum numbers $I_v$ and $I_s$, respectively, for which the corresponding harmonics vanish identically, hence the restrictions listed. In a similar way,
 \begin{equation}
 \Delta_K h_{ir} = 0
 \end{equation}
gives
\begin{eqnarray}
 {1\over r^2} \left(2(n-1) - \lambda^{I_v} \right) b^{I_v} \, Y^{I_v}_i & = & 0 \,, \quad \lambda^{I_v} \neq \lambda^{K_v} = 2(n-1) \label{irv}\,, \\
 \left({1\over r^2} \left(\partial_r - {2\over r} \right) \pi^{I_s} + {1-n \over r} \mathfrak{r}^{I_s} \right) \,\tilde{\nabla}_i Y^{I_s} & = & 0\,, \quad \lambda^{I_s} \neq 0\,,  \label{irs}
\end{eqnarray}
and
\begin{equation}
 \Delta_K h_{rr}=0
\end{equation}
gives
\begin{equation}
 \left({1\over r^2} \left( \partial_r^2 - {2\over r} \partial_r + {2\over r^2} \right) \pi^{I_s} - {1\over r} \left(n \partial_r+ {\lambda^{I_s} \over r} \right) \mathfrak{r}^{I_s} \right) \, Y^{I_s} = 0 \label{rrs}\,.
\end{equation}

\paragraph{Solutions for $\pi$, $\mathfrak{r}\,$:}

\subparagraph{$\lambda^{I_s} \neq 0\,$:}  In this case we have four (three if $\lambda=n$) independent equations (\ref{ijs}, \ref{ijtr}, \ref{irs}, \ref{rrs}) for the two unknowns, $\pi^{I_s}$, $\mathfrak{r}^{I_s}$.  Thus the only solutions are
\begin{equation}
 \left. \begin{array}{ccc} \pi^{I_s}(r) & = & 0 \\ \mathfrak{r}^{I_s}(r) & = & 0  \end{array} \right\} ~~~~ \mathrm{if} ~~ \lambda^{I_s} \neq 0\,.
\end{equation}

\subparagraph{$\lambda^{I_s} = 0\,$:}  Now we have only two equations --- (\ref{ijtr}) and (\ref{rrs}).  We can nevertheless use the residual gauge freedom to set $\pi_0=0$.  Equations (\ref{ijtr}) and (\ref{rrs}) then give $\mathfrak{r}_0 = 0$:
\begin{equation}
 \left. \begin{array}{ccc} \pi_0(r) & = & 0 \\ \mathfrak{r}_0(r) & = & 0  \end{array} \right\}~~~~\mathrm{gauge}~\mathrm{choice}\,.
\end{equation}

\paragraph{Solutions for $b^{I_v}\,$:}

\subparagraph{$\lambda^{I_v} \neq \lambda^{K_v}\,$:}  Equation (\ref{irv}) immediately gives
\begin{equation}
b^{I_v}(r) =0,~~~~\lambda^{I_v}\neq \lambda^{K_v}\,.
\end{equation}

\subparagraph{$\lambda^{I_v} = \lambda^{K_v}\,$:}  We can use the residual gauge symmetry to eliminate the
Killing modes
\begin{equation}
b^{K_v}(r) =0,~~~~\mathrm{gauge}~\mathrm{choice}\,.
\end{equation}

\paragraph{Solution for $\phi^{I_t}\,$:}

The only nontrivial degrees of freedom in the homogeneous case are then the $\phi^{I_t}$, obeying equation (\ref{ijt}).  The two independent solutions are
\begin{equation}\label{apm}
\phi^{I_t}_{\pm}(r) = r^{a_{\pm}(I_t)}\,,~~~~a_{\pm}(I_t) = \frac{1}{2} \left((5-n) \pm \sqrt{4\lambda^{I_t}+(n-1)(n-9)} \right)\,.
\end{equation}
To summarize, the homogeneous solution is given by
\begin{equation}
 h^{\mathcal{H}}_{\{ij\}}(y) = \sum_{I_t}\Bigl(h_+^{I_t}\,r^{a_+(I_t)}+h_-^{I_t}\,r^{a_-(I_t)}\Bigr)\,Y^{I_t}_{\{ij\}}(\Psi) \, , \label{hommetricsol}
\end{equation}
with all other components vanishing, where $h_\pm^{I_t}$ are constants of integration and the $a_\pm(I_t)$ are given by
\begin{equation}\label{apm2}
a_{\pm}(I_t) = \frac{1}{2} \left( (5-n) \pm \sqrt{4\lambda^{I_t} + (n-1)(n-9)} \right).
\end{equation}

\subsubsection{Metric Green's function}
\label{sec:metricgreensappend}

Now we wish to solve
\begin{equation}\label{inhom}
\Delta_K h_{mn} = \mathcal{S}_{mn}\,.
\end{equation}
We continue to impose the same gauge conditions as in the previous subsection, i.e.\ the transverse conditions (\ref{gc1}, \ref{gc2}) as well as the conditions $\pi_0 = 0$ and $b^{K_v}=0$.  The symmetric tensor $S_{mn}$ can in general be expanded as
\begin{eqnarray}
\mathcal{S}_{\{ij\}} & =& \sum_{I_t}\,\mathcal{S}_t^{I_t}(r)\,Y^{I_t}_{\{ij\}}(\Psi)+\sum_{I_v}\,\mathcal{S}_t^{I_v}(r)\,\tilde{\nabla}_{\{i}Y^{I_v}_{j\}}(\Psi)+\sum_{I_s}\,\mathcal{S}_t^{I_s}(r)\,\tilde{\nabla}_{\{i}\tilde{\nabla}_{j\}}Y^{I_s}(\Psi)\label{stexp}\,, ~~~~~~\\
\mathcal{S}_{ir}& =& \sum_{I_v}\,\mathcal{S}_v^{I_v}(r)\,Y^{I_v}_{i}(\Psi)+\sum_{I_s}\mathcal{S}_v^{I_s}(r)\,\tilde{\nabla}_iY^{I_s}(\Psi)\,,\\
\mathcal{S}_k^{\tilde{k}}& =& \sum_{I_s}\,\mathcal{S}_{\mathrm{tr}}^{I_s}(r)Y^{I_s}(\Psi)\,,\\
\mathcal{S}_{rr}& =& \sum_{I_s}\,\mathcal{S}_s^{I_s}(r)Y^{I_s}(\Psi)\label{srexp}\,.
\end{eqnarray}
In the above, the subscripts $t$, $v$, $\mathrm{tr}$, $s$ are used merely to distinguish the various radial functions and should not be interpreted as indices.

We will proceed similarly to the previous section.  We will substitute the expansions for $h_{mn}$ (\ref{texp}--\ref{rexp}) and the expansions for $\mathcal{S}_{mn}$ (\ref{stexp}--\ref{srexp}) into the metric equation of motion (\ref{inhom}) and make use of the decomposition of the operator $\Delta_K$ given in (\ref{ij}--\ref{rr}).  We pick out the coefficient of each independent harmonic to obtain a set of radial equations.\footnote{A subset of these radial equations represent constraints on the source $\mathcal{S}_{mn}$. These constraints must be satisfied in order for  the solution derived below  to be valid, but we will not present the explicit form of the constraints here: we assume that the constraints are automatically obeyed when the stress tensor is well-behaved.}

From the equation $\Delta_K h_{ij}=\mathcal{S}_{ij}$ one obtains the radial equations
\begin{eqnarray}
 \left(\partial_r^2 + {n-4 \over r} \partial_r + {4 - \lambda^{I_t} \over r^2} \right) \phi^{I_t} \, Y^{I_t}_{\{ij\}}
 & = & \mathcal{S}_t^{I_t} \, Y^{I_t}_{\{ij\}} \,, \label{ihijt} \\ \nonumber \\
 -2\left( \partial_r - {2 - n \over r} \right) b^{I_v} \, \tilde{\nabla}_{\{i} Y^{I_v}_{j\}}
 & = & \mathcal{S}_t^{I_v} \, \tilde{\nabla}_{\{i} Y^{I_v}_{j\}} \,, ~~~~ \lambda^{I_v} \neq \lambda^{K_v} \label{ihijv} \\ \nonumber \\
 \left(\frac{1}{r^2} {n-2 \over n} \pi^{I_s} + \mathfrak{r}^{I_s} \right) \, \tilde{\nabla}_{\{i} \tilde{\nabla}_{j\}} Y^{I_s}
 & = & \mathcal{S}_t^{I_s} \, \tilde{\nabla}_{\{i} \tilde{\nabla}_{j\}} Y^{I_s} \,, ~~~~ \lambda^{I_s} \neq 0, n \label{ihijs} \\ \nonumber \\
 \Bigg[ \frac{1}{n} \left( \partial^2_r + {2n-4 \over r} \partial_r - {2n-4 \over r^2} - \frac{\lambda^{I_s}}{r^2} \frac{2n-2}{n} \right) \pi^{I_s}
 & - & \left( r\partial_r - (2-2n) + \frac{\lambda^{I_s}}{n}\right) \mathfrak{r}^{I_s} \Bigg] \, \tilde{g}_{ij} Y^{I_s} \nonumber \\
 & = & \frac{1}{n} \mathcal{S}_{\mathrm{tr}}^{I_s} \, \tilde{g}_{ij} Y^{I_s} \,. \label{ihijtr}
\end{eqnarray}
From the equation $\Delta_K h_{ir}=\mathcal{S}_{ir}$ we get
\begin{eqnarray}
 {1\over r^2} \left(2(n-1) - \lambda^{I_v} \right) b^{I_v} \, Y^{I_v}_i & = & \mathcal{S}_v^{I_v} \,Y^{I_v}_i \,, \quad \lambda^{I_v} \neq \lambda^{K_v} = 2(n-1) \label{ihirv} \,, \\
 \left({1\over r^2} \left(\partial_r - {2\over r} \right) \pi^{I_s} + {1-n \over r} \mathfrak{r}^{I_s} \right) \,\tilde{\nabla}_i Y^{I_s} & = & \mathcal{S}_v^{I_s} \, \tilde{\nabla}_i Y^{I_s}\,, \quad \lambda^{I_s} \neq 0 \label{ihirs}
\end{eqnarray}
and from $\Delta_K h_{rr}=\mathcal{S}_{rr}$ we get
\begin{equation} \label{ihrrs}
 \left({1\over r^2} \left( \partial_r^2 - {2\over r} \partial_r + {2\over r^2} \right) \pi^{I_s} - {1\over r} \left(n \partial_r+ {\lambda^{I_s} \over r} \right) \mathfrak{r}^{I_s} \right) \, Y^{I_s} = \mathcal{S}_s^{I_s} \, Y^{I_s} \,.
\end{equation}

\paragraph{Solutions for $\pi$, $\mathfrak{r}\,$:}
\subparagraph{$\lambda^{I_s} = 0\,$:} Since we have fixed to a gauge where $\pi_0=0$, equations (\ref{ihijtr}, \ref{ihrrs}) give
 \begin{equation}
 \mathfrak{r}_0 = {r^2\,\mathcal{S}_{s}^0- \mathcal{S}_{\mathrm{tr}}^0 \over 2n(n-1)}\,.
 \end{equation}

\subparagraph{$\lambda^{I_s} \neq 0\,$:} Equations (\ref{ihijs}, \ref{ihirs}) give
 \begin{equation}
 \left(\partial_r + \frac{1}{r} {n^2 - 5n + 2 \over n}\right)\,\pi^{I_s} = r^2\,\mathcal{S}_v^{I_s} + (n-1) r \mathcal{S}_t^{I_s}\,.
 \end{equation}
The regular solution to this equation is given by
\begin{equation}
\pi^I_s = \int_0^\infty \dd r' \vartheta(r-r')\Big({r'\over r}\Big)^{n^2 - 5n + 2 \over n} \Bigl(r'^2\,\mathcal{S}_v^{I_s}(r') + (n-1) r' \mathcal{S}_t^{I_s}(r') \Bigr)\,,
\end{equation}
where $\vartheta$ was introduced in equation (\ref{modtheta}). Equation (\ref{ihijs}) then gives the solution for $\mathfrak{r}^{I_s}$,
\begin{equation}
 \mathfrak{r}^{I_s} = \mathcal{S}_t^{I_s} - \frac{1}{r^2} \frac{n-2}{n} \pi^{I_s} \,.
\end{equation}

\paragraph{Solution for $b^{I_v}\,$:}

\subparagraph{$\lambda^{I_v}\neq \lambda^{K_v} = 2(n-1)\,$:}  Equation (\ref{ihirv}) gives
\begin{equation}
b^{I_v} = {r^2 \over 2(n-1) -\lambda^{I_v}} \mathcal{S}_v^{I_v} \,.
\end{equation}

\subparagraph{$\lambda^{I_v} = \lambda^{K_v}\,$:} We take $b^{K_v} = 0$ by gauge choice.

\paragraph{Solution for $\phi^{I_t}\,$:}
Solving (\ref{ihijt}) is practically identical to solving the scalar Poisson % Liam's
equation (\ref{poisson}). Thus we start by considering sources of the form
\begin{equation}
 \mathcal{S}_t^{I_t}(r) = \mathcal{S}_t^{I_t}(\alpha,m) \, r^\alpha \, (\ln r)^m \,,
\end{equation}
with $\mathcal{S}_{t}^{I_t}(\alpha,m)=\text{const.}$, and then generalize to a collection of such sources. For sources with $\alpha \neq -2 + a_\pm$, the solution to equation (\ref{ihijt}) is
\begin{equation}
 \phi^{I_t}(r ; \alpha, m) = \mathcal{S}_t^{I_t} (\alpha, m) \, r^{\alpha+2} \, \big( c_0 + c_1 \ln r + \ldots + c_m (\ln r)^m \big) \,,
\end{equation}
where the coefficients $c_k$ are given by
\begin{equation}
 c_k = (-1)^{m-k} \frac{m!/k!}{a_+ - a_-} \left[(\alpha + 2 - a_+)^{k-1-m} - (\alpha+2-a_-)^{k-1-m} \right] \,, \quad \alpha \neq -2 + a_\pm \,,
\end{equation}
while for sources with $\alpha = -2 + a_\pm$ the solution reads
\begin{equation}
 \phi^{I_t}(r ; \alpha, m) = \mathcal{S}_t^{I_t} (\alpha, m) \, r^{\alpha+2} \, \big( d_0 + d_1 \ln r + \ldots + d_{m+1} (\ln r)^{m+1} \big) \,,
\end{equation}
where the coefficients $d_k$ are given by
\begin{equation}
 d_k = (-1)^{m-k-1} \frac{m!}{k!} (\pm a_+ \mp a_-)^{k-1-m}\,, \quad \alpha = -2 + a_\pm \,.
\end{equation}
For the general case
\begin{equation}
 \mathcal{S}_t^{I_t}(r) \sum_{\alpha,m} \mathcal{S}_t^{I_t}(r;\alpha,m) \, r^\alpha \, (\ln r)^m
\end{equation}
we get a solution
\begin{equation}
 \phi^{I_t}(r) = \sum_{\alpha,m} \phi^{I_t} (r;\alpha,m) \,.
\end{equation}
In this way $\phi^{I_t}$ becomes a function of the source $\mathcal{S}_t^{I_t}$, and we write the solution formally in terms of a Green's function $G^{I_t}$ which we define by
\begin{equation}
 \phi^{I_t}[\mathcal{S}_{t}^{I_t}](r) = \sum_{\alpha,m} \phi^{I_t}[\mathcal{S}_t^{I_t}] (r;\alpha,m) \equiv \int_0^\infty \dd r' G^{I_t}(r,r') \mathcal{S}_t^{I_t}(r) \,.
\end{equation}

\paragraph{Summary:}

In the gauge given by
\begin{eqnarray}
\tilde{\nabla}^{\tilde{k}}h_{\{ik\}} = \tilde{\nabla}^{\tilde{k}}h_{kr} &=& 0\label{metricgauge1}\,,\\
%\tilde{\nabla}^{\tilde{k}}h_{kr}&=&0\,,\\
\int\dd^{n}\Psi\,\sqrt{\tilde{g}}\, \bar{Y}^{i}_{K_v}(\Psi)\,h_{ir}(r,\Psi)& = & 0\,, \\
\int\dd^{n}\Psi\,\sqrt{\tilde{g}}\,h^{\tilde{k}}_k(r,\Psi) &=& 0\label{metricgauge4}\,,
\end{eqnarray}
the general solution to (\ref{inhom}) is
\begin{align}
 h_{ij}(y) & = h^{\mathcal{H}}_{ij}(y) + \int \dd^{n+1} y' \sqrt{g} \, \left( (\mathcal{G}_g)_{ij}^{\phantom{ij}{i'j'}}(y,y') \, S_{i'j'}(y') + 2 (\mathcal{G}_g)^{\phantom{ij}i'r}(y,y')\,S_{i'r}(y') \right) \, , \label{ijsoln} \\
 h_{ir}(y) & = \int \dd^{n+1} y'\sqrt{g} \, 2(\mathcal{G}_g)_{ir}^{\phantom{ir}i'r}(y,y')\,S_{i'r}(y') \, , \\
 h_{rr}(y) & = \int \dd^{n+1} y' \sqrt{g} \, \Big( (\mathcal{G}_g)_{rr}^{\phantom{rr}rr}(y,y') \, S_{rr}(y') + 2 (\mathcal{G}_g)_{rr}^{\phantom{rr}i'r}(y;y') \, S_{i'r}(y') \nonumber \\
 & \phantom{=}+ (\mathcal{G}_g)_{rr}^{\phantom{rr}i'j'}(y,y')\,S_{i'j'}(y') \Big) \, .
\end{align}

The nonzero components of the metric Green's function $(\mathcal{G}_g)_{mn}^{\phantom{mn}m'n'}(y;y')$ are given by
{\allowdisplaybreaks
% Liam canceled \frac{n-1}{n-1} in first equation below
\begin{align} \label{MetricGreens1}
(\mathcal{G}_g)_{ij}^{\phantom{ij}i'j'}(y,y') & =  (r')^{-n} \times \Bigg[ \sum_{I_t} G^{I_t}(r;r') \, Y^{I_t}_{\{ij\}}(\Psi) \, \bar{Y}_{I_t}^{\{i'j'\}}(\Psi')  + \sum_{\lambda^{I_s}>n} \vartheta(r-r') \left( \frac{r'}{r} \right)^{\frac{n^2-5n+2}{n}}  \nonumber \\
  & \times   r' \left( \frac{\lambda^{I_s}}{n} (\lambda^{I_s} - n) \right)^{-1} \left(\frac{1}{n} \tilde{g}_{ij}(\Psi) Y^{I_s}(\Psi) \right) \, \tilde{\nabla}^{\{i'} \tilde{\nabla}^{j'\}} \bar{Y}^{I_s}(\Psi') \Bigg]\,, \\ \nonumber \\
%
% (\mathcal{G}_g)_{ij}^{\phantom{ij}i'j'}(y,y') & =  (r')^{-n} \times \Bigg[ \sum_{I_t} G^{I_t}(r;r') \, Y^{I_t}_{\{ij\}}(\Psi) \, \bar{Y}_{I_t}^{\{i'j'\}}(\Psi') \nonumber \\
%  & + \sum_{\lambda^{I_s}>n} \vartheta(r-r') \left( \frac{r'}{r} \right)^{\frac{n^2-5n+2}{n}} \times (n-1)r' \nonumber \\
%  & \times \left( \frac{n-1}{n} \lambda^{I_s} (\lambda^{I_s} - n) \right)^{-1} \left(\frac{1}{n} \tilde{g}_{ij}(\Psi) Y^{I_s}(\Psi) \right) \, \tilde{\nabla}^{\{i'} \tilde{\nabla}^{j'\}} \bar{Y}^{I_s}(\Psi') \Bigg]\,, \\ \nonumber \\
%%
 2(\mathcal{G}_g)_{ij}^{\phantom{ij}i'r}(y,y') & = (r')^{2-n} \times \Bigg[ \sum_{\lambda^{I_s}>n} \vartheta(r-r') \left( \frac{r'}{r} \right)^{\frac{n^2-5n+2}{n}}
  \left(\lambda^{I_s}\right)^{-1} \left(\frac{1}{n} \tilde{g}_{ij}(\Psi) Y^{I_s}(\Psi) \right) \,\tilde{\nabla}^{i'} \bar{Y}^{I_s}(\Psi') \Bigg]\,, \\ \nonumber \\
 2(\mathcal{G}_g)_{ir}^{\phantom{ir}i'r}(y,y') & = (r')^{-n} \times \sum_{\lambda^{I_v}>2(n-1)} \delta(r-r') \times {r'^2\over 2(n-1)-\lambda^{I_v}} \times Y^{I_v}_i(\Psi)\, \bar{Y}^{i'}_{I_v}(\Psi')\,, \\ \nonumber \\
 (\mathcal{G}_g)_{rr}^{\phantom{rr}i'j'}(y,y') & = (r')^{-n} \times \Bigg[ \sum_{\lambda^{I_s}>n} \left( \delta(r-r') + \vartheta(r-r') \left( \frac{r'}{r} \right)^{\frac{n^2-5n+2}{n}} \times (n-1)r' \left(-\frac{1}{r} \frac{n-2}{n} \right) \right) \nonumber \\
& \times \left( \frac{n-1}{n} \lambda^{I_s} (\lambda^{I_s} - n) \right)^{-1} Y^{I_s}(\Psi) \, \tilde{\nabla}^{\{i'} \tilde{\nabla}^{j'\}} \bar{Y}^{I_s}(\Psi') \nonumber \\
& + \delta(r-r') \times {-1 \over 2n(n-1)} \times Y^{\lambda^{I_s}=0}(\Psi) \, \tilde{g}^{i'j'}(\Psi') \bar{Y}_{\lambda^{I_s}=0}(\Psi') \Bigg]\,, \\ \nonumber \\
 (\mathcal{G}_g)_{rr}^{\phantom{rr}rr}(y,y') & = (r')^{-n} \times \delta(r-r') \times {r'^2 \over 2n(n-1)} \times Y^{\lambda^{I_s}=0}(\Psi)\, \bar{Y}_{\lambda^{I_s}=0}(\Psi') \,.\label{MetricGreens6}
\end{align}}

%%%%%%%%%%%%%%%%%%%%%%%%%%%%%%%%%%%%%%%%%%%%%%%%%%%%%%%%%%%%%%%%%%%%%%

\end{document}